\documentclass[superscriptaddress,twocolumn,prd,preprintnumbers,amsmath,amssymb,nofootinbib]{revtex4}

\usepackage{graphicx}
\usepackage{dcolumn}
\usepackage{bm}
\usepackage{epsfig}
\usepackage{amssymb}
\usepackage{amsmath}
\usepackage{subfigure}

\newcommand{\lsim}{\mathrel{\hbox{\rlap{\lower.55ex\hbox{$\sim$}} \kern-.3em \raise.4ex \hbox{$<$}}}}
\newcommand{\gsim}{\mathrel{\hbox{\rlap{\lower.55ex\hbox{$\sim$}} \kern-.3em \raise.4ex \hbox{$>$}}}}
\newcommand{\beq}{\begin{equation}}
\newcommand{\eeq}{\end{equation}}
\newcommand{\beqa}{\begin{eqnarray}}
\newcommand{\eeqa}{\end{eqnarray}}

\newcommand{\Mpl}{M_\mathrm{Pl}}

\newcommand{\gtil}{\tilde{g}}
\newcommand{\rhoR}{\rho_{*R}}
\newcommand{\initT}{T_{J,i}}
\newcommand{\phimin}{\phi_\mathrm{min}}
\newcommand{\vel}{\dot{\phi}_M}
\newcommand{\phibar}{\bar{\phi}}
\newcommand{\omk}{\omega_k}
\newcommand{\ak}{\alpha_k}
\newcommand{\bk}{\beta_k}
\newcommand{\kp}{k_\mathrm{peak}}
\newcommand{\calD}{{\cal D}}
\newcommand{\kIR}{k_\mathrm{IR}}
\newcommand{\phita}{\bar{\phi}_\mathrm{ta}}
\newcommand{\kex}{k_\mathrm{ex}}
\newcommand{\ph}{\mathrm{phys}}
\newcommand{\Mp}{M_v}

\begin{document}

\title{Chameleons in the Early Universe: Kicks, Rebounds, and Particle Production} 

\author{Adrienne L. Erickcek}\email{erickcek@physics.unc.edu}
\affiliation{Department of Physics and Astronomy, University of North Carolina at Chapel Hill, Phillips Hall CB 3255, Chapel Hill, NC 27599 USA}
\author{Neil Barnaby}\email{n.barnaby@damtp.cam.ac.uk}
\affiliation{
DAMTP, Cambridge University, Wilberforce Road, Cambridge, CB3 0WA, United Kingdom}
\author{Clare Burrage}\email{Clare.Burrage@nottingham.ac.uk}
\affiliation{School of Physics and Astronomy, University of Nottingham, Nottingham, NG7 2RD, UK}
\author{Zhiqi Huang}\email{zqhuang@cita.utoronto.ca}
\affiliation{CITA, University of Toronto, 60 St.~George Street, Toronto, Ontario M5S 3H8, Canada}

\begin{abstract}
Chameleon gravity is a scalar-tensor theory that includes a non-minimal coupling between the scalar field and the matter fields and yet mimics general relativity in the Solar System.   The scalar degree of freedom is hidden in high-density environments because the effective mass of the chameleon scalar depends on the trace of the stress-energy tensor.   In the early Universe, when the trace of the matter stress-energy tensor is nearly zero, the chameleon is very light, and Hubble friction prevents it from reaching the minimum of its effective potential.  Whenever a particle species becomes non-relativistic, however, the trace of the stress-energy tensor is temporarily nonzero, and the chameleon begins to roll.  We show that these ``kicks" to the chameleon field have catastrophic consequences for chameleon gravity.  The velocity imparted to the chameleon by the kick is sufficiently large that the chameleon's mass changes rapidly as it slides past its potential minimum.  This nonadiabatic evolution shatters the chameleon field by generating extremely high-energy perturbations through quantum particle production.  If the chameleon's coupling to matter is slightly stronger than gravitational, the excited modes have trans-Planckian momenta.  The production of modes with momenta exceeding $10^{7}\,\mathrm{GeV}$ can only be avoided for small couplings and finely tuned initial conditions.  These quantum effects also significantly alter the background evolution of the chameleon field, and we develop new analytic and numerical techniques to treat quantum particle production in the regime of strong dissipation.  This analysis demonstrates that chameleon gravity cannot be treated as a classical field theory at the time of Big Bang Nucleosynthesis and casts doubt on chameleon gravity's viability as an alternative to general relativity.
\end{abstract}

\maketitle
\section{Introduction}

Light scalar fields are of great interest in cosmology because they arise in many explanations for the current acceleration of the expansion of the Universe \cite{Wetterich:1994bg, Zlatev:1998tr, Amendola:1999er, Caldwell:2009ix,Copeland:2006wr}.  It is challenging for these models to evade the stringent experimental limits on fifth forces within the Solar System and the laboratory, however, because no theory has been constructed that both explains current cosmological observations and forbids interactions between the scalar field and Standard-Model particles. 
Axionic quintessence models come the closest, because they possess a shift symmetry that forbids a direct coupling to the stress-energy tensor of Standard-Model particles, but they still interact with photons \cite{Carroll:1998zi}.  In all other cases, we must reconcile ourselves to a coupling between matter and the scalar field; unless such a coupling is forbidden, we must include it in our theory as it will be generated by quantum effects.  Problematically, the presence of a new light scalar field coupled to matter usually implies the existence of a new long-range fifth force, and no new forces have been seen in either laboratory experiments or Solar-System tests of general relativity.  The precision of these experiments constrains the strength of any new force to be many orders of magnitude weaker than gravity \cite{Adelberger:2009zz}.  In a simple Yukawa model, this constraint forces the energy scale that controls the strength of the coupling between the scalar field and matter to be many orders of magnitude above the Planck scale.  Such a large energy scale is almost impossible to justify in any reasonable effective field theory.

In 2003 Khoury and Weltman proposed chameleon gravity, which contains a scalar degree of freedom whose potential can provide the vacuum energy required for cosmic acceleration \cite{Khoury:2003aq,Khoury:2003rn}.  The chameleon scalar field is a light field that interacts with the matter fields of the Standard Model, but it possesses a dynamical mechanism to hide these interactions in dense environments.  The chameleon's potential function contains non-linear terms that, when combined with the chameleon's coupling to matter fields, make the chameleon's effective mass dependent on its environment.  The chameleon is heavy in dense environments, which suppresses its ability to mediate a fifth force.  Such dynamical mechanisms for suppressing fifth forces are known as screening mechanisms; the chameleon mechanism is one of only three known screening mechanisms capable of making scalar-tensor gravitational theories compatible with experimental constraints on fifth forces \cite{Jain:2010ka}.  Furthermore, the chameleon mechanism is essential for $f(R)$ theories of modified gravity, which generate cosmic acceleration by making the gravitational Lagrangian a non-trivial function of the Ricci scalar \cite{Carroll:2003wy}.  Such a theory can be rewritten as a metric theory with the standard Einstein-Hilbert action and an additional scalar field that couples to matter in the same way as the chameleon \cite{Chiba:2003ir}.  In order for an $f(R)$ theory to successfully pass observational tests, this scalar field must have a potential function that allows it to employ the chameleon mechanism \cite{Chiba:2006jp,Faulkner:2006ub,Hu:2007nk, Brax:2008hh}.

In the original chameleon theory, and in $f(R)$ gravity, the coupling between the chameleon field and the matter fields was assumed to have gravitational strength.\footnote{The original chameleon proposal allowed different matter fields to have different couplings to the chameleon field.  We restrict our analysis to theories like $f(R)$ gravity that have a universal coupling, but we comment briefly on chameleons that couple exclusively to dark matter in Section \ref{sec:discuss}.}
Later it was found that much stronger couplings are also allowed \cite{Mota:2006ed, Mota:2006fz}; when the coupling to matter is stronger, the screening mechanism is also stronger, and the scalar field can still be hidden from fifth-force experiments.  We can search for strongly coupled chameleons in high-precision low-energy photon experiments \cite{Brax:2007ak,Brax:2007hi,Ahlers:2007st,Gies:2007su,Steffen:2010ze,Brax:2010jk}, with ultra-cold neutrons \cite{Brax:2011hb,Pokotilovski:2012js,Brax:2013cfa}, in precision atomic measurements \cite{Brax:2010gp}, in Casimir force experiments \cite{Brax:2007vm,Brax:2010xx}, with dark-matter direct-detection experiments \cite{Rybka:2010ah}, and in particle colliders \cite{Brax:2009aw,Brax:2009ey}.  It has also been suggested to look for strongly coupled chameleons produced in the Sun \cite{Baker:2012ah,Baker:2012nq} and to seek chameleon signatures in observations of stars and galaxies \cite{Burrage:2009mj,Burrage:2008ii,Levshakov:2010tj,Baker:2012ah,Baker:2012nq,Brax:2013uh}, the cosmic microwave background \cite{Schelpe:2010he,Davis:2009vk,Hu:2013aqa}, and the 21 cm power spectrum \cite{Brax:2012cr}.  These experiments exploit the fact that strongly coupled chameleons interact strongly with matter particles and photons in vacuum, so if an experiment or an astrophysical observation is targeted at a diffuse environment, it has the potential to see a chameleon signal.  A number of these experiments have been purposely designed to look for chameleons \cite{Steffen:2010ze, Brax:2010xx, Rybka:2010ah, Baker:2012nq}, while other results come from exploiting measurements made for other purposes.

Chameleons with gravitational-strength couplings in vacuum are harder to detect directly and are best sought by searching for deviations from general relativity.  Constraints on these theories come from laboratory searches for modifications of gravity \cite{Gannouji:2010fc,Upadhye:2012qu} and from astrophysical observations, including constraining the effects of the chameleon on the formation of structure and the current matter power spectrum \cite{Brax:2012sy,Li:2012by,Hellwing:2013rxa,Brax:2013mua}, on weak lensing \cite{Cardone:2012zn}, and on the evolution of stars \cite{Jain:2012tn}.  We will summarize the best constraints on chameleon theories in Section \ref{sec:constraints}; for our purposes, the essential constraint is that the chameleon potential must have a steep section in which small changes in the chameleon field \mbox{($\Delta \phi \simeq 0.01$ eV)} lead to significant changes in the chameleon potential and its derivatives.  

The chameleon potential was designed to provide the chameleon screening mechanism and does not originate from fundamental physics.  An approach to constructing a chameleon model from a KKLT compactification of string theory was discussed in Ref.~\cite{Hinterbichler:2010wu} and extended in Refs.~\cite{Nastase:2013ik, 2013JHEP...08..053H}; previous attempts to embed the chameleon into UV-complete theories were unsuccessful \cite{Brax:2004ym,Brax:2006np}.  In the absence of a UV-complete theory, chameleon gravity is usually treated as an effective field theory that should only be trusted at relatively low energies.  Quantum corrections to this theory have largely been ignored, even though one-loop corrections to the chameleon potential can be significant in laboratory environments \cite{Upadhye:2012vh}, and oscillations of the curvature scalar in $f(R)$ gravity can lead to particle production \cite{Arbuzova:2011fu, Arbuzova:2013ina}.

In a recent letter \cite{ourprl}, we exposed an additional quantum instability in chameleon gravity: the chameleon's behavior just prior to the time of Big Bang Nucleosynthesis (BBN) triggers catastrophic quantum effects that transfer most of the chameleon's energy to perturbations with momenta greater than $10^{7}$ GeV.  Increasing the strength of the chameleon's coupling to matter increases the energies of the generated perturbations, and chameleons with matter couplings that are moderately stronger than gravitational interactions experience trans-Planckian excitations.  In this work, we provide a more detailed treatment of this phenomenon, including the derivations that were omitted from Ref.~\cite{ourprl}.  We also extend our analysis to power-law chameleon potentials and find that they generate even more energetic perturbations than the exponential potentials studied in our earlier work.  

The chameleon's behavior in the early Universe was first investigated in Ref.~\cite{Brax:2004qh}.  Of particular concern is how much the chameleon scalar field evolves between the time of BBN and the present day.  Large variations in the chameleon's value can be interpreted as large variations in particle masses, and yet we know that particle masses at the time of BBN do not significantly differ from the masses that we measure today \cite[e.g.][]{2012PhRvD..86d3529C, 2013PhRvD..87h5018B}.  Ref.~\cite{Brax:2004qh} found that the chameleon is driven toward its current value prior to BBN, thus ensuring that the nucleon masses are sufficiently close to their observed values that BBN is unaffected, regardless of the chameleon's initial value (but also see Ref.~\cite{Mota:2011nh}).  Although the chameleon is usually light while the Universe is radiation dominated, the field is able to overcome Hubble friction and approach its present-day value because it becomes momentarily heavy whenever the Universe's temperature equals the mass of a particle species in equilibrium with the radiation bath.  We will discuss how these mass thresholds dramatically perturb the dynamics of the chameleon field in Section \ref{sec:review}; in summary, they kick the chameleon scalar field closer to the minimum of its effective potential, thus enabling it to approach the value it holds today.  

We will show that these kicks are generally too effective; the chameleon reaches the minimum of its effective potential with a large velocity ($\dot{\phi} \gg \mathrm{MeV}^2$) and climbs up the steep part of its potential.  Since the chameleon potential changes significantly when the chameleon value changes by 0.01 eV, these large velocities lead to rapid changes in the chameleon's effective mass, which generate perturbations via quantum particle production.  These perturbations have sufficiently high energies that they push chameleon gravity outside its low-energy regime of validity, and quantum corrections dominate the chameleon's potential.  Therefore, the chameleon's evolution during BBN cannot be understood using only a low-energy effective field theory, which casts doubt on chameleon gravity's viability.  Previous studies of the chameleon's evolution prior to and during BBN \cite{Brax:2004qh, Mota:2011nh} treated the chameleon purely classically and consequently missed these important quantum effects.  

We begin by reviewing chameleon gravity in Section \ref{sec:review}; we focus on the shape of the chameleon potential and the chameleon's dynamics in a radiation-dominated Universe.  In Section \ref{sec:surfing}, we present a novel solution to the chameleon's equations of motion in the presence of the aforementioned mass-threshold kicks.  We apply this solution in Section \ref{sec:classicalkicks}, where we consider how the chameleon responds to the kicks generated by Standard-Model particles, and we calculate the chameleon's velocity when it reaches the minimum of its effective potential.  In Section \ref{sec:quantum}, we show that these velocities lead to non-adiabatic changes in the chameleon's effective mass, and we investigate the resulting particle production both numerically and analytically.  Finally, we discuss the implications of our results and conclude in Section \ref{sec:discuss}.  Appendices \ref{App:Sigma} and \ref{App:TA} provide further details about the kicks from Standard-Model particles and the chameleon field's evolution at high temperatures, and we review the fundamental theory of quantum particle production in an expanding Universe in Appendix \ref{App:pp}.

\section{Chameleon Gravity}
\label{sec:review}

In chameleon gravity, the spacetime metric $\gtil_{\mu\nu}$ that appears in the matter Lagrangian is a conformal rescaling of the metric $ g^*_{\mu\nu}$ that solves Einstein's equations:
\beq
\gtil_{\mu\nu} = e^{2\beta\phi/\Mpl}\ g^*_{\mu\nu},
\label{eq:coupling}
\eeq
where $\beta$ is a dimensionless coupling constant, $\phi$ is the chameleon field, and $\Mpl = (8\pi G)^{-1/2}$.  The action for this theory can be written as
\beqa
S&=& \int d^4x \sqrt{-g_*} \left[\frac{\Mpl^2}{2}R_* - \frac{1}{2}(\nabla_* \phi)^2 - V(\phi)\right]  \nonumber \\
&&+S_m\left[\gtil_{\mu\nu}, \psi_m\right],
\label{EframeAction}
\eeqa
where $g_*$ is the determinant of the metric $g^*_{\mu\nu}$, $R_*$ is its Ricci scalar, $V(\phi)$ is the chameleon potential, and $S_m$ is the action for the matter fields.  If we define $T^*_{\mu\nu}\equiv(-2/\sqrt{-g_*})\delta S_m/\delta g_*^{\mu\nu}$, then varying this action with respect to $g^*_{\mu\nu}$ yields the Einstein field equation with a stress-energy tensor equal to the sum of $T^*_{\mu\nu}$ and the stress-energy tensor for the chameleon field.  Consequently, $g^*_{\mu\nu}$ and $T^*_{\mu\nu}$ are respectively called the Einstein-frame metric and stress-energy tensor.  Meanwhile, the matter fields couple to the Jordan-frame metric $\gtil_{\mu\nu}$.  We can also define a Jordan-frame stress-energy tensor: $\tilde{T}_{\mu\nu}\equiv(-2/\sqrt{-\gtil})\delta S_m/\delta \gtil^{\mu\nu}$.  The conformal relationship between $\gtil_{\mu\nu}$ and $g^*_{\mu\nu}$ implies that ${T_*^\mu}_\nu = e^{4\beta\phi/\Mpl}\tilde{T}^\mu{}_\nu$.  Therefore, if the matter fields are perfect fluids with density $\rho$ and pressure $p$, $\rho_* = e^{4\beta\phi/\Mpl}\tilde{\rho}$, and the equation of state parameter $w\equiv p/\rho$ is the same in both frames.  

\subsection{The chameleon potential}
\label{sec:thinshell}

The choice of the potential $V(\phi)$ is crucial to the success of the chameleon mechanism, as it is the non-linearities of the potential that allow the chameleon field to hide from fifth-force experiments.  Here we give a brief review of the properties required of a chameleon potential.  
Due to the coupling between the chameleon field and matter, the chameleon explores a wide region of its potential, and its evolution is driven by the ambient energy distribution.  First, it is necessary for there to be a region of the potential that is close to being flat; if we want the chameleon to be cosmologically relevant today, then the field must have a very small mass in cosmological environments.  Second, the potential must have a steep section to provide a barrier that limits the chameleon's excursion from its cosmological value in the interior of the objects used in fifth-force searches.  Such objects include the Earth, the Moon, the Sun, and laboratory test masses.  

To study such situations it is easiest to assume that the source object is spherically symmetric, static, and composed of non-relativistic matter.  Then the equation of motion governing the behavior of the chameleon in the Einstein frame is
\begin{equation}
\frac{d^2\phi}{dr^2}+\frac{2}{r}\frac{d\phi}{dr}=\frac{dV}{d\phi}+\frac{\beta\rho_*}{\Mpl}.
\end{equation}
Therefore, the chameleon's evolution is governed by an effective potential
\begin{equation}
V_{\rm eff}(\phi)=V(\phi)+\frac{\beta \phi\rho_*}{\Mpl}.
\label{Veff}
\end{equation}
This effective potential has a minimum at $\phi_{\rm min}$, where $V^\prime(\phi_{\rm min}) =-\beta \rho_*/\Mpl$, which implies that $\phi_{\rm min}$ depends on the ambient matter density $\rho_*$.  At this minimum, the effective mass of the chameleon field is
\begin{equation}
m^2=\left.\frac{d^2V}{d\phi^2}\right|_{\phi=\phi_{\rm min}}.
\end{equation}
For the chameleon mechanism to operate, $m^2$ must be a positive and monotonically increasing function of the density $\rho_*$ for all relevant densities.  Since $\phi_{\rm min}$ increases as the density decreases, $V^{\prime\prime\prime}(\phi)$ must be negative over the range of accessible $\phi$ values. 

It is not sufficient for the effective mass of the chameleon to simply increase as the density increases; screening the fifth force mediated by the chameleon requires the mass to increase sharply.
As discussed above, $V(\phi)$ must be chosen so that, when the density approaches the cosmological background density, the chameleon's Compton wavelength approaches cosmological distance scales.
To show that the mass must be much greater inside a test mass, we consider the condition for the scalar potential well generated by a source to be shallower than that source's gravitational well.  This condition, known as the thin-shell condition, guarantees that the fifth force mediated by the scalar field will be much weaker than the gravitational force \cite{Khoury:2003aq,Khoury:2003rn}.  If the source object has mass $M_c$ and radius $R$,  then the thin-shell condition is
\begin{equation}
\frac{(\phi_{\infty}-\phi_c)}{\beta \Mpl}\frac{8\pi \Mpl^2 R}{M_c} \ll 1,
\label{thinshell}
\end{equation}
where $\phi_c$ and $\phi_{\infty}$ are the positions of the minimum of the effective potential inside and far outside the source object, respectively.  Since $V^{\prime\prime}(\phi)$ must be positive over the relevant range of $\phi$ values,
\begin{equation}
-\frac{\beta\rho_{\infty}}{\Mpl} \ll V^{\prime}\left(\phi_c +\frac{\beta M_c}{8\pi  \Mpl R}\right).
\end{equation}
To get a rough estimate of a bound on the mass of the chameleon inside the source object [$m_c^2 = V^{\prime\prime}(\phi_c)$], we Taylor expand the right-hand side of this inequality.  On rearranging, and assuming $\rho_{\infty} \ll\rho_c$,  we find
\begin{equation}
\frac{6}{R^2}\ll m_c^2.
\end{equation}
Although this inequality only approximates the thin-shell condition, it illuminates the essential component of the chameleon screening mechanism: inside a source, the chameleon is too massive to carry a force beyond the source's boundary.
Since $1/R \gg H$ for all sources, there must be a region of the chameleon potential that is much steeper than the region probed by cosmological densities.  This feature of the chameleon potential will be crucial in the discussion that follows.

\subsection{Current constraints on chameleon models}
\label{sec:constraints}
Chameleon gravity is constrained by laboratory experiments, gravitational tests in the Solar System, and astrophysical observations. The best current bounds on the coupling parameter $\beta$ come from laboratory experiments that study diffuse systems in a vacuum.  The chameleon behaves as a very light scalar field within a laboratory vacuum, and none of its effects are screened.  Consequently, precision measurements of very diffuse systems can detect signatures of the chameleon.  The best current constraint comes from measurements of atomic energy levels in hydrogen \cite{Brax:2010gp}, which would be perturbed by the existence of a new chameleon force.  These measurements set an upper bound on the coupling parameter: $\beta \lesssim 10^{14}$.  For reference, gravitational-strength coupling corresponds to $\beta \sim {\cal O}(1)$, and $f(R)$ gravity theories have $\beta = 1/\sqrt{6}$.  

Constraints on the energy scale that controls the chameleon potential, which we will denote $M$, are more model dependent.  
The best constraints come from laboratory searches for fifth forces and from Casimir experiments.  
The E\"{o}t-Wash experiment currently provides the best constraints on weakly coupled chameleon theories ($\beta \lsim 20$) \cite{Gannouji:2010fc,Upadhye:2012qu}. Over a wide range of the parameter space, however, these fifth-force experiments are not sensitive to the chameleon.   In these experiments, thin plates are often used to shield electromagnetic forces, but unfortunately, these plates often shield the chameleon forces too.  Experimental searches for Casimir effects do not use such shielding; since they look for forces between parallel plates held very closely together, they can be very constraining for chameleon theories.   The constraints on $M$ typically depend on the value of $\beta$ and the precise form of the chameleon potential \cite{Mota:2006fz}.  It is possible to make some general statements, however: for power-law potentials of the form $V(\phi)= M^{4+n}\phi^{-n}$ and $\beta \gtrsim 0.1$, searches for Casimir effects and fifth forces constrain $M\leq 0.01 \mbox{ eV}$.  For $0.1\lesssim \beta \lesssim 10^5$ the constraints on $M$ are typically much stronger in specific models; for full details of the constraints on specific choices of the chameleon potential, we refer the reader to Refs.~\cite{Mota:2006fz,Gannouji:2010fc,Upadhye:2012qu}.   

In our analysis, we will consider both the power-law potential and the exponential potential considered in previous studies of the chameleon's cosmological evolution \cite{Brax:2004qh}:
\beq
V(\phi) = M^4 \exp\left[\left(\frac{M}{\phi}\right)^n\right] 
\label{pot}
\eeq
with $n>0$.  If $M\sim 0.001\,\mathrm{eV}$, this potential provides the vacuum energy required to drive cosmic acceleration at late times.  In low-density environments, $\phi \gg M$, and this potential is effectively a power-law potential plus a constant.  Therefore, it is subject to the same laboratory constraints as power-law potentials: $M < 0.01 \mbox{ eV}$.  This potential meets all the requirements discussed in the previous subsection: it is nearly flat when $\phi \gg M$; it is steep when $\phi \lsim M$; $V^{\prime\prime}(\phi)$ is always positive; and $V^{\prime\prime\prime}(\phi)$ is always negative.   Constraints on this potential were analyzed in detail in Refs.~\cite{Gannouji:2010fc,Upadhye:2012qu}, where it was found that if $M$ is chosen to be the dark energy scale, constraints from the E\"{o}t-Wash experiment demand fairly large values for $n$ and $\beta$: roughly $n \gsim 10$ for $\beta\sim 1$, and  $n\gtrsim 4$ for $\beta\sim 10$.  

Finally, there is an astrophysical constraint on the present-day cosmological mass of the chameleon ($m_\infty$ in the notation of the previous section) \cite{Brax:2011aw,Wang:2012kj}.  
Since the chameleon force must be screened inside galaxies, galaxies must satisfy the thin-shell condition given by Eq.~(\ref{thinshell}).
Satisfying this constraint requires $m_\infty/H_0 \gtrsim 10^3$, which corresponds to $1/m_\infty \lesssim 10 \mbox{ Mpc}$.  For both the power-law potential [$V(\phi)= M^{4+n}\phi^{-n}$] and the potential given by Eq.~(\ref{pot}), this constraint implies that $M\lsim0.07\beta^{2/3}$~MeV for $n=2$, $M\lsim2\beta^{3/4}$~GeV for $n=4$, and $M\lsim1000\beta^{4/5}$~GeV for $n=6$.

\subsection{A cosmological chameleon}

If the Universe is homogeneous and isotropic, then the Einstein-frame and Jordan-frame metrics are FRW metrics with conformally related scale factors ($\tilde{a} = e^{\beta\phi/\Mpl}a_*$) and proper times ($d\tilde t = e^{\beta\phi/\Mpl}dt_*$).  In the Jordan frame, the matter fields do not interact with the scalar fields, so the matter stress-energy is conserved: $\tilde{\nabla}_\mu \tilde{T}^\mu{}_\nu = 0$ and $\tilde{\rho} \propto \tilde{a}^{-3(1+w)}$.   It follows that $\rho_* e^{({3w-1})\beta\phi/\Mpl} \propto a_*^{-3(1+w)}$: the energy density in radiation is proportional to $a_*^{-4}$ in the Einstein frame, but the Einstein-frame energy density in matter is not proportional to $a_*^{-3}$.  While ${T_*^\mu}_\nu$ is not conserved in the Einstein frame, the sum of ${T_*^\mu}_\nu$ and the stress-energy tensor for the scalar field $\phi$ is conserved.  

Varying Eq.~(\ref{EframeAction}) with respect to $\phi$ gives the chameleon equation of motion
\beqa
\ddot{\phi}+3H_* \dot{\phi} &=& - \frac{dV}{d\phi}+\frac{\beta}{\Mpl}{T_*^\mu}_\mu ,\\
&=& - \frac{dV}{d\phi} - \frac{\beta}{\Mpl}\rhoR(\Sigma + f_m)
\label{eom}
\eeqa
where a dot represents differentiation with respect to Einstein proper time $t_*$ and $H_* \equiv \dot{a}_*/a_*$.  In the second line, we have evaluated the trace of the stress-energy tensor ${T_*^\mu}_\mu$: $\rhoR$ is the Einstein-frame energy density of the cosmic radiation bath, $f_m$ is the Einstein-frame density of nonrelativistic matter divided by $\rhoR$, and $\Sigma \equiv (\rhoR-3p_{*R})/\rhoR$, where $p_{*R}$ is the Einstein-frame pressure of the radiation bath.   Since both $f_m$ and $\Sigma$ are ratios of elements of the stress-energy tensor, the conformal relationship between $\tilde{T}^\mu{}_\nu$ and ${T_*^\mu}_\nu$ implies that these ratios are the same in the Einstein frame and the Jordan frame, so $f_m$ and $\Sigma$ may be evaluated using Jordan-frame energy densities and pressures.  As in Eq.~(\ref{Veff}), we can use Eq.(\ref{eom}) to define an effective potential for the chameleon, which is minimized when $\phi=\phimin$.  If $M\sim 0.001$ eV, $\phimin < M$ in the pre-BBN Universe.

If the radiation bath only consisted of photons, then $\Sigma$ would be zero.  In the early Universe, however, several massive particles were in thermal equilibrium with the photons, and we include the energy densities of these particles in $\rhoR$.  When the temperature of the radiation is much larger than the mass of the particle, these particles are relativistic, and their contribution to $\Sigma$ is zero.  As the radiation cools, the particles' pressure decreases faster than their energy density and their contribution to $\Sigma$ increases.  When the temperature is much less than the mass of the particle, the particles are Boltzmann suppressed and their contribution to $\Sigma$ decreases again.  Therefore, each species of massive particles makes a contribution to $\Sigma$ that peaks when the temperature of the radiation bath is nearly equal to the mass of the particle.
In Appendix \ref{App:Sigma}, we evaluate $(\tilde{\rho}_i-3\tilde{p}_i)/\tilde{\rho}_R$ for a particle with mass $m_i$ and $g_i$ degrees of freedom that is in thermal equilibrium with a radiation bath at  temperature $T_J$ \cite{Damour:1992kf, Damour:1993id, Coc:2006rt}:
\beq
\Sigma_i(T_J) = \frac{15}{\pi^4}\frac{g_i}{g_*(T_J)} \left(\frac{m_i}{T_J}\right)^2\int_{m_i/T_J}^\infty \frac{\sqrt{u^2 - (m_i/T_J)^2}}{e^u \pm 1} du,
\label{singlekick}
\eeq
where $g_*(T_J) \equiv \tilde{\rho}_R[(\pi^2/30)T_J^4]^{-1}$ is the number of relativistic degrees of freedom.  In the denominator of the integrand, the $+$ sign applies to fermions, and the $-$ sign applies to bosons.  While $m_i \simeq T_J$ and $\Sigma\neq0$, the chameleon experiences a force that drives it to smaller $\phi$ values; $\Sigma$ effectively ``kicks" the chameleon.  

To numerically solve the chameleon's equation of motion, we will need to specify how the Jordan-frame temperature depends on $a_*$ and $\phi$.  It is useful to consider the entropy density of the radiation bath, $s_R = (\tilde{\rho}_R+\tilde{p}_R)/T_J$, and to define $g_{*S} \equiv s_R[(2\pi^2/45)T_J^3]^{-1}$.  Entropy conservation in the Jordan frame implies that $g_{*S}(T_J) \tilde{a}^3 T_J^{3}$ is constant, which gives us an (implicit) expression for $T_J$ in terms of Einstein-frame variables, including the values of $a_*$ and $\phi$ at some fixed time [$\phi(a_{*,i})\equiv \phi_i$], and the Jordan-frame temperature $\initT$ at that same time:
\beq
T_J \left[{g_{*S}(T_{J})}\right]^{1/3}= \left[{g_{*S}(\initT)}\right]^{1/3} \initT e^{\beta(\phi_i-\phi)/\Mpl} \frac{a_{*,i}}{a_*}.
\label{T_J}
\eeq
We evaluate $g_{*S}(T_J)$ for the Standard Model particle spectrum (see Appendix \ref{App:Sigma} for details), and then we numerically invert the function $f(T_J) = T_J \times [g_{*S}(T_J)]^{1/3}$ to obtain $T_J$.  The initial temperature $\initT$ is chosen so that $\Sigma+f_m \ll 1$ for $T_J > \initT$.  We expect that the chameleon is at rest prior to the onset of the kicks, because any velocity it may have obtained during reheating would be damped by Hubble friction \cite{Brax:2004qh}.  In this case, the chameleon will remain at rest while $\Sigma+f_m \ll 1$, so its subsequent evolution does not depend on the specific value of $\initT$.

We assume a rather generic initial condition for $\phi$: \mbox{$M\ll\phi_i \lsim \Mpl$}.  For $\phi \lsim M$, the chameleon potential is very steep, and if $\phi$ is less than $\phimin$ in the early Universe, the chameleon will quickly roll to larger field values, where it will eventually stick due to Hubble friction. This evolution was demonstrated explicitly in Ref.~\cite{Brax:2004qh}; if $\phi_i$ is less than $\phimin$ prior to BBN, then the driving term from the chameleon potential dominates over the frictional Hubble term, and the field rolls until it stops at a value 
\begin{equation}
\phi_{\rm stop} \simeq \sqrt{6\Omega_{\phi}^{(i)}}\Mpl,
\end{equation}
where $\Omega_{\phi}^{(i)}<1$ is the initial fraction of the Universe's energy density in the chameleon field.  For the purposes of our analysis, $\phi_i = \phi_{\rm stop}$ in this scenario, and thus we expect $\phi_i\lsim \Mpl$ if $\phi \simeq \phimin$ during inflation, when $\phimin \ll M$.  It is also conceivable that $\phi$ had an initial value that far exceeded $M$.  We still restrict our analysis to $\phi_i < \Mpl$ because we expect that matter loop corrections will significantly renormalize the bare chameleon potential for larger field values.  We note, however, that our analysis is applicable to larger values of $\phi_i$ if one is willing to consider $\phi_i > \Mpl$ in spite of these concerns.  In fact, we will frequently set $\phi_i > \Mpl$ in Section \ref{sec:classicalkicks} in order to illustrate how far the chameleon rolls during the kicks. 

A change of variables facilitates our analysis of the chameleon's evolution.  We define a new time variable: $p \equiv \ln (a_*/a_{*,i})$, and we use a prime to denote differentiation with respect to $p$.  We also define a dimensionless scalar field $\varphi \equiv \phi/\Mpl$.  With these definitions, and using the Friedmann equations in the Einstein frame, the equation of motion for $\phi$ [Eq.~\ref{eom}] becomes
\begin{widetext}
\beq
\varphi^{\prime\prime}+\varphi^\prime\left\{\left(1+\frac{\Sigma+f_m}2 \right) \left[1-\frac{(\varphi^\prime)^2}6  \right]+ \frac{2V}{3H_*^2\Mpl^2}\left(1-\frac{\Sigma+f_m}4\right) \right\}= \frac{-dV/d\varphi}{H_*^2\Mpl^2} - 3\beta\left[1-\frac{(\varphi^\prime)^2}6 -\frac{V}{3H_*^2\Mpl^2}\right]\left(\Sigma + f_m - \Sigma f_m\right),
\label{longeom}
\eeq
where we have dropped ${\cal O}(f_m^2)$ terms.   The first Friedmann equation also implies that
\beq
\frac{1}{H_*^2\Mpl^2} = \frac{3}{\rhoR(1+f_m)+V}\left[1-\frac{(\varphi^\prime)^2}6\right],
\eeq
which we can use to eliminate $H_*^2$ from Eq.~(\ref{longeom}).  We can further simplify the chameleon's equation of motion by noting that $\Sigma \lsim 0.1$ and $f_m \lsim 10^{-6}$ prior to BBN (see Section \ref{sec:classicalkicks}).  Therefore, we can approximate $1+f_m \simeq 1+\Sigma \simeq 1$, which leaves
\beq
\varphi^{\prime\prime}+\varphi^\prime\left[1-\frac{(\varphi^\prime)^2}6\right]\left(1+\frac{2V}{\rhoR+V}\right) = \left[1-\frac{(\varphi^\prime)^2}6\right]\left[-3\frac{ dV/d\varphi}{\rhoR+V}-3\beta\left(1-\frac{V}{\rhoR+V}\right)(\Sigma + f_m)\right].
\label{fulleom}
\eeq
\end{widetext}
Finally, to close the system of equations, we note that Eq.~(\ref{T_J}) implies that
\beq
\rhoR =  \frac{\pi^2}{30} g_*(T_J)  \left[\frac{g_{*S}(\initT)}{g_{*S}(T_{J})}\right]^{4/3}  \initT^4 e^{4(\beta\varphi_i - p)},
\eeq
which explicitly shows that, if there are no changes in the number of relativistic degrees of freedom, $\rhoR \propto a_*^{-4}$ as expected.  

In the next two sections, we use this system of equations to examine the evolution of the chameleon field during the radiation-dominated era.  In the absence of massive particles, so that both $\Sigma$ and $f_m$ are zero, analytically solving Eq.~(\ref{fulleom}) for $M\ll\phi_i \lsim \Mpl$ reveals that any initial field velocity possessed by the chameleon will quickly damp away to zero, and the field will freeze at a value $\phi>\phimin$ that is fixed by its initial conditions \cite{Brax:2004qh}.  We will see that the kicks dislodge the chameleon and send it rolling toward the minimum of its effective potential.

\section{The Surfing Solution}
\label{sec:surfing}

In this section, we analyze the chameleon's response to the kicking function $\Sigma$ analytically, and we expose a new solution to its equation of motion.  First, we simplify Eq.~(\ref{fulleom}) by noting that $V(\phi) \ll \rhoR$ while $\phi \gsim M$ in the early Universe.  Second, we assume that $\phi \gg \phimin$, as is the case for the range of initial conditions that we consider ($M \ll \phi_i \lesssim \Mpl$), so that the $dV/d\varphi$ driving term in Eq.~(\ref{fulleom}) is negligible compared to the driving term from $\Sigma$.  Finally, we neglect the background density of non-relativistic matter $f_m$.  These approximations simplify the system and allow us to analytically examine the dynamics of the kicks.  However, it is not necessary to make such assumptions in order to study the evolution of the chameleon kicks numerically, and the numerical results of the next section do not require such assumptions.  Working with the simplified system, the chameleon equation of motion (\ref{fulleom}) is approximately
\begin{equation}
\frac{\varphi^{\prime\prime}}{1-{(\varphi^\prime)^2}/6}+\varphi^\prime = -3\beta\Sigma(T_J).
\label{eomsimp}
\end{equation}
In deriving this equation, we have assumed that $\varphi^\prime < \sqrt{6}$, which is equivalent to assuming that the chameleon's kinetic energy is less than the critical density of the Universe.

Recall that the Jordan-frame temperature $T_J$ is given by Eq.~(\ref{T_J}) as a function of the Einstein-frame scale factor, the scalar field value, and the number of relativistic degrees of freedom:  
\begin{equation}
T_J [g_{*S}(T_J)]^{1/3} = \initT [g_{*S}(\initT)]^{1/3} e^{\beta (\varphi_i-\varphi)}e^{-p}\;.
\label{T_J2}
\end{equation}
The dependence of the Jordan-frame temperature on the chameleon field is important because it allows the existence of a novel solution to the chameleon equation of motion, which we call the surfing solution:
\begin{equation}
\varphi=\varphi_s -\frac{(p-p_s)}{\beta},
\label{eq:surfer}
\end{equation}
where $p_s$ is the time at which the surfing behavior begins, and the value of the field at this time is
\begin{equation}
\varphi_S=\varphi_i-\frac{p_s}{\beta}-\frac{\lambda}{\beta},
\end{equation}
where $\lambda$ is a constant. Inserting this ansatz for $\varphi$ into Eq.~(\ref{T_J2}) yields
\begin{equation}
T_J [g_{*S}(T_J)]^{1/3} = \initT [g_{*S}(\initT)]^{1/3} e^\lambda,
\label{T_Jlambda}
\end{equation}
which implies that the Jordan-frame temperature is constant while the chameleon follows the surfing solution.  We call this constant value of $T_J$ the surfing temperature ($T_s$).  Since the surfing solution has $\varphi^{\prime\prime} = 0$ and $\varphi^\prime = -1/\beta$, Eq.~(\ref{eq:surfer}) solves Eq.~(\ref{eomsimp}) provided that
\beq
\Sigma(T_s)=\frac{1}{3\beta^2}.
\label{eq:surfcondition}
\eeq
Therefore, for a given kick function $\Sigma$, the surfing temperature is determined by $\beta$, and then $\lambda$ is set by Eq.~(\ref{T_Jlambda}).  Notice that the existence of the surfing solution is independent of the form of $\Sigma$ and the temperature of the Universe at the time of the kick.  Variations in $\Sigma$ and $T_{J,i}$ only vary the parameter $\lambda$ for the surfing solution.  As $\Sigma(T_J)$ is a bounded function, a value of $T_s$ that solves Eq.~(\ref{eq:surfcondition}) does not exist for all values of $\beta$.  The maximum value for $\Sigma$ in the Standard Model is $\Sigma_{\rm max} \sim \mathcal{O}(0.1)$ (see Appendix \ref{App:Sigma} and Fig.~\ref{Fig:sigma}), so we can expect surfing solutions for $\beta \gtrsim \mathcal{O}(1)$.  

On the surfing solution, the value of the scalar field decreases with time $p$, so eventually the field value approaches $\phimin$, and the bare scalar potential is no longer negligible.  At this point our approximations break down, and the surfing solution ceases to exist.  It is important to stress, however, that once the chameleon reaches the surfing solution, it will remain on that solution until the scalar field gets close to the minimum of its effective potential.  The name ``surfing solution" was chosen because the chameleon field surfs the wave of the kick function all the way to the minimum of its effective potential.

While the chameleon is surfing, the Jordan-frame scale factor $\tilde{a}=e^{\beta \varphi} a_*$ remains constant, and the Jordan-frame Universe is static.  The Einstein-frame scale factor does continue to increase, but this expansion has no observable effects on particles or Jordan-frame energy densities.  The surfing solution effectively pauses the evolution of the Universe from the time the Jordan-frame temperature reaches the surfing temperature to the time when $\phi\simeq\phimin$.  Despite the interruption in the Jordan-frame expansion, we stress that time in both the Jordan and Einstein frames continues to evolve forward during the surfing phase.

The surfing solution is relevant for the cosmological evolution of the chameleon field only if it is an attractor in the space of solutions.  Otherwise, the field only surfs the kicks if an unlikely fine-tuning of initial conditions occurs.  It can be seen that the surfing solution is reached regardless of the initial value of $\varphi$ by noting that, while $V(\varphi)$ can be neglected, the equations of motion are invariant under the transformations
\begin{eqnarray}
\varphi & \rightarrow& \varphi +C\\
p & \rightarrow & p-\beta C
\end{eqnarray}
for constant $C$.  Therefore, all initial values for $\varphi$ are equivalent up to a time translation; changing $\varphi_i$ or, equivalently, changing $\initT$ changes the value of $p_s$ and $\lambda$ but does not change the existence of the surfing solution or the surfing temperature.  

\begin{figure}
\centering
\includegraphics[width=3.4in]{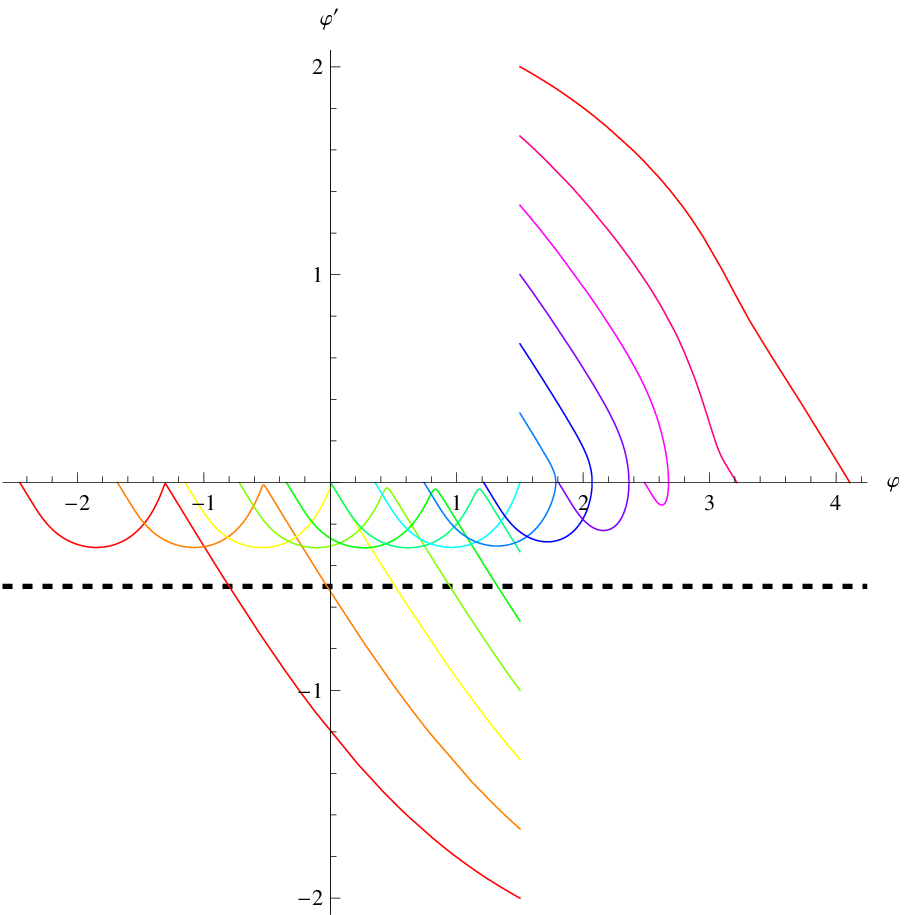}\\
\includegraphics[width=3.4in]{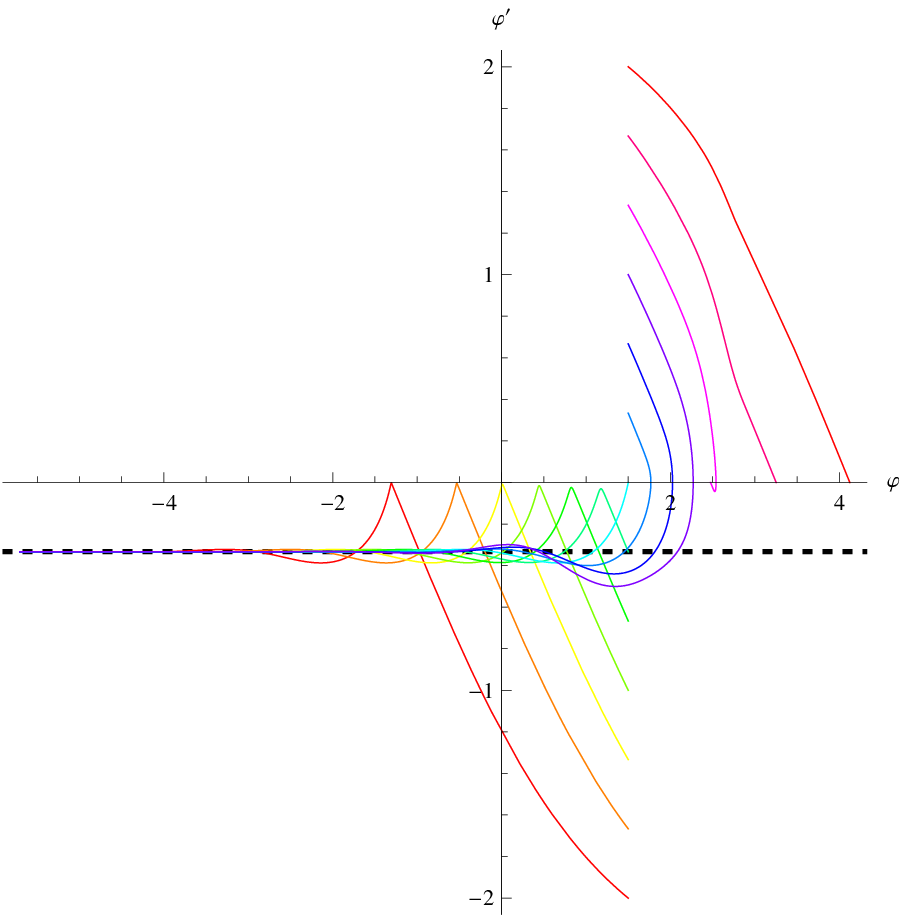}
\caption{Phase portraits for the chameleon's response to a single kick.  The top plot shows non-surfing evolution, with $\beta=2$, and the bottom plot shows the surfing solution with $\beta=3$.  Other parameters are kept the same between plots; $g_i/g_*=4/10.75$ and $m_i/\initT=1/25$.  The black dashed line in each plot shows $\varphi^{\prime}=-1/\beta$, which is the surfing solution when such a solution exists. The colored solid lines show the evolution of the chameleon field for different initial values of $\varphi^{\prime}$; the colors of the lines are to help the reader distinguish between curves and have no physical meaning.  We take the initial value of the field to be $\varphi_i= 1.5$ and show twelve solutions to the chameleon equation of motion with different initial values of $\varphi^{\prime}$ equally spaced between -2 and 2.}
\label{fig:phaseportraits}
\end{figure}

To see that the surfing solution is an attractor as the initial field velocity is also varied, it is easiest to solve the field equations numerically and plot the phase portraits, as shown in Fig.~\ref{fig:phaseportraits}.    We take numerical values of the parameters that approximate the kick coming from the electron and positron (discussed in more detail in Appendix \ref{App:Sigma}): $g_i = 4$ and $g_{*S} = g_*= 10.75$.  With these parameters, the maximum value of the kick function is $\mbox{Max}[\Sigma(T)] \approx 0.062$, so surfing solutions exist for all $\beta \gtrsim 2.3$.    In Fig.~\ref{fig:phaseportraits}, we show phase portraits both for $\beta=2$, for which a surfing solution does not exist, and $\beta=3$, which has a surfing solution.  We consider initial field velocities with $|\varphi_i^{\prime}|\lsim\sqrt{3}$, as is required to prevent the scalar field's kinetic energy from dominating the Universe ($\Omega_{\dot{\phi}} \lsim 0.5$).  When generating these figures, we have omitted $V(\phi)$, so there is nothing preventing $\varphi$ from going negative.  

The top panel of Fig.~\ref{fig:phaseportraits}, with $\beta=2$, shows that the kick moves the chameleon field to smaller values if $\varphi_i^{\prime}\lsim 1$, and in all cases, the field's velocity goes to zero after the kick passes.  In the bottom panel, we see that all chameleons with $\varphi_i^{\prime}\lsim 1$ (corresponding to $\Omega_{\dot{\phi}} \lsim 1/6$) end up on the surfing solution, with $\varphi^\prime = -1/\beta$.  These plots show that, when the surfing solution exists, it is an attractor in the space of solutions for all except the largest positive values of $\varphi_i^{\prime}$.  Although we only display phase portraits for one particular choice of parameter values, this behavior persists for surfing solutions over the whole range of parameter space for $\Sigma$. 

\section{Kicks from The Standard Model}
\label{sec:classicalkicks}
\begin{figure}
 \centering
 \resizebox{3.4in}{!}
 {
      \includegraphics{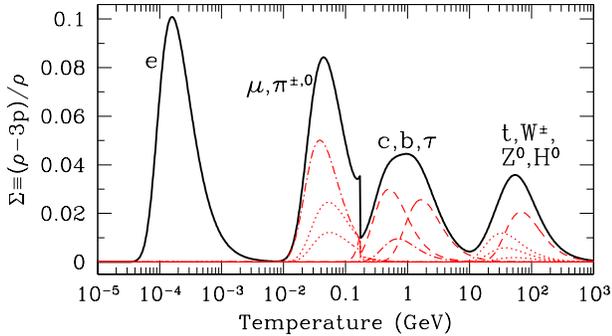}
 }
\caption{The kicking function $\Sigma \equiv (\rhoR-3p_{*R})/\rhoR$ as a function of the Jordan-frame temperature.  $\Sigma$ deviates from zero when the temperature falls below the mass of a particle that is in thermal equilibrium with the radiation bath: see Eq. (\ref{singlekick}).  The kicking function shown here includes contributions from all the Standard-Model particles, including a Higgs particle with a mass of 125 GeV.  The dashed curves show the contributions to $\Sigma$ from individual quark species; the dotted curves show the contributions from individual boson species; and the dot-dashed curves show the contributions from individual lepton species.  The discontinuity at a temperature of 170 MeV corresponds to the QCD phase transition.}
\label{Fig:sigma}
\end{figure}

We now consider how the chameleon responds to the kicks generated by the Standard-Model particle spectrum.  As described in Appendix \ref{App:Sigma}, we evaluate $\Sigma$ by summing contributions from the particles in the Standard Model, including a Higgs particle with a mass of 125 GeV.  Figure \ref{Fig:sigma} shows the resulting $\Sigma(T_J)$.  We see that the individual kicks from different particles are not distinct events; instead, the Standard-Model particles produce four ``combo-kicks."  Each combo-kick has a larger amplitude than the previous kicks because each particle's contribution to $\Sigma$ is suppressed by a factor of $1/g_*(T_J)$ (see Eq.~\ref{singlekick}), and the number of relativistic degrees of freedom decreases as the Universe cools.  The discontinuity between the second and third combo-kicks arises from the QCD phase transition, which we assume happens instantaneously at a temperature of $170$ MeV.  

The longest pause between kicks occurs prior to the last kick, when $\Sigma$ reaches a minimum value of $0.00026$ at a temperature of $7.4$ MeV.  Even at this temperature, $\Sigma$ is much larger than the matter fraction $f_m$.  Moreover, even if we assume that the current matter content of the Universe, including dark matter, is decoupled and nonrelativistic at all temperatures, $f_m \ll \Sigma$ for all temperatures greater than 50 keV.  Therefore, $f_m$ does not affect the chameleon's evolution during the kicks, and we do not consider it further.  

Our calculation of $\Sigma(T)$ at temperatures above 100 MeV is an incomplete treatment that provides a minimal value for the kick function.  First, it underestimates $\Sigma(T)$ during the QCD phase transition.  Lattice QCD calculations indicate that contributions from other hadrons and interactions between fields cause $\Sigma$ to increase sharply during the QCD phase transition, reaching values between 0.2 and 0.4 \cite{Bazavov:2009zn,Borsanyi:2010cj,Caldwell:2013mox}.  Due to the discrepancies between different lattice QCD calculations of $\Sigma$, we choose to neglect these additional contributions.  We note, however, that this additional peak in the kick function would lower the minimal value of $\beta$ required for the surfing solution, and it would enhance the impact velocity of the chameleon field for models that reach the minimum of their effective potential at temperatures below 400 MeV.  Second, we do not include contributions from particles beyond the Standard Model; if nothing else, the dark matter particle should contribute to $\Sigma$.  Third, we neglect the electroweak phase transition and use the particle spectrum given in Appendix \ref{App:Sigma} at high temperatures.  Ref. \cite{Caldwell:2013mox} showed that this approximation differs only slightly from a one-loop treatment of electroweak thermodynamics \cite{Arnold:1992rz} for temperatures less than 100 GeV, and the approximation is accurate within an order of magnitude for temperatures between 100 GeV and 300 GeV.  At these high temperatures, the potential contributions from beyond-Standard Model particles dwarf our calculation of $\Sigma(T)$, making our neglect of the electroweak phase transition irrelevant.  Finally, we do not include the QCD trace anomaly \cite{Kajantie:2002wa,Davoudiasl:2004gf}.  In the perturbative regime of QCD (i.e. energies above 100 GeV), the QCD trace anomaly implies that $(\rhoR-3p_{*R})/\rhoR \simeq 0.001$ even if all components of the plasma are relativistic.  At temperatures less than the electroweak phase transition, this contribution to $\Sigma$ is much smaller than the contributions from the Standard-Model kicks, and it does not significantly affect the chameleon's evolution if $\beta \lsim 5$.  For more strongly coupled chameleons, we show in Appendix \ref{App:TA} that the trace anomaly significantly increases the chameleon's velocity toward the minimum of its effective potential.

In Section \ref{sec:surfing}, we showed that chameleons with $\beta > \sqrt{1/(3\Sigma_\mathrm{max})}$, where $\Sigma_\mathrm{max}$ is the maximum value of $\Sigma$ during the kick, will ``surf" the kick and approach the minimum of the effective potential with a velocity $d\phi/d\ln a_* = -\Mpl/\beta$.  The amplitude of the first combo-kick (due to the top quark and the W, Z, and Higgs bosons) implies that all chameleons with $\beta>3.05$ can surf this kick, but numerically solving Eq.~(\ref{fulleom}) reveals that the chameleon reaches the surfing solution during the first kick only if $\beta \geq 3.07$.  If $\beta<3.07$, then the first combo-kick will push the chameleon toward the potential minimum, but as the Jordan-frame temperature cools, $\Sigma$ will decrease, and the chameleon will eventually come to a halt at a new position.  When we consider non-surfing chameleons later in this section, we will compute how far the chameleon moves during a kick that it cannot surf.  For now, let us assume that the value of the chameleon prior to the first combo-kick was sufficiently large that $\phi \gg \phimin$ after the passage of all the kicks that the chameleon cannot surf.  In that case, the chameleon will surf subsequent kicks if $\beta > \sqrt{1/(3\Sigma_\mathrm{max})}$ for these kicks.  Chameleons with $\beta > 1.82$ can surf the final kick, which occurs when the electrons and positrons become non-relativistic.  Since this kick has the largest amplitude of the four combo-kicks we consider, chameleons with $\beta< 1.82$ cannot surf.  As previously mentioned, however, the peak in $\Sigma$ due to the QCD phase transition would extend the surfing solution to smaller values of $\beta$.

\begin{figure}
\centering
 \resizebox{3.4in}{!}
 {
      \includegraphics{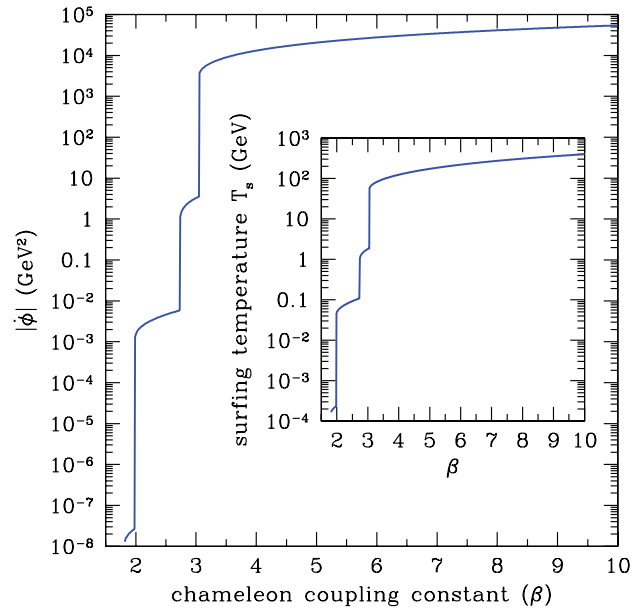}
 }
\caption{The value of $|\dot{\phi}|$ and $T_s$ when $\phi \ll \Mpl$ for chameleons that follow the surfing solution described in Section \ref{sec:surfing}.  This scenario is unavoidable if $\beta \geq 3.07$, and it also applies to smaller values of $\beta$ if the initial value of $\phi$ is sufficiently large. The computation of these velocities does not include the contribution to $\Sigma$ from the QCD trace anomaly and also ignores the electroweak phase transition.} \label{Fig:surfphidot}
\end{figure}


The existence of the surfing solution guarantees that all chameleons with $\beta>1.82$ will be kicked to $\phi \lsim \phimin$, regardless of their initial field values.  If $\beta\geq3.07$, then the chameleon surfs the first combo-kick, and it approaches the minimum of its effective potential with a constant velocity.  If $1.82<\beta<3.07$, then the chameleon's velocity at $\phi \simeq \phimin$ depends on its value prior to the kicks ($\phi_i$).  If $\phi_i$ is greater than the displacement caused by the previous kicks, then the chameleon will reach the minimum of its effective potential by surfing the first kick for which $\beta > \sqrt{1/(3\Sigma_\mathrm{max})}$, and it too will approach $\phimin$ with a constant velocity.  For all surfing chameleons, the chameleon's velocity depends only on $\beta$: $d\phi/d\ln a_* = -\Mpl/\beta$ and 
\beq
\left.\dot{\phi}\right|_{\phi\ll \Mpl} =  T_s^2\sqrt{ \frac{\pi^2}{15}\left[\frac{g_*(T_s)}{6\beta^2-1}\right]},
\label{surfingvelocity}
\eeq
where $T_s$ is the temperature in the Jordan frame during the surfing phase: \mbox{$\Sigma(T_s)={1/(3\beta^2)}$}.  In deriving this equation, we assumed that $\exp[4\beta\phi/\Mpl] \simeq 1$ so that we could equate $\rhoR$ to the radiation density in the Jordan frame.  Therefore, this equation is only applicable when $\phimin \lsim \phi \ll \Mpl$.  Figure \ref{Fig:surfphidot} shows $\dot{\phi}$ and $T_{s}$ for the surfing solution given the Standard-Model particle spectrum and no contribution from the QCD trace anomaly.  We see that $\dot{\phi}(\beta)$ is a smooth function for $\beta \geq 3.07$; these chameleons surf the first kick.  Chameleons with $1.82<\beta<3.07$ values must wait for $\Sigma(T_J)$ to equal ${1/(3\beta^2)}$ near the peaks of subsequent kicks.  Since there are no surfing solutions in the gaps between the kicks, $T_{s}(\beta)$ and $\dot{\phi}(\beta)$ are discontinuous for $\beta<3.07$.  If $\beta>4.5$, then the surfing temperature $T_{s}$ exceeds 150 GeV, and the chameleon is sensitive to the details of the electroweak phase transition and the QCD trace anomaly.

Since $\Sigma$ decreases as the temperature increases for $T_J>100$ GeV, adding the trace anomaly will increase the value of $T_{s}$ that satisfies $\Sigma(T_{s})={1/(3\beta^2)}$ for a given value of $\beta$.  
From Eq.~(\ref{surfingvelocity}), we see that $\dot\phi \propto T_{s}^2$, so the trace anomaly increases the value of $|\dot\phi|$ during the surfing phase.  The trace anomaly can have a more profound impact if its contribution implies that $\Sigma > {1/(3\beta^2)}$ at all temperatures.  This scenario is discussed in Appendix \ref{App:TA}, where we show that a constant high-temperature plateau in $\Sigma$ with $\Sigma > {1/(3\beta^2)}$ implies that the temperature in the Jordan frame \emph{increases} as the chameleon rolls toward the minimum of the effective potential, leading to a drastic increase in $|\dot\phi|$ compared to the values shown in Fig.~\ref{Fig:surfphidot}.  Furthermore, the inclusion of kicks from massive particles beyond the Standard Model will also increase $\Sigma$ at high temperatures, and consequently, $T_{s}$ and $|\dot\phi|$.   Therefore, the $|\dot\phi|$ values shown in Fig.~\ref{Fig:surfphidot} should generally be considered lower bounds.   

\begin{figure}
 \centering
 \resizebox{3.4in}{!}
 {
      \includegraphics{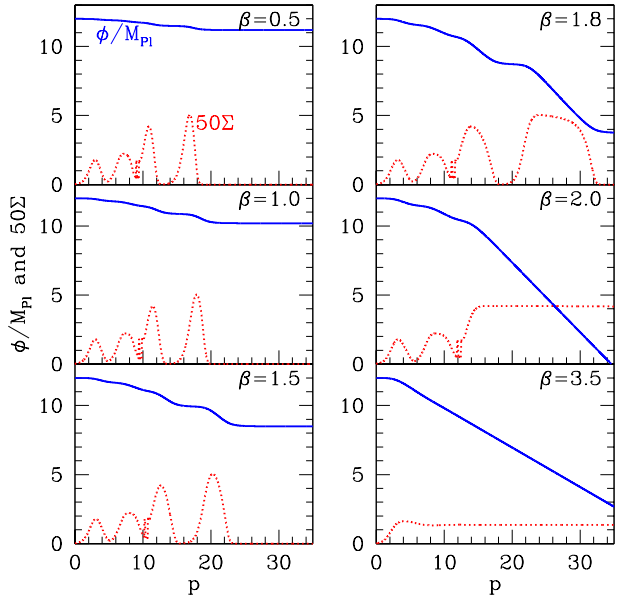}
 }
\caption{The evolution of the chameleon field for several values of the coupling constant $\beta$.  The solid curves show $\phi(p)/\Mpl$, where $p=\ln(a_*/a_i)$ marks the expansion in the Einstein frame.  The dotted curves show $50\Sigma(p)$.  If $\beta < 1.82$, then the chameleon cannot surf, and each kick displaces $\phi$ by a finite amount.  If $\beta=2.0$, then the chameleon surfs the third kick after being displaced by the first two kicks.  If $\beta\geq3.07$, then the chameleon surfs the first kick.  During the surfing phase, $\phi^\prime(p) = -\Mpl/\beta$ and $\Sigma$ is constant.}
\label{Fig:phitrajs}
\end{figure}

As previously mentioned, chameleons with $1.82< \beta <3.07$ can only surf the second, third, or fourth kicks if the earlier kicks leave $\phi\gg\phimin$.  The numerical solution to the chameleon equation of motion for $\beta=1.83$ confirms that the chameleon surfs the last kick; it rolls $3.5\Mpl$ toward $\phimin$ during the first three kicks, and then its velocity reaches a value of $d\phi/d\ln a_* = -\Mpl/\beta = - 0.54\Mpl$ near the peak of the last kick.  The chameleon maintains that velocity until the surfing solution is no longer valid ($\phi \lsim \phimin$).  If $\beta=1.99$, just below the threshold for surfing the third kick, then the chameleon rolls $13.5 \Mpl$ toward $\phimin$ before it begins to surf the last kick.  Chameleons with $2.00<\beta<2.73$ will surf the third kick, and they roll between $1.5\Mpl$ (for $\beta = 2.0$) and $25.5\Mpl$ (for $\beta=2.73$) during the first two kicks.  Finally, chameleons with $2.74\leq\beta\leq3.06$ will surf the second kick.  The displacement of the chameleon due to the first kick depends on the electroweak phase transition; the higher the temperature at which the top quark becomes massive, the larger the displacement.  For all $2.74\leq\beta\leq3.06$, however, the chameleon will roll more than $\Mpl$ during the first kick, even if $\Sigma=0$ at temperatures greater than 150 GeV.  In summary, chameleons with $1.83\leq \beta <3.07$ can only surf if the initial value of the chameleon significantly exceeds $\Mpl$.  For smaller values of $\phi_i$, the chameleon will reach $\phi \simeq \phimin$ before it can surf.  

Figure~\ref{Fig:phitrajs} shows the evolution of the chameleon field for several values of $\beta$.  If $\beta<1.82$, the chameleon experiences four rolling episodes, corresponding to the four combo-kicks produced by the Standard-Model particles, and then it stops rolling when $T_J \lsim 10^{-5}$ GeV and $\Sigma \simeq 0$.  In Fig.~\ref{Fig:phitrajs}, we purposefully chose large values for $\varphi_i$ so that $\varphi \gg \varphi_\mathrm{min}$ during all four kicks.   In this case, the total displacement in the chameleon field produced by $\Sigma(T_J)$ does not depend on $\varphi_i$.  We refer to this displacement as $\Delta \varphi$, and it is a function of $\beta$ alone.  

Earlier treatments of the chameleon's response to the kicks \cite{Brax:2004qh} obtained an analytic estimate of $\Delta \varphi$ by neglecting the Jordan-frame temperature's dependence on $\varphi$.  If we assume that $e^{\beta(\varphi_i-\varphi)} \simeq1$ in Eq.~(\ref{T_J}), then Eq.~(\ref{eom}) may be integrated twice to obtain $\varphi(p)$:
\beq
\varphi_1(p) - \varphi_i= -3 \beta \int_1^{e^p} \frac{dx}{x^2} \int_1^x  \Sigma(T_J[\phi= \phi_i, a]) da,
\label{delphi1}
\eeq
where the subscript ``1" indicates that this is the first-order solution, derived assuming that $\beta(\varphi_i-\varphi)\simeq 0$.  In deriving this expression, we assumed that $\varphi \gg \varphi_\mathrm{min}$ during all four kicks so that we can neglect $V^\prime(\phi)$ in Eq.~(\ref{eom}).  If we also neglect changes in the number of relativistic degrees of freedom and take $T_J = \initT e^{-p}$, we can analytically evaluate $\Delta\varphi$.  First consider a single kick, with $\Sigma(T_J)$ given by Eq.~(\ref{singlekick}) for one species with mass $m_i$ and $g_i$ degrees of freedom.  If the chameleon is initially at rest,
\beqa
\Delta \varphi_{s} &=& -3 \beta \frac{15 g_i}{\pi^4 g_*} \int_0^\infty \frac{dy}{y^2} \int_0^y \tau_{\pm} (z) dz,\\
\tau_{\pm} (z)  &=& z^2 \int_z^\infty \frac{\sqrt{u^2-z^2}}{e^u\pm1} du,
\eeqa
where it is best to evaluate $g_*$ at $T_J = m_i/2.45$ for fermions and $T_J = m_i/2.30$ for bosons because $\tau_\pm$ reaches its maximum at these temperatures.  Remarkably, this integral can be evaluated analytically:
\beq
\Delta \varphi_s = -\beta \frac{g_i}{g_*\left[\frac{m_i}{T_J}= \left(\begin{array}{c} 2.45 \\ 2.30\end{array}\right)\right]} \left(\begin{array}{c} 7/8 \\ 1\end{array}\right),
\eeq
where the top number applies to fermions, and the bottom number applies to bosons.  Since $\varphi_1(p)$ is linearly dependent on $\Sigma(T_J)$, we can obtain the total field displacement by summing over all the contributions from the Standard Model.  We find that $\Delta \varphi = -1.45 \beta$.  Meanwhile, numerically evaluating Eq.~(\ref{delphi1}) gives $\Delta \varphi_1 = -1.57 \beta$; the difference arises because the evaluation of  $\Delta \varphi_1$ fully accounts for the changes in the number of relativistic degrees of freedom when evaluating $\Sigma(T_J)$ and $T_J(\phi= \phi_i, a)$.  Since the cooling of the Jordan-frame radiation is slowed by the energy injected by annihilating particles ($T_J(p)>\initT e^{-p}$ after the first kick), each kick lasts a little longer (in terms of the Einstein clock $p$) than it does if changes in $g_*$ are neglected.  The extra duration of the kicks in the Einstein frame leads to a slightly larger displacement of the chameleon field. 

\begin{figure}
 \centering
 \resizebox{3.4in}{!}
 {
      \includegraphics{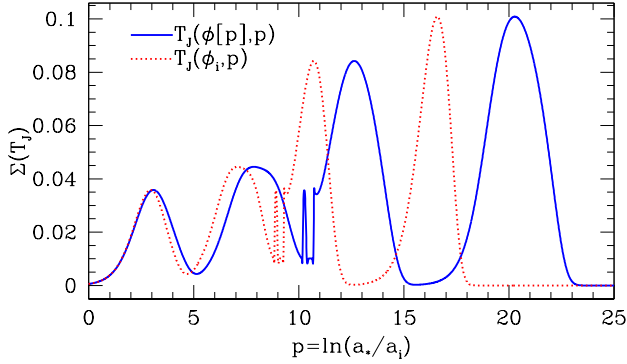}
 }
\caption{The evolution of $\Sigma(p)$ for $\beta = 1.5$.  For both curves, $\Sigma(T_J)$ is shown in Fig.~\ref{Fig:sigma}, but the two curves evaluate $T_J$ differently.  The solid curve uses the numerical solution $\phi(p)$ when evaluating $T_J(\phi, p)$, while the dotted curve evaluates $T_J$ assuming that $\phi = \phi_i$.  Neglecting the displacement $\phi$ underestimates $T_J(p)$, which shortens the duration of the kicks in the Einstein frame.}
\label{Fig:sigdilate}
\end{figure}

\begin{figure}
 \centering
 \resizebox{3.4in}{!}
 {
      \includegraphics{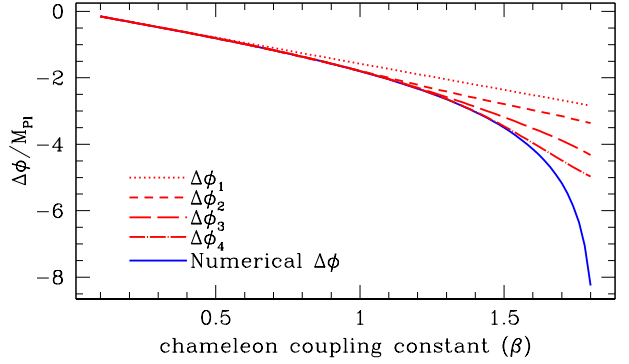}
 }
\caption{The total displacement of the chameleon field as a function of $\beta$.  The bottom solid curve shows the numerical results.  The dotted line shows the first-order approximation: $\Delta\varphi_1 = -1.57\beta$, and other curves show higher-order analytical solutions obtained by iterating Eq.~(\ref{delphin}).}
\label{Fig:DeltaPhi}
\end{figure}

A similar effect implies that $\Delta\varphi_1$ will not accurately describe the chameleon's displacement if $\beta\Delta\varphi \gsim 1$. In this case, $\exp[\beta(\varphi_i-\varphi)]$ is significantly greater than one during the kicks.  Consequently,  $T_J(p) >\initT e^{-p}$ after the first kick [see Eq.~(\ref{T_J})], and $\Sigma$ is nonzero for a larger range of $p$ values.  This dilation of $\Sigma$ is illustrated in Fig.~\ref{Fig:sigdilate}, which shows $\Sigma(p)$ using both $T_J[\phi(p), p]$ and $T_J(p) = T_J(\phi=\phi_i, p)$ for $\beta=1.5$.  The longer duration of the kicks in the Einstein frame increases $\Delta\varphi$; for example, if $\beta = 1.5$, $\Delta \varphi_1 = -2.355$, but Fig.~\ref{Fig:phitrajs} shows that $\Delta\varphi = -3.5$.  
A more accurate analytical estimate for $\Delta\varphi$ can be obtained by iterating the solutions to Eq.~(\ref{delphi1}):
\beq
\varphi_{n+1}(p) - \varphi_i= -3 \beta \int_1^{e^p} \frac{dx}{x^2} \int_1^x  \Sigma(T_J[\phi= \phi_n(p)],a) da.
\label{delphin}
\eeq
Figure~\ref{Fig:DeltaPhi} shows $\Delta\varphi_1, \Delta\varphi_2, \Delta\varphi_3,$ and $\Delta\varphi_4$ for different values of $\beta$ and compares them to the full numerical solution for $\Delta\varphi$.  We see that the analytical approximations always underestimate $\Delta\varphi$, but the first-order approximation $\Delta\varphi_1$ is accurate to within 10\% for $\beta \lsim 0.9$, and $\Delta\varphi_4$ is accurate to within 10\% for $\beta \lsim 1.7$.  As $\beta$ approaches 1.83, $|\Delta\varphi |$ increases rapidly and higher-order analytical approximations are required.  The surfing solution, which corresponds to $\Delta\varphi = -\infty$, is the extension of this pattern.  

We have shown that all chameleons with $\beta\geq1.83$ or $\varphi_i \leq \Delta\varphi$ will reach $\varphi \lsim \varphi_\mathrm{min}$ during the kicks.  There is an additional constraint, however: the chameleon must satisfy $\varphi < 0.1/\beta$ prior to the last kick to ensure that Einstein-frame particle masses do not vary by more than 10\% between now and BBN \cite{Brax:2004qh}.  The last kick is the most powerful kick; it alone displaces the chameleon by at least $|\Delta\varphi_1| = 0.56\beta$.  Therefore, if $\beta > 0.43$, then requiring that $\varphi < 0.1/\beta$ at the onset of BBN implies that the last kick will take the chameleon to $\varphi < \varphi_\mathrm{min}$.  For smaller values of $\beta$, the chameleon can avoid reaching $\phi \lsim \phimin$ while satisfying the BBN constraint only if
\beq
(x+0.56) \beta < \varphi_i < x \beta+\frac{0.1}{\beta},
\label{initrange}
\eeq
where 
\beq
x \simeq \sum_{m_i>3 \, \mathrm{MeV}}  \frac{g_i}{g_*\left[\frac{m_i}{T_J}= \left(\begin{array}{c} 2.45 \\ 2.30\end{array}\right)\right]} \left(\begin{array}{c} 7/8 \\ 1\end{array}\right).
\eeq
(If we only include the usual ensemble of Standard-Model particles, $x \simeq 1$, but there may be additional particles that we have not considered.) If $\beta = 1/\sqrt{6}$, which corresponds to $f(R)$ gravity, then this condition implies $0.23 \lsim (\varphi_i-x\beta) \lsim 0.245$, which is a very finely tuned initial condition!  We conclude that the chameleon can only avoid being kicked to $\varphi \lsim \varphi_\mathrm{min}$ if $\beta<0.43$ and $\varphi_i$ is within the limited range given by Eq.~(\ref{initrange}).

\begin{figure}
 \centering
 \resizebox{3.4in}{!}
 {
      \includegraphics{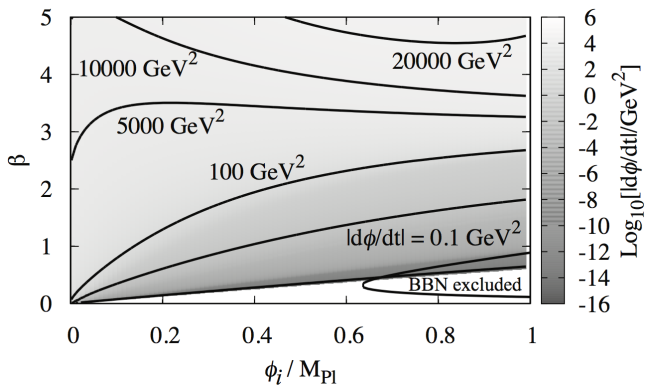}
 }
\caption{The velocity of the chameleon field when it reaches the minimum of its effective potential as a function of its initial value $\phi_i$ and the coupling constant $\beta$.  The white region at small $\beta$ values depict models that are not kicked to the minimum of the effective potential.  Chameleons that lie within the ``BBN excluded" region do not satisfy the constraint $\phi/\Mpl < 0.1/\beta$ after the first three kicks and therefore endanger the success of BBN.}
\label{Fig:VelContour}
\end{figure}

Having established that the kicks almost always take the chameleon to $\phimin$, we now consider the chameleon's velocity when it reaches the minimum of its effective potential: \mbox{
$\dot{\phi} = \Mpl H_* \varphi^\prime \simeq\varphi^\prime\sqrt{\rhoR/[3-0.5(\varphi^\prime)^2]}$} for $\rhoR \gg V(\phi)$.  Since $\phimin \ll \Mpl$, $\rhoR$ at impact is nearly equal to the Jordan-frame radiation density: \mbox{$\tilde{\rho} \equiv (\pi^2/{30})g_*(T_J) T_J^4$}.  During the kicks, $\varphi^\prime$ never exceeds 0.55, so $\dot{\phi} \sim T_J^2\varphi^\prime$, where $T_J$ is evaluated when $\phi = \phimin$.  Figure~\ref{Fig:VelContour} shows how the chameleon's velocity when $\phi \ll \Mpl$ depends on $\beta$ and the chameleon's initial value $\phi_i$.  In most cases, chameleons with larger initial values will reach $\phimin$ with smaller velocities because the Jordan frame will have longer to cool prior to impact.  Surfing chameleons are an exception to this rule because $T_J\gsim T_s$ when they reach $\phimin$, regardless of the initial field value.  For a given value of $\varphi_i$, increasing $\beta$ always increases the impact velocity; for nonsurfing chameleons, increasing $\beta$ generally increases both $|\varphi^\prime|$ and $T_J$ at impact, while for surfing chameleons, increasing $\beta$ increases the surfing temperature $T_s$ (see Fig.~\ref{Fig:surfphidot}), which more than compensates for the reduction in $|\varphi^\prime|$.  The key result of this section is that $|\dot{\phi}| \gg M^2$ when $\phi \simeq \phimin$: at impact, $T_J \gsim 0.5 \,\mathrm{MeV}$ and $\varphi^\prime$ is nearly always $>0.02$, so $|\dot{\phi}| > 5\times 10^{-9}\, \mathrm{GeV}^2$ in all but a few finely tuned cases. Moreover, $|\dot\phi|$ is usually much larger; for instance, a surfing chameleon with $\beta\geq3.07$ has $|\dot{\phi}| > 4000\, \mathrm{GeV}^2$ when it reaches the minimum of its effective potential.  

\section{Particle Production}
\label{sec:quantum}

In the previous section, we derived the velocity imparted to the chameleon by the kicks, and we showed that $|\dot\phi|\gg M^2$ when the chameleon reaches the minimum of its effective potential ($\phi \simeq \phimin)$.  While $\phi \gsim \phimin$, we can neglect the chameleon's bare potential $V(\phi)$, but $V(\phi)$ becomes important when the chameleon rolls to smaller values.  The chameleon will climb up its bare potential until it exhausts its kinetic energy, and then it will roll back toward the minimum of its effective potential.  During this rebound, $V''(\phi)$ changes rapidly, and the chameleon condensate cannot adjust its mass adiabatically.  Instead, we will show that the rapid changes in mass excite high-energy perturbations in the chameleon field.

\subsection{A first look at the rebound}
\label{firstlook}

Since the rebound occurs when $\phi \lsim \phimin$, we must first determine the value of $\phimin$ during the kicks.  From Eq.~(\ref{eom}), we see that the value of $\phimin$ in the early Universe is determined by the value of $\rhoR\Sigma$.  Specifically,
\beq
V^\prime(\phimin) = - \frac{\beta}{\Mpl}\rhoR\Sigma \simeq -\frac{\beta}{\Mpl}\left(\frac{\pi^2}{30}g_*T_J^4\right)\Sigma;
\eeq  
we can make the approximation $\rhoR \simeq \tilde{\rho}$ because $\phimin$ is always much smaller than $\Mpl$.  After the electroweak phase transition, $\tilde{\rho}\Sigma$ monotonically decreases as the Universe cools, so $\phimin$ moves to larger values as the kicks progress.  At the peak of the final kick, $T_J = 0.16$ MeV and \mbox{$\tilde{\rho}\Sigma = 32 \mathrm{\,g\, cm^{-3}}$}, which exceeds the mean density of the Earth.  Since the Earth must satisfy the thin-shell condition discussed in Section \ref{sec:thinshell}, the chameleon's rebound after it is kicked past $\phimin$ will sample the steep part of the chameleon's potential.  For the exponential potential given by Eq.~(\ref{pot}) and inverse-power-law potentials [$V(\phi) =  M^{n+4}\phi^{-n}$] with $M=10^{-3}$ eV, $\phimin \lsim M$ while $T_J \gsim 0.16$ MeV.   For example, if $n=2$ and $\beta=2$, then $\phimin$ increases from 0.14$M$ to 0.62$M$ between the peaks of the first and the last kicks if the potential is exponential, and it increases from $(5.3 \times 10^{-9})M$ to 0.26$M$ if the potential is an inverse power law.  Since the chameleon potential diverges as $\phi$ approaches zero, the change in $\phi$ during the rebound must be less than $M$.  Later in our analysis, we will see that quantum particle production starts while $\phimin<\phi \lsim M$, so we will broaden our definition of the rebound to include all times while $\phi \lsim M$.

We can estimate the duration of the rebound as \mbox{$\Delta t \sim M/\dot{\phi}$}, where $\dot{\phi}$ is the velocity imparted by the kicks. In the previous section we showed that $\dot{\phi} \sim T_J^2$; it follows that the duration of the rebound is much smaller than the Hubble time: $H_*\Delta t \sim M/\Mpl$.  Consequently, the expansion of the Universe will not affect the evolution of the chameleon during the rebound, and $\rhoR \Sigma$ will not change significantly during the rebound.  During the rebound, the matter coupling induces a change in the chameleon's velocity $\Delta\dot{\phi} \sim - (\beta/\Mpl)\rhoR\Sigma \Delta t$.  Since $\rhoR \sim T_J^4$ and $\Delta t \sim M/T_J^2$, the fractional change in the chameleon's velocity is minuscule: $\Delta\dot{\phi}/\dot{\phi} \sim M/\Mpl$.  A more detailed analysis of the chameleon's equation of motion while $\phi > \phimin$ for both surfing and nonsurfing chameleons confirms these estimates; while the chameleon moves a distance $\Delta \phi \sim M$, neither Hubble friction nor the chameleon's coupling to matter changes its velocity significantly ($\Delta\dot{\phi}/\dot{\phi} \sim M/\Mpl$ in all cases).  Therefore, we will neglect both Hubble friction and the chameleon's coupling to matter while analyzing the rebound.  We assume that the chameleon starts at $\phi \sim M$ with a velocity $\vel$, which is determined by the chameleon's value prior to the kicks and the value of $\beta$, as shown in Fig.~\ref{Fig:VelContour}.  

In Appendix \ref{App:pp}, we show that a plane-wave perturbation in the chameleon field with a comoving wavenumber $k$ has an effective mass
\beq
\omega_k^2(\tau) = k^2 + a^2 V_\mathrm{eff}^{\prime\prime}\left(\phibar\right) - \frac{a^{\prime\prime}(\tau)}{a} \label{omegamain},
\eeq
where $\tau$ is conformal time, $\phibar$ is the spatially averaged value of the chameleon field, and we have dropped the subscript $*$ on the scale factor $a$ because we will work exclusively in the Einstein frame throughout this section.  We will also no longer use the variable $p = \ln(a_*/a_i)$, and primes will denote differentiation with respect to the function's argument.  We show in Appendix \ref{App:pp} that perturbations in the chameleon field are excited when $\omega^\prime_k(\tau)/\omega_k^2 \gsim 1$.   While the Universe is radiation dominated, $a^{\prime\prime}(\tau) = 0$, and
\beq
\omega_k^\prime(\tau) = \frac{a^3}{2\omega_k}\left[V_\mathrm{eff}^{\prime\prime\prime}(\phibar)\dot{\phibar} + 2 H V_\mathrm{eff}^{\prime\prime}(\phibar)\right],
\label{omegaprime}
\eeq
We recall that $V^\prime_\mathrm{eff}(\phi) \equiv V^\prime (\phi)+ (\beta/\Mpl)\rhoR\Sigma(T_J)$, and the contribution from the bare potential dominates when $\phi \ll \phimin$, by definition.  The matter coupling contributes to higher derivatives of the effective potential through the dependence of $T_J$ on $\phi$: from Eq.~(\ref{T_J}), we see that $dT_J/d\phi \simeq - (\beta/\Mpl) T_J$.   Since differentiation of the bare potential introduces a factor of $M^{-1}$ while differentiation of the matter coupling term introduces a factor of $\Mpl^{-1}$, the relative importance of the matter coupling to $V_\mathrm{eff}^{\prime\prime}(\phi)$ is suppressed by a factor of $M/\Mpl$ compared to its relative contribution to $V^\prime_\mathrm{eff}(\phi)$.  Therefore, the bare potential dominates $V_\mathrm{eff}^{\prime\prime}(\phi)$ even if $\phi \gsim \phimin$; for our fiducial exponential potential with $n=2$ and $M=10^{-3}$ eV, the bare potential dominates $V_\mathrm{eff}^{\prime\prime}(\phi)$ for $\phi \lsim M$ and $T_J \lsim 1$ TeV.  The matter coupling's contribution to $V_\mathrm{eff}^{\prime\prime\prime}(\phi)$ is suppressed by an additional factor of $M/\Mpl$, so our fiducial bare potential dominates $V_\mathrm{eff}^{\prime\prime\prime}(\phi)$ for $\phi \lsim 10^6M$ and $T_J\lsim 1$ TeV. 

\begin{figure}
 \centering
 \resizebox{3.4in}{!}
 {
      \includegraphics{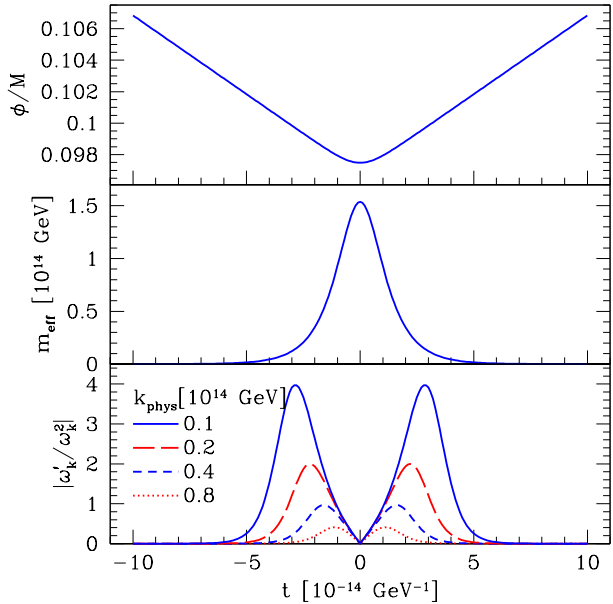}
 }
\caption{The classical evolution of the chameleon field as it rebounds off its bare potential $V(\phi)$ after starting at $\phi = 2M$ with a velocity $\dot{\phi} = -0.1\, \rm{GeV}^2$.  The potential $V(\phi)$ given by Eq.~(\ref{pot}) with $n=2$ and $M=10^{-3}$ eV.  The top panel shows the field value as it turns around.  The middle panel shows that the effective mass of the field, $m_{\rm{eff}}^2 = V^{\prime\prime}(\phi)$, rapidly changes during the rebound.  The bottom panel shows the evolution of the adiabatic ratio $\omega^\prime_k(\tau)/\omega_k^2$ during the rebound for perturbations with physical wave numbers $k_\mathrm{phys} = \{0.1,0.2,0.4, 0.8\} \times 10^{14}$ GeV.  We expect particle production when the adiabatic ratio exceeds unity.}
\label{Fig:ClassicalBounce}
\end{figure}

Since the bare potential dominates both terms in Eq.~(\ref{omegaprime}), we can neglect the matter coupling when evaluating the relative importance of these terms.  For $\phi \sim M$, $HV^{\prime\prime}(\phi)/[V^{\prime\prime\prime}(\phi)\dot{\phi}] \sim (M/\Mpl)(T_J^2/\dot{\phi}) \sim M/\Mpl$, so the $HV^{\prime\prime}(\phi)$ term is negligible.  It follows that
\beq
\frac{\omega_k^\prime(\tau)}{ \omega_k^2} \simeq \frac{a^3}{2\omega^3_k} V_\mathrm{eff}^{\prime\prime\prime}(\phibar)\dot{\phibar} = \frac{V_\mathrm{eff}^{\prime\prime\prime}(\phibar)\dot{\phibar} }{2[(k/a)^2+V^{\prime\prime}_\mathrm{eff}]^{3/2}}.
\eeq
We will see that the physical wavenumbers ($k_\mathrm{phys} = k/a$) of the perturbations that are excited when the chameleon rebounds are much larger than $\sqrt{V^{\prime\prime}_\mathrm{eff}(\phi)}$ while $\phi \gsim M$, which implies that $V^{\prime\prime}_\mathrm{eff}(\phi)$ only significantly contributes to the denominator when $V^{\prime\prime}_\mathrm{eff}(\phi) \simeq V^{\prime\prime}(\phi)$.  Since $V^{\prime\prime\prime}_\mathrm{eff}(\phi) \simeq V^{\prime\prime\prime}(\phi)$ throughout the rebound, the chameleon's coupling to matter has no impact on the adiabaticity condition and cannot affect particle production.  Furthermore, since we have already shown that the matter coupling has a negligible effect on $\dot{\phi}$ while the chameleon rolls $\Delta\phi \lsim M$, we can neglect the chameleon's coupling to matter entirely while analyzing particle production during the rebound.  

Figure \ref{Fig:ClassicalBounce} illustrates how the chameleon field rebounds off its bare potential, with $V(\phi)$ given by Eq.~(\ref{pot}) with $n=2$ and $M=10^{-3}$ eV.  The initial condition is $\vel= -0.1\, \rm{GeV}^2$ when $\phi = 2M$, and we have neglected both Hubble friction and the chameleon's coupling to matter when solving Eq.~(\ref{eom}).  As expected, the chameleon rolls toward zero until $V(\phi) = \vel^2/2$ and then it quickly turns around and rolls back with the same speed.   During this rebound, the effective mass of the field [$m_{\rm{eff}}^2 = V^{\prime\prime}(\phi)$] changes dramatically over a time $\Delta t \simeq  0.005 M/|\vel|$, momentarily reaching values greater than $10^{14}$ GeV.  Figure \ref{Fig:ClassicalBounce} also shows how the adiabatic ratio evolves during the rebound for several values of the physical wavenumber $k_\mathrm{phys}$; we see that $\omega^\prime_k(\tau)/\omega_k^2 \gsim 1$ for $k_\mathrm{phys} \lsim 400 |\vel|/M$.  Therefore, we expect particle production for modes with $k_\mathrm{phys} \lsim (\Delta t)^{-1}$, where $\Delta t$ is the timescale over which $V^{\prime\prime}(\phi)$ changes significantly.  

When we formulate an analytic model for the rebound in Section \ref{sec:analytics}, we will derive an expression for $\Delta t$.  For now, we simply note that $\Delta t < M/|\vel|$, so we expect modes with $k_\mathrm{phys} \gsim |\vel|/M$ to be excited during the rebound.  From Eq.~(\ref{rhofluc}), we see that the energy density in perturbations per logarithmic interval in $k$ is 
\beq
\rho_k = \frac{k^3n_k\omega_k}{2\pi^2 a^4}  \simeq \frac{k_\mathrm{phys}^3 n_k }{2\pi^2}\sqrt{k_\mathrm{phys}^2 + V^{\prime\prime}(\phi)},
\label{Ek}
\eeq
where $n_k$ is the mode occupation number defined in Eq.~(\ref{nkdef}).  In Appendix \ref{App:pp}, we show how $\omega^\prime_k(\tau)/\omega_k^2 \gsim 1$ implies that $n_k \sim 1$.  Therefore, we expect that the rebound will excite modes with $k_\mathrm{phys} \gsim |\vel|/M$ with $\rho_k \sim (\vel/M)^4$.  If we compare this energy density to the initial energy in the chameleon field, $\rho_i \simeq \vel^2/2$, we find that $\rho_k/\rho_i \sim \vel^2/M^4$.   In the previous section, we found that the kicks impart a velocity to the chameleon that greatly exceeds $M^2$, so this naive calculation yields $\rho_k/\rho_i \gg 1$.  Therefore, the modes with $\omega^\prime_k(\tau)/\omega_k^2 \gsim 1$ are too energetic to have $n_k \sim 1$ given the energy available to the chameleon field.  If these modes are excited during the rebound, this heuristic treatment implies that they must have $n_k \ll 1$, and even then, we expect them to absorb a significant fraction of the chameleon's energy.  Consequently, we must include the backreaction of the perturbations when analyzing the chameleon's evolution.   In the next section, we consider this backreaction in detail, and we show that it significantly alters the chameleon's trajectory during the rebound.

\subsection{Analytical model for the rebound}
\label{sec:analytics}

To analyze the excitation of perturbations during the chameleon field's rebound off its bare potential, we first write the chameleon field as
\beq
\phi(\tau,\vec{x}) =\phibar(\tau)+ \delta\phi(\tau,\vec{x}),
\eeq
where $\phibar(\tau)$ is the spatial average of the field.  We insert this expression into the chameleon's equation of motion and Taylor expand $V^\prime (\phi)$ around $\phibar$ to obtain
\beq
(\partial_t^2 + 3H\partial_t-\frac{\nabla^2}{a^2})(\phibar +\delta\phi) + V^\prime(\phibar) + \sum_{n=1}^\infty \frac{1}{n!} V^{(n+1)}(\phibar) \delta\phi^n=0.
\label{fullexpand}
\eeq
As discussed in the previous section, the chameleon's coupling to matter is irrelevant during the rebound.  Therefore, we neglect the matter coupling in this equation, but we note that it could be reinserted by replacing $V(\phibar)$ with $V_\mathrm{eff}(\phibar)$.  To find the equation of motion for $\phibar$, we take the spatial average of this equation.  Since $\langle \delta \phi \rangle = \langle \delta \phi^3 \rangle = 0$, we are left with
\beq
\ddot{\phibar }+3H\dot{\phibar}+ V^\prime(\phibar) + \frac{1}{2} V^{\prime\prime\prime}(\phibar)\langle \delta\phi^2 \rangle + {\cal O}(\langle\delta \phi^4\rangle)=0.
\label{spatialavg}
\eeq
The $\langle \delta\phi^2 \rangle$ term in this equation represents the first-order backreaction of the perturbations on the spatially averaged field.  We will show that the inclusion of this term is sufficient to ensure that energy is conserved during the rebound.  Since our primary aim is to understand how the transfer of energy to the perturbations affects the evolution of the spatially averaged field, we neglect the higher-order terms in Eq.~(\ref{spatialavg}).  We note, however, that these terms represent higher-order corrections to the chameleon's evolution and may not be negligible, especially if the first-order backreaction term becomes large compared to $V^\prime(\phibar)$.  

To evaluate $\langle \delta\phi^2 \rangle$, we expand $ \delta\phi(\tau,\vec{x})$ in terms of creation and annihilation operators as shown in Eq.~(\ref{operatorquant}), and then we take the vacuum expectation value of $\delta\phi^2$.  As discussed in Appendix \ref{App:pp}, we regularize the resulting expression by subtracting terms associated with the vacuum state \cite{Kofman:2004yc}, which leaves
\beq
\langle\delta\phi^2 \rangle =  \frac{1}{a^2}\int \frac{d^3k}{(2\pi)^3}\left( |\phi_k|^2 - \frac{1}{2\omk}\right).
\eeq
In Appendix \ref{App:pp}, we show how the mode functions $\phi_k(\tau)$ may be expressed in terms of Bogoliubov coefficients $\ak$ and $\bk$.  Inserting Eq.~(\ref{WKB1}) into the expression for $\langle\delta\phi^2 \rangle$ gives
\beq
\langle\delta\phi^2 \rangle =  \frac{1}{a^2}\int \frac{d^3k}{(2\pi)^3} \frac{1}{\omk} \left(|\bk|^2 +\mathrm{Re}\left[\ak\bk^* e^{-2i\int^\tau \omk(\tau^\prime) d\tau^\prime}\right]\right).
\label{delsqrd}
\eeq

We can use this expression to show how the first-order backreaction term ensures conservation of energy.  The regularized energy density in perturbations  $\langle\rho_{\mathrm{fluct}}\rangle$ is given by Eq.~(\ref{rho_fluct}).  If we neglect the expansion of the Universe and take $a$ to be constant, then 
\beq
\frac{d}{dt} \langle\rho_{\mathrm{fluct}}\rangle = \frac{1}{a^4}\int \frac{d^3k}{(2\pi)^3} (\dot{\omega}_k n_k + \omk\dot{n}_k).
\label{defdt}
\eeq
Since $n_k = |\bk|^2$, we can use Eq.~(\ref{beta}) to evaluate $\dot{n}_k$:
\beq
\dot{n}_k  = \frac{\dot{\omega}_k}{\omk}\, \mathrm{Re}\left[\ak\bk^* e^{-2i\int^\tau \omk(\tau^\prime) d\tau^\prime}\right].
\eeq
In the limit that $a$ is constant, Eq.~(\ref{omegamain}) implies that \mbox{$\dot{\omega}_k = a^2 V^{\prime\prime\prime}(\phibar)\dot{\phibar}/(2\omk)$}.  Inserting both of these expressions into Eq.~(\ref{defdt}) and comparing with Eq.~(\ref{delsqrd}) yields
\beq
\frac{d}{dt} \langle\rho_{\mathrm{fluct}}\rangle = \frac{1}{2} V^{\prime\prime\prime}(\phibar) \dot{\phibar}  \langle\delta\phi^2 \rangle.
\eeq
The energy density in the spatially averaged field is \mbox{$\bar{\rho} = \dot{\phibar}^2/2 + V(\phibar)$}, which implies that
\beqa
\frac{d\bar{\rho}}{dt} &=& \dot{\phibar}\left[\ddot{\phibar} + V^\prime(\phibar)\right];\nonumber \\
&=& -\frac{1}{2} V^{\prime\prime\prime}(\phibar) \dot{\phibar}  \langle\delta\phi^2 \rangle,
\eeqa
where the last line follows from Eq.~(\ref{spatialavg}) with $H=0$.  We see that the sum $\langle\rho_{\mathrm{fluct}}\rangle+\bar{\rho}$ is constant on time scales that are short compared to $H^{-1}$.   Therefore, when energy is transferred to perturbations, the first-order backreaction term in Eq.~(\ref{spatialavg}) ensures that an equal amount of energy is extracted from the spatially averaged field.  

To probe the backreaction of the perturbations on the spatially averaged field further, we return to Eq.~(\ref{delsqrd}), which expresses $\langle\delta\phi^2 \rangle$ in terms of the Bogoliubov coefficients $\ak$ and $\bk$.  In the previous section, we showed that the perturbation modes that we expect to be excited during the rebound ($k\sim |\vel|/M$) must have $n_k \ll 1$.  Since $n_k = |\bk|^2$, the normalization condition for the Bogoliubov coefficients ($|\ak|^2-|\bk|^2 = 1$) demands that $|\ak| \simeq 1$ when $n_k \ll 1$.  In this regime of perturbative particle production, we can obtain an approximate solution for $\bk$ by taking $\ak = 1$ in Eq.~(\ref{beta}) \cite{Braden:2010wd}:
\beq
\beta_k(\tau) = \int_0^\tau \frac{\omega_k'(\tilde{\tau})}{2\omega_k(\tilde{\tau})} e^{-2i\int^{\tilde{\tau}} \omega_k(\tau')d\tau'} d\tilde{\tau}, \label{betasol} 
\eeq
where we have chosen $\tau=0$ to correspond to some time before particle production begins.  Inserting this expression into Eq.~(\ref{delsqrd}) and taking $\ak = 1$ yields
\beqa
\langle\delta\phi^2 \rangle &=&  \frac{1}{a^2}\int \frac{d^3k}{(2\pi)^3} \frac{1}{\omk(\tau)} \label{delexp} \\
&&\times \left[|\bk|^2 + \int_0^\tau \frac{\omega_k'(\tilde{\tau})}{2\omega_k(\tilde{\tau})} \cos \left( 2 \int_{\tilde{\tau}}^\tau \omega_k(\tau')d\tau' \right) d\tilde{\tau}\right]. \nonumber
\eeqa
Given that $|\bk| \ll 1$, we expect the $|\bk|^2$ term to be much smaller than the $\ak\bk^*$ term that generates the second term in the integrand of Eq.~(\ref{delexp}).  One may be concerned, however, that this second term involves the integral of an oscillating function and may therefore be suppressed relative to the $|\bk|^2$ term.  However, our approximate solution for $\bk(\tau)$ implies that
\beq
|\bk^2| = \int_0^\tau \int_0^\tau \frac{\omega_k'(\tilde{\tau}_1)\omega_k'(\tilde{\tau}_2)}{4\omega_k(\tilde{\tau}_1)\omega_k(\tilde{\tau}_2)} \cos \left( 2 \int_{\tilde{\tau}_1}^{\tilde{\tau}_2} \omega_k(\tau')d\tau' \right) d\tilde{\tau_1} d\tilde{\tau_2},
\eeq
so $|\bk|^2$ also contains a cosine integral.   Therefore, we may safely assume that the $|\bk|^2$ term makes a negligible contribution to $\langle\delta\phi^2 \rangle$.   

Since we are only interested in time scales that are much shorter than the Hubble time, we can further simplify this expression for $\langle\delta\phi^2 \rangle$ by taking the scale factor $a$ to be constant and using Eq.~(\ref{omegaprime}) to evaluate $\omega_k'$.  With these simplifications, we obtain
\beqa
&&\langle\delta\phi^2 \rangle = \frac{1}{8\pi^2}\int_0^t dt'\, V'''[\phibar(t')]\dot{\phibar}(t') \times \label{delsqrdfinal} \\
&&\left[ \int \frac{k_\mathrm{phys}^2 dk_\mathrm{phys}}{ \omega_{k,\mathrm{phys}}(t)\omega_{k,\mathrm{phys}}^2(t')} \cos \left(2 \int_{t'}^t \omega_{k,\mathrm{phys}}(t'')dt''\right) \right],
\nonumber
\eeqa
where we have defined $\omega_{k,\mathrm{phys}}^2 \equiv (\omk/a)^2 = k_\mathrm{phys}^2 + V''(\phibar)$.  For the rest of this section, we will neglect the expansion of the Universe and deal only with physical wavenumbers evaluated at the time of the chameleon's rebound.  To make the expressions less cluttered, we will omit the ``phys" subscript on both $k$ and $\omk$ in subsequent equations.  For consistency, we will also drop the Hubble friction term from Eq.~(\ref{spatialavg}); recall from Section\ref{firstlook} that this friction term has a negligible impact on the evolution of $\phibar$ during the rebound.

We now have a new equation for the evolution of the spatially averaged $\phibar$ field that includes the first-order backreaction:
\beq
\ddot{\phibar}+ V^\prime(\phibar) + {\cal D}(t)  = 0,
\label{phibareom}
\eeq
where ${\cal D}(t) \equiv (1/2) V'''[\phibar(t)] \langle\delta\phi^2\rangle$ with $\langle\delta\phi^2\rangle$ given by Eq.~(\ref{delsqrdfinal}) is the ``dissipation" term that expresses how energy is transferred from the spatially averaged field to the perturbations.  The dissipation term ${\cal D}(t)$ is non-Markovian; it depends on the entire history of the chameleon's evolution up to $t$ and therefore has ``memory."  The non-Markovian nature of the dissipation term was also highlighted in an earlier analysis of particle production in scalar field theory \cite{Boyanovsky:1994me}, which used the ``in-in" formalism to calculate dissipation from particle production in $\phi^4$ theory.  If $\omk$ is assumed to be constant, then our ${\cal D}(t)$ is identical to their dissipation term (see Eq.~(27) in Ref.~\cite{Boyanovsky:1994me}), which demonstrates how our perturbative Bogoliubov technique can simplify particle production calculations.   Ref.~\cite{Boyanovsky:1994me} showed that there is no Markovian limit for the dispersion term in $\phi^4$ theory.  However, the fact that $V'''(\phi)$ sharply increases as $\phi$ decreases for chameleon potentials implies that, prior to the rebound, the non-Markovian integral in Eq.~(\ref{delsqrdfinal}) is dominated by values of $t'$ that are just slightly smaller than $t$, and we will show that this feature allows us to derive a local approximation for $\calD(t)$.  

The dominance of the $t' \lsim t$ portion of the time integral in Eq.~(\ref{delsqrdfinal}) also ensures that the cosine integral in this expression is positive, which allows us to extract the qualitative behavior of $\calD(t)$.  As $\phibar$ rolls to smaller values prior to the rebound, $\dot{\phibar}$ is nearly constant (as shown in Fig.~\ref{Fig:ClassicalBounce}), and ${\cal D}(t)$ is proportional to $\dot{\phibar}$ with a positive coefficient.  At this stage, ${\cal D}(t)$ acts like a drag term, and it slows the chameleon down.  However, unlike a drag term, ${\cal D}(t)$ is actually an integral over $\dot{\phibar}$, which implies that it does not vanish when $\dot{\phibar} = 0$.  Instead, ${\cal D}(t)$ will remain negative during and shortly after the rebound.  Consequently, ${\cal D}(t)$ acts like a new potential term during the rebound: we will show that ${\cal D}\simeq V'_{\cal D}(\phi)$ for some $V_\calD (\phi)$.  

To derive this new ``dissipative potential," we must further simplify our expression for ${\cal D}(t)$.  First, we assume that the modes that are excited have $k^2 \gg V''(\phi)$ throughout the rebound,\footnote{Although the non-adiabatic modes had $k^2 \lsim V''(\phi)$ in the classical solution for the evolution of $\phibar$, as seen in Fig.~\ref{Fig:ClassicalBounce}, we will show that the backreaction term prevents $V''(\phi)$ from exceeding $k^2$ by forcing the chameleon to turn around at a larger value of $\phibar/M$.} which allows us to set $\omk(t) \simeq k$ in Eq.~(\ref{delsqrdfinal}).  Second, we impose an infrared (IR) cut-off on the cosine integral in Eq.~(\ref{delsqrdfinal}) and consider only $k \geq \kIR$.  This IR cut-off makes the separation of $\phibar$ and $\delta \phi$ explicit; modes with $k<k_\mathrm{IR}$ are absorbed into $\phibar$, while modes with $k\geq k_\mathrm{IR}$ are considered perturbations.  Clearly, we must choose $\kIR$ to be smaller than the wavenumbers of the modes that we expect to be excited during the rebound.  However, we do not want to make $\kIR$ arbitrarily small because we only solve a linearized equation for $\delta \phi$ [Eq.~(\ref{linear_no_br})], while the equation of motion for $\phibar$ is nonlinear.  Therefore, decreasing $\kIR$ implies that we are neglecting more nonlinear effects.  When we calculate particle production numerically in the next section, we will see that taking different values of $\kIR$ can be used to determine the importance of nonlinear field interactions.  For now though, we simply assume that $\kIR \lsim (\Delta t)^{-1}$, where $\Delta t \lsim M/|\vel|$ is the duration of particle production during the rebound.  

With these simplifications, we find that
\beq
\calD(t) = - \frac{V^{\prime\prime\prime}[\phibar(t)]}{16\pi^2}  \int_0^t V^{\prime\prime\prime}[\phibar(t^\prime)] \dot{\phibar}(t^\prime) \mathrm{Ci}\left[{2 k_\mathrm{IR}(t-t^\prime)}\right]dt^\prime,
\label{DiwithCi}
\eeq
where $\mathrm{Ci}(x) \equiv - \int_x^\infty (\cos{y})/{y} \, dy$.  We now take advantage of the fact that $V^{\prime\prime\prime}(\phi)$ sharply increases as $\phi$ decreases, which implies that the integral over $t'$ will be dominated by a limited range of values that are just slightly smaller than $t$.  Moreover, $\mathrm{Ci}(x)$ is divergent for small $x$, which further enhances the contribution to the integral from small values of $(t-t')$.   Numerical evaluation of Eq.~(\ref{DiwithCi}) using the $\phibar(t)$ solution found numerically in the next section confirms that restricting $(t-t') \lsim 0.1 M/|\vel|$ does not significantly change $\calD(t)$ near the rebound.  Therefore, we can assume that $k_\mathrm{IR}(t-t^\prime) < 1$ and use the approximation
\beq
 \mathrm{Ci}[k_\mathrm{IR}(t-t^\prime)]  \simeq \gamma_E + \ln[k_\mathrm{IR}(t-t^\prime)],
 \eeq
where $\gamma_E \simeq 0.577$ is Euler's constant, to obtain
\beqa
\calD(t) = - \frac{V^{\prime\prime\prime}[\phibar(t)]}{16\pi^2} \int_{t_\mathrm{min}}^t &dt^\prime&\left[ \frac{d}{dt^\prime}V^{\prime\prime}[\phibar(t^\prime)]\right] \label{DiInt}
\\
&&\times\{ \gamma_E + \ln\left[{2 k_\mathrm{IR}(t-t^\prime)}\right]\},  \nonumber
\eeqa
where $t_\mathrm{min} = t - 0.1 M/|\vel|$.  We then integrate Eq.~(\ref{DiInt}) by parts, and we make the approximation
\beq
\int_{t_\mathrm{min}}^t \frac{V^{\prime\prime}\left[\phibar(t^\prime)\right]}{t-t^\prime}dt^\prime \simeq V^{\prime\prime}[\phibar(t)]\int_{t_\mathrm{min}}^t  \frac{dt^\prime}{t-t^\prime},
\eeq
again taking advantage of the fact that contributions from $t'\simeq t$ dominate the integral, to obtain
\beqa
\calD(t) &=& - \frac{V^{\prime\prime\prime}[\phibar(t)]}{16\pi^2} \left\{V^{\prime\prime}\left[\phibar(t)\right]-V^{\prime\prime}\left[\phibar(t_\mathrm{min})\right]\right\} \nonumber \\
&&\times\left\{ \gamma_E + \ln\left[{2 k_\mathrm{IR}(t-t_\mathrm{min})}\right]\right\}.
\eeqa
Since $V''(\phi)$ increases sharply as $\phi$ decreases, $V''\left[\phibar(t)\right] \gg V^{\prime\prime}\left[\phibar(t_\mathrm{min})\right]$ as the chameleon climbs its bare potential and turns around.  Also, during the rebound, $\ln\left[{2 k_\mathrm{IR}(t-t_\mathrm{min})}\right]$ changes only slightly, so we may approximate it as constant.  We then find that 
\beq
\calD(t) \simeq \kappa V^{\prime\prime\prime}(\phibar)V^{\prime\prime}\left(\phibar\right),
\label{approxD}
\eeq
for some constant $\kappa$.  Comparing the numerical evaluation of $\calD(t)$ as given by Eq.~(\ref{DiwithCi}) to the approximation given by Eq.~(\ref{approxD}) indicates that $\kappa$ increases from $\sim\!\!0.02$ to $\sim\!\!0.05$ during the rebound for both exponential and power-law potentials.  Therefore, we expect that a value of $\kappa$ in this range will accurately approximate ${\cal D}(t)$ during the rebound; we will see in the next section that this is indeed the case. 

The approximate expression for $\calD(t)$ during the rebound given by Eq.~(\ref{approxD}) shows that the backreaction of the perturbations on the evolution of $\phibar$ effectively adds a new term to the chameleon potential: $\calD(t) \simeq V_\calD'(\phi)$, where
\beq
V_\calD(\phi) = \frac{\kappa}{2} \left[V^{\prime\prime}\left(\phibar\right)\right]^2,
\label{VDdef}
\eeq
with $0.02 < \kappa <0.05$.  For both exponential and power-law potentials, $V_\calD'(\phi) > V'(\phi)$ while $\phi < M$, so the dissipative potential will dominate the field's evolution during the rebound.  This dominance of the first-order backreaction term is concerning, for it indicates that higher-order contributions to the backreaction are probably not negligible and signals a breakdown of perturbation theory.  Our primary aim in this Section, however, is to understand how the extraction of energy from the spatially averaged field affects particle production during the rebound.  Since the first-order backreaction captures this energy transfer, we can use our analysis of the first-order backreaction to gain insight into how the evolution of $\phi$ is affected by particle production.   

Since the dissipative potential is dominant when \mbox{$\phi < M$}, it determines the minimum value of $\bar{\phi}$ during the rebound, which we denote $\phita$: $V_\calD(\phita) \equiv \vel^2/2$.  It follows from Eq.~(\ref{VDdef}) that the maximum value of the chameleon's effective mass during the rebound is much smaller than the wave numbers of the excited modes ($k \gsim |\vel|/M$): \mbox{$V''(\phita) = |\vel|/\sqrt{\kappa} \ll \vel^2/M^2$} given that $|\vel|\gg M^2$.   Therefore, the backreaction prevents $m_\mathrm{eff}$ from reaching the extremely large values seen in Fig.~\ref{Fig:ClassicalBounce}, and our approximation that $\omk \simeq k$ in Eq.~(\ref{delsqrdfinal}) is justified.  Furthermore, the field will turn around before the adiabatic ratio $\omega^\prime_k(\tau)/\omega_k^2$ exceeds unity, which will keep $n_k \ll 1$ for the excited modes.

Next, we consider how the dissipative potential affects the duration of the rebound.  Thus far, we have used $\Delta t \lsim {M/|\vel|}$ to estimate the duration of the rebound.  However, Fig.~\ref{Fig:ClassicalBounce} shows that this definition of $\Delta t$ overestimates the duration of the change in $m_\mathrm{eff}$, which is the timescale that determines which perturbation modes will be excited.  For the parameters shown in Fig.~\ref{Fig:ClassicalBounce}, $M/|\vel| = 10^{-11}\, \mathrm{GeV}^{-1}$, but $m_\mathrm{eff}$ changes significantly and $|\omk^\prime|/\omk^2$ exceeds unity for a much shorter time: $\sim 5\times  10^{-14}\, \mathrm{GeV}^{-1}$.  Therefore, we need to refine our calculation of $\Delta t$.  We also need to incorporate our new understanding of the evolution of the spatially averaged field.  While $\phibar \sim \phita$ during the rebound, the equation of motion for $\phibar$, including the first-order backreaction, is approximately 
\begin{align}
\ddot{\phibar} &\simeq - V_\calD'(\phibar) \nonumber \\
&\simeq  -  V_\calD'(\phita) - m_\calD^2(\phibar-\phita), \label{approxeom}
\end{align}
where we have defined $m_\calD^2 \equiv V_\calD''(\phita)$.  If we set $t=0$ when the field turns around ($\phibar= \phita$), the solution to Eq.~(\ref{approxeom}) is 
\beqa
\phibar - \phita &=& -\frac{V_\calD'(\phita)}{m_\calD^2}\left[1 - \cos(m_\calD t)\right] \nonumber \\
&\simeq& -\frac12 V_\calD'(\phita) t^2 \quad \mathrm{for} \,\, m_D t \ll1. \label{phitsol}
\eeqa
We can now calculate how long it takes for $m_\mathrm{eff}^2$ to change significantly:  $|V''(\phibar) - V''(\phita)|/V''(\phita) = 1$ when 
\beqa
\phibar - \phita &=& {\left|\frac{V''(\phita)}{V'''(\phita)}\right|}; \nonumber \\
t &=& \sqrt{\frac{2 V''(\phita)}{V_\calD'(\phita)V'''(\phita)}}, \label{tchange}
\eeqa
where the last line follows from Eq.(\ref{phitsol}).  (We are interested in the change in $m_\mathrm{eff}^2$, as opposed to $m_\calD^2$, because $m_\mathrm{eff}^2$ governs the behavior of the perturbations.)  To account for the change in $m_\mathrm{eff}^2$ as $\phibar$ approaches $\phita$ and as $\phibar$ rolls away from $\phita$, we multiply Eq.~(\ref{tchange}) by 2 when evaluating $\Delta t$.  Our estimated value of the wave number of the most energetic excited mode ($\kex$) is then
\beq 
\kex = (\Delta t)^{-1} = \frac{1}{2} \sqrt{\frac{V_\calD'(\phita)V'''(\phita)}{2 V''(\phita)}},
\label{kex}
\eeq
where $\phita$ satisfies $V_\calD(\phita) \equiv \vel^2/2$.

If the chameleon has an exponential bare potential,
\beq
V(\phi) = \Mp^4 \exp\left[\left(\frac{M_s}{\phi}\right)^n\right],
\label{Vexp}
\eeq
then evaluating Eq.~(\ref{kex}) gives
\beq
\kex = \frac{n}{2}\left(\frac{M_s}{\phita}\right)^{n+1} \frac{\sqrt{V_\calD(\phita})}{M_s}\left[1+{\cal O}\left(\frac{\phita^n}{M_s^n}\right)\right].
\label{kexexpfull}
\eeq
This potential is more general than those we have considered previously, because it allows for two mass scales.  To avoid fine-tuning, we will assume that $M_s/\Mp$ is of order unity and we will continue to assume that $\Mp \sim 10^{-3} \, \mathrm{eV}$ to maintain a connection to dark energy.
Numerically solving $V_\calD(\phita) \equiv \vel^2/2$ for $2 \leq n \leq 10$ and $10^{-6} < |\vel|/\mathrm{GeV}^2 < 10^6$ reveals that $(\phita/M_s)^n \lsim 0.05$ in all cases, so we may neglect the ${\cal O}\left({\phita^n}/{M_s^n}\right)$ term in Eq.~(\ref{kexexpfull}).  We then use $V_\calD(\phita) \equiv \vel^2/2$ to obtain
\beq
\kex \simeq \frac{n}{2\sqrt{2}}\left(\frac{M_s}{\phita}\right)^{n+1} \frac{|\vel|}{M_s}.
\label{kexexp}
\eeq
Since $V_\calD(\phita) \equiv \vel^2/2$ with an exponential potential is a transcendental equation, we cannot obtain an exact algebraic expression for $\phita/M_s$.  For a limited range of $|\vel|$ values, however, we can approximate $\phita$ as 
\beq
\frac{\phita}{M_s}\simeq \frac{1}{c_n} \left(\frac12 \ln\left[\frac{\vel^2(M_s/\Mp)^4}{\kappa n^4 \Mp^4}\right]\right)^{-1/n},
\label{phitaexp}
\eeq
where $c_n$ is an constant of order unity that depends on the range of $|\vel|$ values under consideration.  Inserting this expression into Eq.~(\ref{kexexp}) yields 
\beq
\kex \simeq \frac{n b_n|\vel|}{2\sqrt{2}M_s} \ln^{\frac{n+1}n}\left[\frac{\vel^2(M_s/\Mp)^4}{n^4\kappa \Mp^4}\right],
\label{kexexpapprox}
\eeq
where $b_n = (c_n/2^{1/n})^{n+1}$ is also an order-unity constant.\footnote{Equation (\ref{kexexpapprox}) differs from the equation for $\kex$ given in our earlier work \cite{ourprl} by a factor of $1/\sqrt{2}$ because we initially used the time required for $|V''(\phibar) - V''(\phita)|/V''(\phita) = 0.5$ as the starting point of our derivation of $\kex$.  We later realized that $|V''(\phibar) - V''(\phita)|/V''(\phita) = 1$ gave a better fit to the numerical results obtained Section \ref{sec:numerics}, so we modified our expression for $\kex$.}  For $10^{-6} < |\vel|/\mathrm{GeV}^2 < 10^6$, $b_n$ ranges from 0.25 for $n=2$ to $0.38$ for $n=10$, and Eq.~(\ref{kexexpapprox}) differs from Eq.~(\ref{kexexp}) by less than 7\%.  Equation~(\ref{kexexpapprox}) shows that $\kex$ is rather insensitive to both $\kappa$ and $n$; varying $\kappa$ within the range $0.01 \leq \kappa \leq 0.06$ has nearly no impact on $\kex$, and increasing $n$ from 2 to 10 changes $\kex$ by less than 25\% for $10^{-6} < |\vel|/\mathrm{GeV}^2 < 10^6$.

If the chameleon has a power-law bare potential,
\beq
V(\phi) = \Mp^4\left[1+\left(\frac{M_s}{\phi}\right)^n\right],
\label{Vpow}
\eeq
then evaluating Eq.~(\ref{kex}) gives
\beq
\kex =  \frac{(n+2)}{2\sqrt{2}} \frac{|\vel|}{M_s}\left(\frac{M_s}{\phita}\right).
\label{kexpow}
\eeq
For this potential, we can algebraically solve \mbox{$V_\calD(\phita) \equiv \vel^2/2$} for $\phita$ to obtain
\beqa
\kex = \frac{(n+2)}{2\sqrt{2}} \frac{|\vel|}{M_s} \left[\frac{|\vel|/ \Mp^2}{n(n+1)\sqrt{\kappa}}\left(\frac{M_s}{\Mp}\right)^2\right]^{1/(n+2)}.\label{kexpowfull} 
\eeqa
Fortunately, Eq.~(\ref{kexpowfull}) indicates that $\kex$ is still relatively insensitive to order-unity changes in $\kappa$.  For power-law potentials, however, $\kex$ depends very strongly on $n$, with larger $n$ values giving smaller values for $\kex$.  Furthermore, comparing Eq.~(\ref{kexpowfull}) to Eq.~(\ref{kexexpapprox}) reveals that exponential potentials have smaller $\kex$ values than power-law potentials.  In both cases, our earlier estimate, \mbox{$\kex \simeq |\vel|/M$}, significantly underestimates $\kex$ by missing factors related to ($M_s/\phita$), which is much greater than unity.  Steeper potentials generally have smaller $\kex$ values because $\phibar$ turns around at a larger value.  

To summarize the key results of this section, we evaluated the first-order backreaction of the perturbations on the spatially averaged field $\phibar$ and found that the dynamics of $\phibar$ during the rebound are governed by a new ``dissipative" potential given by Eq.~(\ref{VDdef}).  This new potential forces the chameleon field to turn around much earlier than the solution without backreaction predicted, which prevents $\omk^\prime/\omk^2$ from exceeding unity.  Consequently, the occupation numbers of the excited modes remain very small.  We then used the dissipative potential to estimate the duration of the rebound $\Delta t$.  Assuming that the rebound will excite perturbation modes with $k\lsim (\Delta t)^{-1}$, we derived predictions for the wave number of the most energetic excited mode ($\kex$) for both exponential and power-law potentials.  We found that $\kex \gg |\vel|/M$ in both cases, which indicates that the backreaction does not prevent the transfer of energy to extremely energetic modes.  Therefore, we now have two reasons to expect that chameleon gravity will suffer a computational breakdown during the rebound; the rebound will excite modes that lie far beyond the expected limits of effective field theory, and if the theory can be trusted during the rebound, the backreaction of these modes will also significantly alter the evolution of the spatially averaged field.

\subsection{Numerical computation of particle production}
\label{sec:numerics}
We test our analytical analysis of the rebound by numerically solving the linearized perturbation equations and the spatially averaged equation with first-order backreaction.  In our numerical analysis, we neglect the expansion of the Universe and take the scale factor to be constant.  We define ``physical" creation and annihilation operators $\hat{a}_{\bf k, \ph} = a^{3/2} \hat{a}_{\bf k}$ that obey the commutation relation
\begin{equation}
  \left[\hat{a}_{\bf k,\ph} , \hat{a}_{\bf k', \ph}^\dagger \right] = (2\pi)^3\delta^{(3)}\left({\bf k_\ph}-{\bf k'_\ph}\right),
\end{equation}
and express $\delta\phi(t,\vec{r}=a\vec{x})$ in terms of these operators:
\begin{align}
\delta\phi(t,{\bf r}) = \int \frac{d^3k_\ph}{(2\pi)^{3}} &\bigg[\hat{a}_{{\bf k},\ph} \phi_{k,\ph}(t) e^{i{\bf k}_\ph\cdot{\bf r}} \label{delphiNE} \\
&+ \hat{a}^{\dagger}_{{\bf k}, \ph} \phi^*_{k,\ph}(t) e^{-i {\bf k}_\ph \cdot{\bf r}} \bigg]. \nonumber
\end{align}
Comparing Eq.~(\ref{delphiNE}) to Eq.~(\ref{operatorquant}) reveals that \mbox{$\phi_{k,\ph} = \sqrt{a} \phi_k$}.  If we take $a$ to be constant, then Eq.~(\ref{mode_eq}) implies 
\beq
\ddot{\phi}_{k, \ph}+\omega_{k,\ph}^2\phi_{k, \ph}= 0, \label{lineq}\\ 
\eeq
where $\omega_{k,\ph}^2 \equiv k_\mathrm{phys}^2 + V''[\phibar(t)]$.  We solve this equation for $\phi_{k, \ph}$ for several logarithmically spaced $k_\ph$ values with $\kIR < k_\ph < k_\mathrm{max}$.  As in the previous section, $\kIR$ separates the long-wavelength perturbations that are included in $\phibar$ from the shorter-wavelength perturbations that compose $\delta \phi$, and it is chosen to be smaller than the wavenumbers of the modes we expect to be excited during the rebound ($\kIR \ll \kex$).  We also do not expect modes with $k_\ph \gg \kex$ to be excited, so we choose a value of $k_\mathrm{max}$ that is much larger than $\kex$.  The number of $k$ values we sample depends on the ratio $k_\mathrm{max}/\kIR$ and is chosen so that the interval between $\log{k}$ values is $\sim$0.05.  

To evaluate $\omega_{k,\ph}^2$ in Eq.~(\ref{lineq}), we have to solve Eq.~(\ref{phibareom}) for $\phibar(t)$:  
\beq
\ddot{\phibar} + V^\prime(\phibar) + \frac{1}{2} V^{\prime\prime\prime}(\phibar)\langle \delta\phi^2 \rangle = 0. \label{bkgrd} \\
\eeq
We evaluate $\langle\delta\phi^2 \rangle$ at each time step by converting Eq.~(\ref{delsqrd}) into physical variables and then using the $\phi_{k, \ph}$ solutions to compute the integral
\beq
\langle\delta\phi^2 \rangle =  \int_{\kIR}^{k_\mathrm{max}} \frac{k_\ph^2dk_\ph}{2\pi^2}\left( |\phi_{k,\ph}|^2 - \frac{1}{2\omega_{k,\ph}}\right).
\eeq
With the $\phi_{k, \ph}$ solutions, we can also evaluate the occupation number $n_k = |\beta_k|^2$:
\begin{equation}
  n_k = \frac{1}{2\omega_{k, \ph}} \left[  |\dot{\phi}_{k, \ph}|^2 + \omega_{k,\ph}^2 |\phi_{k,\ph}|^2  \right] - \frac{1}{2},
  \label{nkphys}
\end{equation}
which corresponds to Eq.~(\ref{nkdef}) if $a$ is constant.  We also use Eq. (\ref{Ek}) to evaluate $\rho_k$ for each $\phi_{k, \ph}$ solution.

\begin{figure}
 \centering
 \resizebox{3.4in}{!}
 {
      \includegraphics{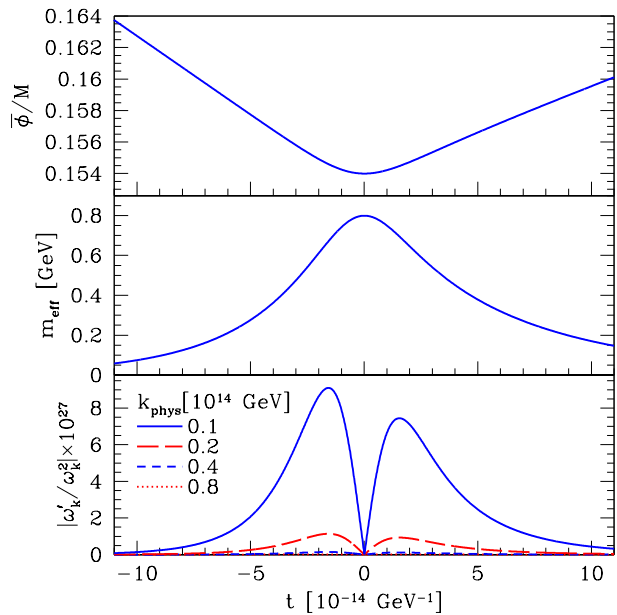}
 }
\caption{The evolution of the spatially averaged chameleon field, including the backreaction from particle production, as it rebounds off its bare potential for the same parameters as in Figure \ref{Fig:ClassicalBounce}.  The top panel shows that the field value turns around at a much larger value than predicted by the classical solution, and the middle panel shows that the effective mass is confined to much smaller values.  The bottom panel shows that the earlier turn-around ensures that the adiabatic ratio is always much less than unity.}
\label{Fig:NumBounce}
\end{figure}
\begin{figure}
 \centering
 \resizebox{3.4in}{!}
 {
      \includegraphics{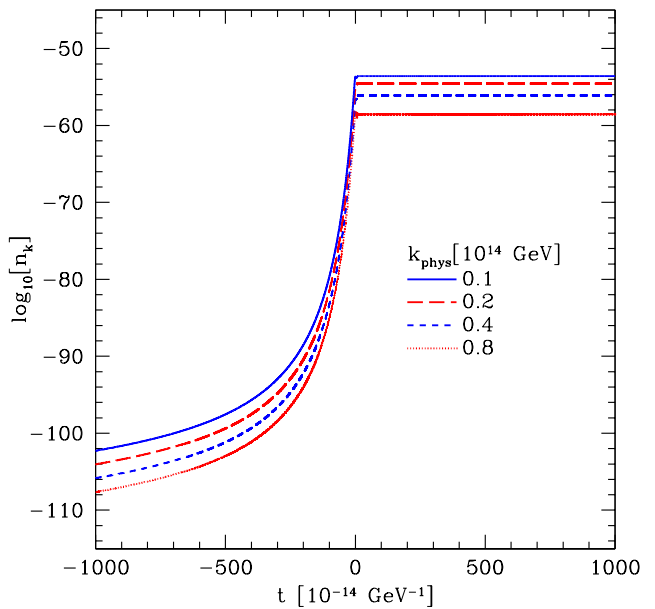}
 }
\caption{The evolution of the occupation number $n_k$ for four wave numbers: $k_\mathrm{phys} = \{0.1,0.2,0.4, 0.8\} \times 10^{14}$ GeV.  As in Figures \ref{Fig:ClassicalBounce} and \ref{Fig:NumBounce}, $\vel = -0.1\, \rm{GeV}^2$ and $V(\phi)$ is given by Eq.~(\ref{pot}) with $n=2$ and $M=10^{-3}$ eV.  The occupation number is initially zero, and then it increases sharply just before $\phibar$ turns around at $t=0$.}
\label{Fig:nkevol}
\end{figure}

When we numerically solve Eqs. (\ref{lineq}) and (\ref{bkgrd}), we must choose initial conditions for $\phibar(t)$ and all the $\phi_{k, \ph}(t)$ functions.  We initially set $\phibar = 2M$ with $\dot{\phibar} = \vel$, where $\vel$ is chosen from the range of velocities shown in Figure \ref{Fig:VelContour}.  Since most of the particle production occurs near $\phita \lsim 0.1M$, the final spectrum of perturbations is insensitive to the initial value of $\phibar$, provided that it is greater than 0.5$M$.  We neglect the tiny amount of particle production that occurs before $\phibar = 2M$ and initially set $n_k = 0$.  If $n_k = 0$ at some initial time $t_i$, Eqs.~(\ref{WKB1}) and (\ref{alphbeta}) imply that 
\begin{subequations}\label{vacphik}
\begin{align}
\phi_{k, \ph}(t_i) &= \frac{1}{\sqrt{2\omega_{k,\ph}(t_i)}} e^{-i\int^{t_i} \omega_{k,\ph}(t')dt'}; \\
\dot{\phi}_{k, \ph}(t_i) &= -i \omega_{k,\ph}(t_i) \phi_{k, \ph}(t_i).
\end{align}
\end{subequations}

We do not solve directly for $\phi_{k, \ph}(t)$ in our numerical analysis because it is impossible to accurately evaluate Eq.~(\ref{nkphys}) when $n_k \ll 1$.  Instead, we express $\phi_{k, \ph}(t)$ as
\beq
\phi_{k, \ph}(t) = \frac{1}{\sqrt{2W_k(t)}} e^{-i \theta_k(t)}.
\eeq
Equation~(\ref{lineq}) with an initial condition given by Eq.~(\ref{vacphik}) requires that $\dot{\theta}_k(t) = W_k(t)$. Equation~(\ref{lineq}) also provides an evolution equation for $W_k(t)$, and demanding that $n_k = 0$ at the initial time requires $W_k(t_i) = \omega_{k,\ph}(t_i)$ and $\dot{W}_k(t_i) = 0$.  We then define a new function $\varpi(t) \equiv W_k(t) - \omega_{k,\ph}(t)$ that describes the deviation of $\phi_{k, \ph}(t)$ from Eq.~(\ref{vacphik}), and we numerically solve Eq.~(\ref{lineq}) for the evolution of $\varpi(t)$. When the $\phi_{k, \ph}$ terms in Eq.~(\ref{nkphys}) are expressed as functions of $\varpi$, the result is $1/2+f(\varpi, \dot{\varpi})$.  Therefore, we can evaluate $n_k$ directly from $\varpi(t)$ without having to numerically subtract the vacuum contribution (the $1/2$ term in Eq.~\ref{nkphys}), thus avoiding the numerical errors introduced by subtracting two numbers with values that far exceed their difference.  

Figure \ref{Fig:NumBounce} shows the numerically computed evolution of the spatially averaged chameleon field after starting at $\phibar = 2M$ with a velocity $\vel = -0.1\, \rm{GeV}^2$.  The potential $V(\phi)$ is given by Eq.~(\ref{pot}) with $n=2$ and $M=10^{-3}$ eV.  Comparing this figure to the classical solution shown in Fig.~\ref{Fig:ClassicalBounce} for the same $V(\phi)$ and initial conditions reveals how profoundly the evolution of $\bar{\phi}$ is affected by the transfer of energy to perturbations.  As predicted in the previous section, the field turns around when $V_{\cal D}(\phibar) = \vel^2/2$, which gives a much larger value for $\phita$ than the classical $V(\phita) = \vel^2/2$.  Consequently, $\phibar$ turns around before its effective mass exceeds a GeV and before the adiabatic ratio for $k\simeq (\Delta t)^{-1}$ exceeds unity, in stark contrast to the classical evolution depicted in Fig.~\ref{Fig:ClassicalBounce}.  Since the adiabatic ratio is always very small, we expect $n_k \ll 1$ as well.  Figure \ref{Fig:nkevol} shows the evolution of $n_k$ for the same wavenumbers; we see that $n_k$ increases dramatically as the field rebounds, and then it maintains a constant value as the field rolls out to larger values.  We also see that $n_k \ll 1$; as expected, the excited modes are so energetic that $n_k \ll 1$ is required to conserve energy.

\begin{figure}
 \centering
 \resizebox{3.4in}{!}
 {
      \includegraphics{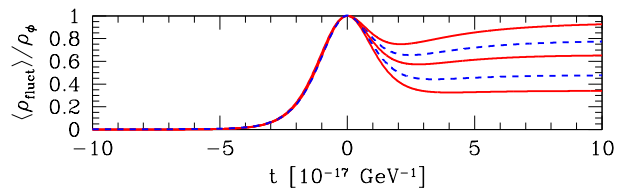}
 }
\caption{The evolution of the energy density in fluctuations, $\langle{\rho_{\mathrm{fluct}}}\rangle$, as a fraction of the total energy density of the chameleon field $\rho_\phi \simeq \vel^2/2$.  In this figure, $\vel = -100\, \mathrm{GeV}^2$, and $V(\phi)$ is given by Eq.~(\ref{pot}) with $n=2$ and $M=10^{-3}$ eV.  
The different curves show different $\kIR$ values: from bottom to top, $\kIR = 10^{13}, 10^{14}, 10^{14.7}, 10^{15},$ and $10^{15.3}$ GeV.  In all cases, $k_\mathrm{max} = 10^{18}$ GeV.  The spatially averaged field turns around at $t=0$.
}
\label{Fig:Epert}
\end{figure}

Even though $n_k \ll 1$, the fluctuations still contain a significant fraction of the chameleon's energy.  Figure \ref{Fig:NumBounce} illustrates that the rebound is not elastic; $\phibar$ rolls out with a smaller velocity because some energy has been transferred to the fluctuations.  Figure \ref{Fig:Epert} shows the evolution of $\langle{\rho_{\mathrm{fluct}}}\rangle$, as defined by Eq.~(\ref{rho_fluct}), for $\vel = -100\, \mathrm{GeV}^2$.  We see that all of the chameleon's energy is transferred to fluctuations at the rebound (at $t=0$), but then some of that energy is returned to the spatially averaged field; all values of $\vel$ share this basic behavior.  The post-rebound transfer of energy from the fluctuations to $\phibar$ is a manifestation of the dissipative potential $V_{\cal D}(\phi)$; since the backreaction of the perturbations on $\phibar$ acts as a new potential immediately after the rebound, it can accelerate $\phibar$ as it rolls out to larger values.   The amount of energy returned to $\phibar$ after the rebound depends on $\kIR$; Figure \ref{Fig:Epert} shows that smaller values of $\kIR$ lead to less final energy in fluctuations.  Since, $\kIR$ determines which modes are treated linearly, this dependence on $\kIR$ indicates that the final value of $\langle{\rho_{\mathrm{fluct}}}\rangle$ depends on nonlinear interactions that are not included in our analysis.  Therefore, we cannot determine how much energy is transferred to perturbations during the rebound.  This limitation is disappointing, but not surprising; as we discussed in the previous section, the dominance of the dissipative potential $V_{\cal D}(\phi)$ over $V(\phi)$ during the rebound foretold that our linear analysis with only a first-order backreaction would be insufficient to determine the chameleon's final state.  However, the fact that increasing $\kIR$ (and thus including more nonlinear effects) increases the final value of $\langle{\rho_{\mathrm{fluct}}}\rangle$ indicates that nonlinear interactions are unlikely to prevent the transfer of energy to fluctuations.  

\begin{figure}
 \centering
 \resizebox{3.4in}{!}
 {
      \includegraphics{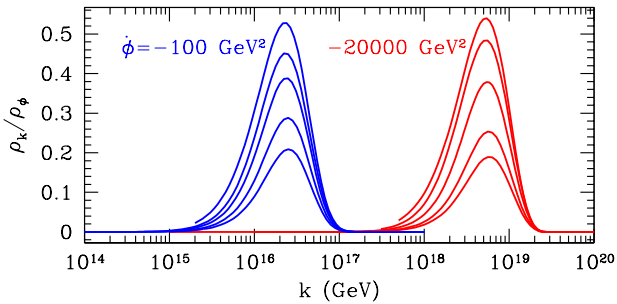}
 }
\caption{The fluctuation energy density per logarithmic interval in $k$ (see Eq.~\ref{Ek}) as a fraction of the total energy density in the chameleon field.  The left set of peaks, with maxima at $k=2.5\times10^{16}$ GeV, have $\vel = -100\,\mathrm{GeV}^2$ and $k_\mathrm{max} = 10^{18}$ GeV, while the right set of peaks, with maxima at $k=5.5\times10^{18}$ GeV, have $\vel=-20000\,\mathrm{GeV}^2$ and $k_\mathrm{max} = 10^{20}$ GeV.  For all spectra, $V(\phi)$ is given by Eq.~(\ref{pot}) with $n=2$ and $M=10^{-3}$ eV.  The different curves for each value of $\vel$ correspond to different values of $\kIR$; from bottom to top, $\kIR = 10^{13}, 10^{14}, 10^{14.7}, 10^{15},$ and $10^{15.3}$ GeV for $\vel= -100\,\mathrm{GeV}^2$ and $\kIR = 10^{15}, 10^{16}, 10^{17}, 10^{17.5}$ and $10^{17.7}$ for $\vel=-20000\,\mathrm{GeV}^2$.   As $\kIR$ increases, more energy is transferred to the fluctuations, and the amplitude of the spectrum increases, but the value of $\kIR$ does not affect which modes are excited.
}
\label{Fig:EkIR}
\end{figure}

Although the total energy transferred to fluctuations depends on $\kIR$, the energy spectrum of the fluctuations is more robust.  Figure \ref{Fig:EkIR} shows the post-rebound fluctuation energy density per logarithmic interval in $k$, as defined in Eq.~(\ref{Ek}), for both $\vel = -100\, \mathrm{GeV}^2$ and $\vel = -20000\, \mathrm{GeV}^2$.  For both cases, the spectra are shown for several values of $\kIR$.  As expected, the rebound generates a spectrum of fluctuations that is rather sharply peaked at a specific wavelength, with larger $|\vel|$ values exciting more energetic fluctuations.  
Figure \ref{Fig:EkIR} shows that the amplitude of the fluctuation spectrum depends on the value of $\kIR$, but the shape of the spectrum does not.  We conclude that the basic characteristics of the fluctuation spectrum, particularly which wave numbers receive the most energy, does not depend on nonlinear effects.  As discussed in Section \ref{sec:analytics}, the duration of the rebound determines which fluctuation modes are excited, and we see no evidence that nonlinear effects change the rebound's basic timescale.  On the contrary, Fig.~\ref{Fig:NumBounce} illustrates that even the first-order backreaction does not significantly alter the duration of the rebound.  Furthermore, the analytic calculation of $\kex$ in Section \ref{sec:analytics} successfully predicts the peak in the fluctuation spectrum; Eq.~(\ref{kexexp}) with $\kappa = 0.03$ gives $\kex = 2.4\times10^{16}$ GeV for $\vel=-100\, \mathrm{GeV}^2$ and $\kex = 5.6\times10^{18}$ GeV for $\vel=-20000\, \mathrm{GeV}^2$.  The turn-around value of $\phibar$ is also independent of $\kIR$; for all the spectra shown in Fig.~\ref{Fig:EkIR}, $V_{\cal D}(\phita) \simeq \vel^2/2$.   We conclude that nonlinear effects only become important after the rebound, when the generated fluctuations begin to interact.  Since both $\phita$ and $\kex$ are determined by the dynamics of $\phibar$ prior to the rebound, they are insensitive to $\kIR$.

\begin{figure}
 \centering
 \resizebox{3.4in}{!}
 {
      \includegraphics{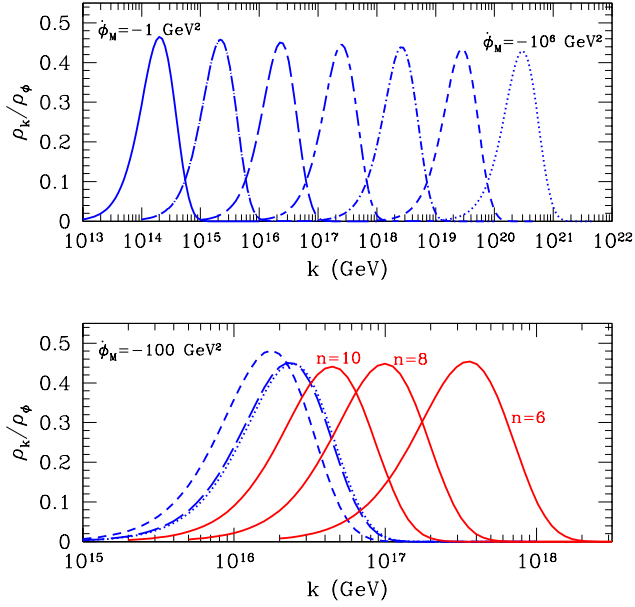}
 }
\caption{The fluctuation energy density per logarithmic interval in $k$ (see Eq.~\ref{Ek}) as a fraction of the total energy density in the chameleon field.  In the top panel, $V(\phi)$ is given by an exponential potential (Eq.~\ref{pot}) with $n=2$ and $M=10^{-3}$ eV.  From left to right, the different spectra correspond to $\vel = -1, -10, -100, -1000, - 10^4, -10^5$ and $-10^6\, \mathrm{GeV^2}$.   In the bottom panel, $\vel=-100\, \mathrm{GeV}^2$ for all spectra.  The long-dashed curve is the same as the long-dashed curve in the top panel, and the short-dashed and dotted spectra correspond to the same $V(\phi)$, but with $n=4$ and $n=10$, respectively.  The solid curves correspond to power-law potentials (Eq.~\ref{Vpow}) with $M_s=\Mp=10^{-3}$ eV and $n=6, 8,$ and 10.
In all cases with exponential potentials, $\kIR = 10 |\vel|/M$, which implies that $0.03\kex \lsim \kIR\lsim0.06\kex$, and for the power-law spectra $\kIR\simeq 0.05 \kex$.
}
\label{Fig:EkPhiDot}
\end{figure}

Having established that the fluctuation spectra are insensitive to nonlinear effects, we begin a more extensive comparison of the analytic model developed in Section \ref{sec:analytics} to the numerical results.  Figure \ref{Fig:EkPhiDot} shows how the fluctuation spectra depend on both $\vel$ and $V(\phi)$.  The top panel shows spectra for an exponential potential with $n=2$, while the bottom panel shows spectra for both exponential and power-law potentials with various values of $n$.  In all cases, the fluctuation spectrum is sharply peaked at a wave number we call $\kp$, and $\kp$ increases as $\vel$ increases.  We also see that $\kp$ does not vary much with $n$ for exponential potentials, but $\kp$ is very sensitive to $n$ for power-law potentials, with smaller $n$ values producing larger $\kp$ values.  Finally, we see that power-law potentials produce larger $\kp$ values than exponential potentials.  All of these findings are consistent with the predictions of Section \ref{sec:analytics}, and we see that the rebound does indeed excite fluctuations with extremely high energies. 

\begin{figure}
 \centering
 \resizebox{3.4in}{!}
 {
      \includegraphics{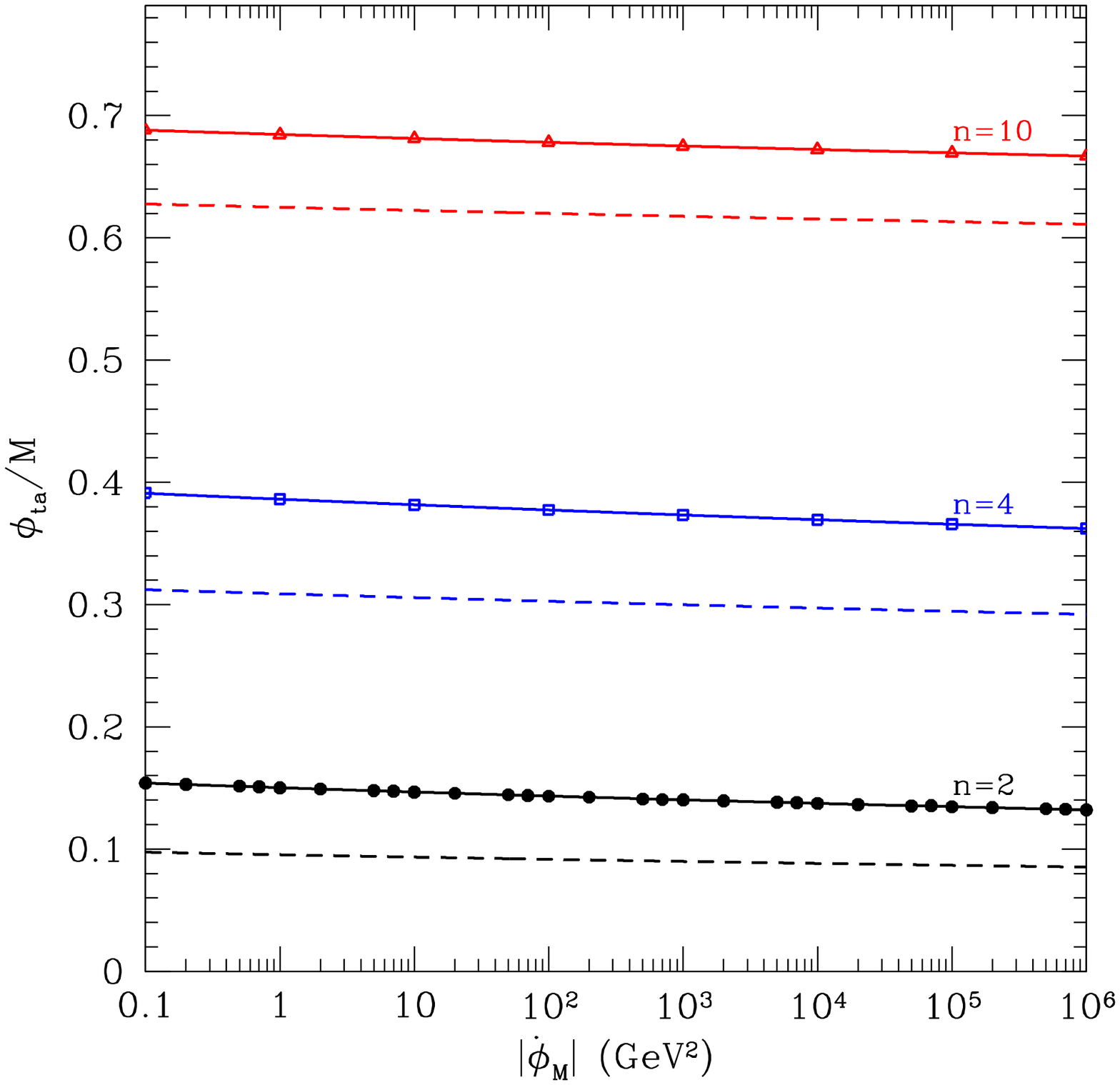}
 }
\caption{The minimum value of $\phibar$ during the rebound ($\phita$) as a function of $\vel$ for an exponential potential (Eq.~\ref{pot}) with $M=10^{-3}$ eV and different values of $n$.  The triangles, squares, and circles show the numerical values of $\phita$ for $n=10, 4$ and 2, respectively.   For each value of $n$, the solid curve is $V_{\cal D}(\phita) = \vel^2/2$ with $\kappa=0.03$ for $n=2$, $\kappa=0.022$ for $n=4$, and $\kappa=0.025$ for $n=10$.   The dotted curve directly beneath each solid curve is $V(\phita) = \vel^2/2$.  Thus we see that the dissipative potential, not $V(\phi)$, governs the evolution of $\phibar$ during the rebound.
}
\label{Fig:PhiMinExp}
\end{figure}
\begin{figure}
 \centering
 \resizebox{3.4in}{!}
 {
      \includegraphics{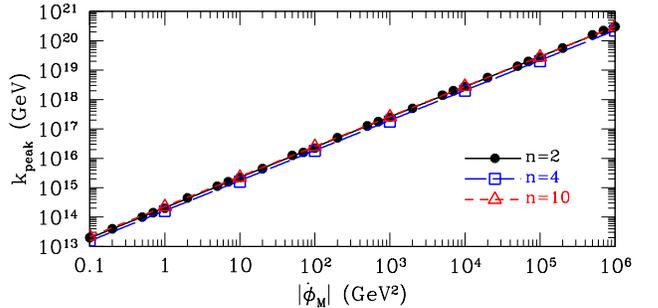}
 }
\caption{The wave number $k_\mathrm{peak}$ that maximizes $\rho_k$ as a function of $\vel$ for exponential potentials (Eq.~\ref{pot}) with $M=10^{-3}$ eV and $n=2$ (circles), $n=4$ (squares), and $n=10$ (triangles).  The lines show $\kex$ as given by Eq.~(\ref{kexexp}) with $\kappa=0.03$ for these potentials.  In all cases, $\kex$ matches $\kp$ to within 10\%.
}
\label{Fig:kpeakExp}
\end{figure}

We quantitatively test our analytic model for the rebound in Fig.~\ref{Fig:PhiMinExp}, which shows the turn-around value of $\phibar$ for exponential potentials with $n=2, 4$ and 10 as a function of $\vel$.  The solid lines show $V_{\cal D}(\phita) = \vel^2/2$; the values of $\kappa$ were chosen to match the numerical results, and all lie within the range predicted in Section \ref{sec:analytics} ($ 0.02\lsim \kappa \lsim 0.05$).  The dotted curve directly beneath each solid line shows $V(\phita) = \vel^2/2$.  We see that particle production does force $\phibar$ to turn around before $V(\phita) = \vel^2/2$, and the value of $\phita$ can be predicted by the dissipative potential (up to the choice of $\kappa$).  Figure \ref{Fig:kpeakExp} shows $\kp$ for the same potentials and compares it to $\kex$ given by Eq.~(\ref{kexexp}); in all cases $\kex$ is within 10\% of $\kp$.  This level of agreement between $\kex$ and $\kp$ is rather shocking; the derivation of $\kex$ was based on the rough estimate that the rebound would excite modes with $k \sim (\Delta t)^{-1}$, where $\Delta t$ was the duration of the rebound, and we expected it to differ from $\kp$ by a factor of order unity.  Instead, we see that Eq.~(\ref{kexexp}) and Eq.~(\ref{kexexpapprox}) accurately predict $\kp$, so we can use these expressions to predict the fluctuation spectra generated by other exponential potentials.  We see from Eq.~(\ref{kexexp}) that the value of $n$ has a mixed impact on $\kex$; $\kex$ is proportional to $n$, but increasing $n$ also slightly increases $\phita$, which decreases $\kex$.  As a result, $n$ does not significantly affect $\kp$, as seen in Fig.~\ref{Fig:kpeakExp}.  For $n\leq 10$, changing $n$ changes $\kex$ by less than 25\% for $10^{-6} \,\mathrm{GeV}^2 < |\vel| < 10^{6} \,\mathrm{GeV}^2.$

\begin{figure}
 \centering
 \resizebox{3.4in}{!}
 {
      \includegraphics{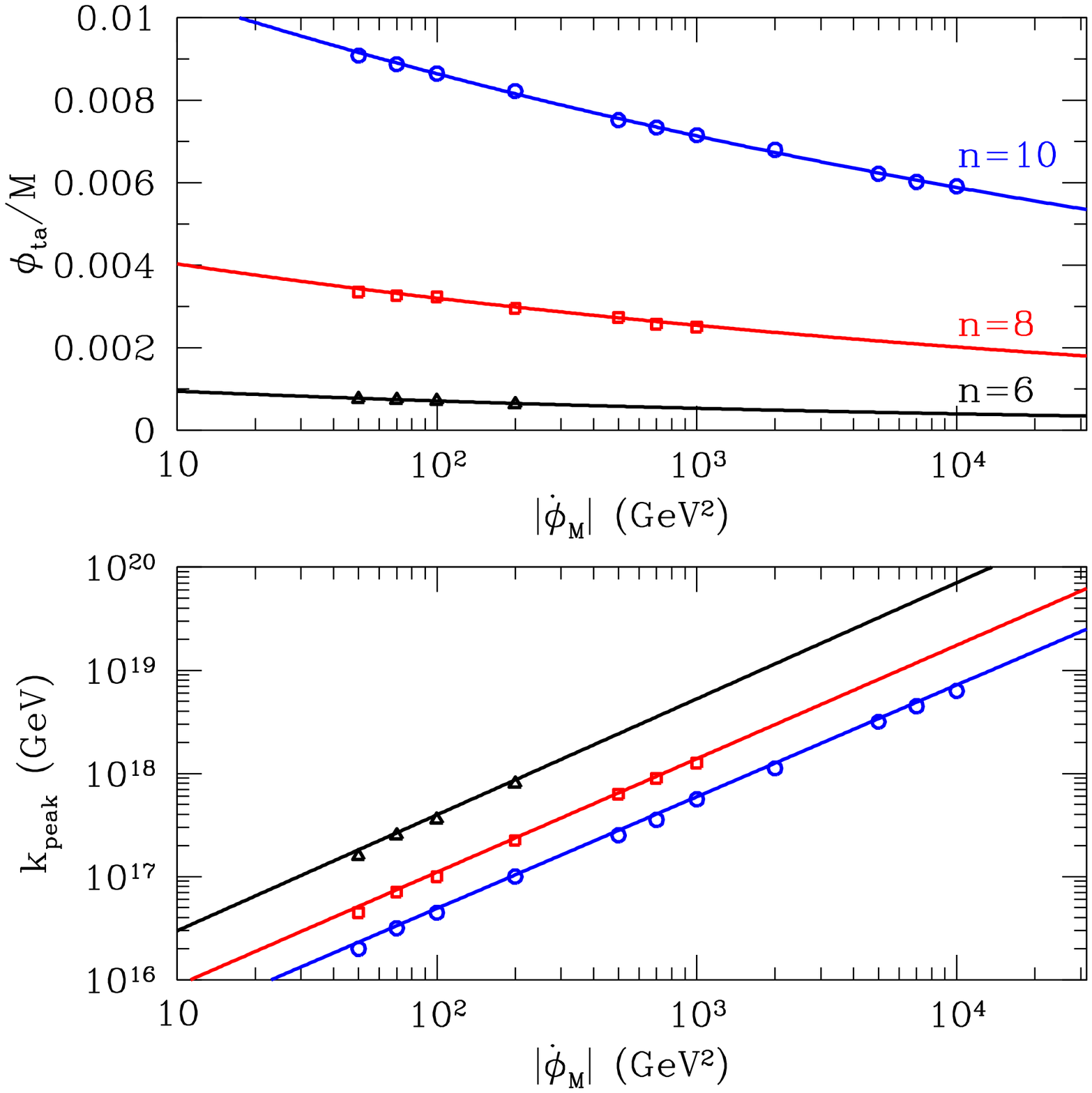}
 }
\caption{The minimum value of $\phibar$ during the rebound ($\phita$) and the wave number $k_\mathrm{peak}$ that maximizes $\rho_k$ as a function of $\vel$ for power-law potentials (Eq.~\ref{Vpow}) with $M_s=\Mp=10^{-3}$ eV and $n=6, 8,$ and 10.  The solid curves in the top panel show $V_{\cal D}(\phita) = \vel^2/2$ with $\kappa=0.025$, and the solid curves in the bottom panel show $\kex$ as given by Eq.~(\ref{kexpow}).  In all cases, $\kex$ matches $\kp$ to within 16\%.
}
\label{Fig:PLmodel}
\end{figure}

Figure \ref{Fig:PLmodel} shows that the analytic model developed in Section \ref{sec:analytics} is equally successful for power-law potentials.  We see that $V_{\cal D}(\phita) = \vel^2/2$ with $\kappa=0.025$ successfully predicts the turn-around value of $\phibar$, and $\kex$ given by Eq.~(\ref{kexpow}) matches $\kp$ to within $16\%$.  Power-law potentials are more numerically challenging than exponential potentials because of the small values of $\phita$.  Consequently, we can only numerically study potentials with $n\geq 6$, and we are restricted to a smaller range of $\vel$ values. However, the agreement between $\kex$ and $\kp$ shown in Fig.~\ref{Fig:PLmodel} indicates that Eq.~(\ref{kexpow}) provides an accurate estimate of the wave numbers of perturbations excited by rebounds off more general power-law potentials.  In particular, potentials with smaller values of $n$ give significantly larger values for $\kex$; for example, if $n=2$, then $\kex$ exceeds the reduced Planck mass for $|\vel| \gsim 2\, \mathrm{GeV}^2$.  In summary, numerically solving Eqs. (\ref{lineq}) and (\ref{bkgrd}) for both power-law and exponential potentials confirms the predictions of Section \ref{sec:analytics}; the evolution of $\phibar$ during the rebound is governed by the dissipative potential $V_{\cal D}(\phi)$, and the rebound excites extremely energetic fluctuations, with shallower potentials generating higher-energy fluctuations.  When combined with the values of $\dot{\phi}$ obtained in Section \ref{sec:classicalkicks}, both of these facts indicate that chameleon gravity experiences a catastrophic breakdown of calculability just prior to BBN.

\section{Summary and Discussion}
\label{sec:discuss}

Chameleon gravity runs into trouble in the early Universe because it attempts to unite two radically different energy scales: the MeV--GeV scale of the Standard Model and the meV scale of dark energy.  The same coupling to the trace of the stress-energy tensor that enables the chameleon to evade astronomical and laboratory constraints on fifth forces also makes the chameleon susceptible to excitation whenever the trace of the stress-energy tensor changes, as occurs when particles become non-relativistic in the early Universe.  Meanwhile, the chameleon's ability to evade detection is also contingent on the presence of a different energy scale in the chameleon's potential: $V(\phi/M)$ where $M \lsim 0.01$ eV.  The possibility of using the chameleon's potential to drive the current epoch of cosmic acceleration leads us to consider even smaller values for $M$: $M\simeq 0.001$ eV.  We have shown that the combination of these disparate energy scales leads to a breakdown of calculability just prior to BBN. The chameleon's coupling to matter accelerates the field to GeV-scale velocities, causing its effective mass to change rapidly and leading to the production of extremely energetic fluctuations that violate the limits of Effective Field Theory.  Moreover, the production of these fluctuations significantly alters the chameleon's evolution, leaving its state during BBN unknown.  

The chameleon's difficulties begin when the temperature of the radiation bath in the early Universe falls below the mass of a particle that is in thermal equilibrium.  At that time, the trace of the stress-energy tensor momentarily increases because the pressure of the massive particles decreases faster than their density.  During this transition, the trace of the stress-energy tensor is sufficiently large that the chameleon's coupling to it overcomes Hubble friction and forces the chameleon field to roll down the slope of its effective potential.  Earlier treatments of chameleon cosmology used these kicks to the chameleon field to move the chameleon field from its expected initial value ($M \ll \phi_i \lsim \Mpl$) to the minimum of its effective potential ($\phimin \simeq M$) prior to BBN \cite{Brax:2004qh,Mota:2011nh}.   

We have shown that the chameleon field does not just reach the minimum of its effective potential; the field rolls past it with a very large velocity.  Unlike earlier work, our analysis includes the effect the chameleon's evolution has on the expansion of the Universe in the Jordan frame.  As the chameleon rolls, it slows the Jordan-frame expansion, which extends the duration of the kicks and enhances their impact.  If the chameleon coupling to matter is slightly stronger than gravitational ($\beta \gsim 1.8$), this effect halts the expansion of the Jordan frame entirely until the chameleon reaches the minimum of its effective potential.  We call this novel solution to the chameleon's equation of motion the ``surfing solution," and it guarantees that all chameleon fields with $\beta\gsim2$ will reach the minima of their effective potentials with velocities $|\dot\phi | \gsim 10^{-3}\,\mathrm{GeV}^2$, regardless of their initial value.  In general, nearly all chameleons with sub-Planckian initial values reach their potential minima with $|\dot\phi | \gsim \mathrm{MeV}^2$;  only weakly coupled chameleons ($\beta \lsim 0.4$)  with finely tuned initial conditions avoid this fate.  Moreover, our calculation of $|\dot\phi|$ uses a minimal prescription for the kicks that only includes contributions from Standard-Model particles.  The inclusion of the QCD trace anomaly \cite{Kajantie:2002wa,Davoudiasl:2004gf}, interactions during the QCD phase transition \cite{Caldwell:2013mox}, or additional particles would give the chameleon field an even larger velocity when it reaches the minimum of its effective potential.

When the chameleon rolls past the minimum of its effective potential with $|\dot\phi| \gg M^2$, it climbs up the steep portion of its bare potential, where $\phi \lsim M$.  This steep region is required to give the chameleon a large mass in dense environments, thus screening the fifth force it generates.  After the kicks, however, the steepness of $V(\phi)$ becomes a severe liability.  Since  $V''(\phi)$ changes sharply for small displacements ($\Delta \phi < M$), the chameleon's velocity after the kicks causes its effective mass to change rapidly.  These non-adiabatic changes in mass trigger the excitation of fluctuations with wave numbers $k\simeq (\Delta t)^{-1}$, where $\Delta t < M/|\dot\phi|$ is the timescale over which the mass changes.  Since $|\dot\phi| \gg M^2$, these fluctuations are extremely energetic, with $k \gg 10^7$ GeV in most cases.  The classical evolution of the chameleon field as it rolls up and rebounds off the steep part of its potential predicts that these fluctuations should be generated with occupation numbers of order unity, but there is insufficient energy in the chameleon field to generate such high-energy particles.  Therefore, the backreaction of the fluctuations must significantly alter the evolution of the chameleon field.

\begin{figure}
 \centering
 \resizebox{3.4in}{!}
 {
      \includegraphics{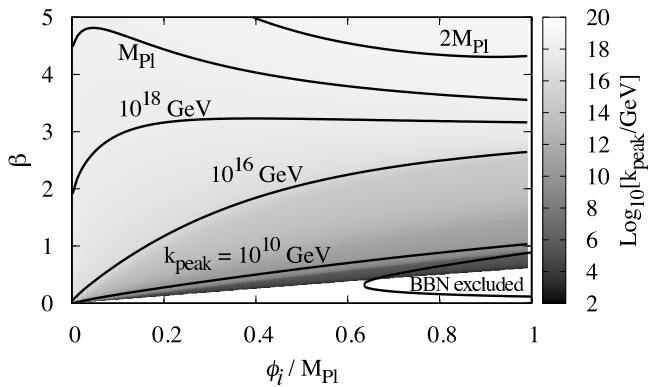}
 }
\caption{The fluctuation wave number that receives the most energy during the rebound as a function of the chameleon coupling constant $\beta$ and the chameleon's initial value $\phi_i$, as predicted by Eq.~(\ref{kexexp}) for an exponential potential (Eq.~\ref{pot}) with $M=10^{-3}$ eV and $n=2$.  Increasing $n$ does not significantly alter the fluctuation spectrum.   In the white region,  $\phi_i$ is sufficiently large that the field does not reach the minimum of its effective potential. In the region marked ``BBN excluded," $\phi > 0.1 \Mpl/\beta$ when the temperature is 1 MeV, which spoils the success of BBN.
}
\label{Fig:kpeakContour}
\end{figure}

To account for the transfer of energy to the fluctuations, we added a first-order backreaction term to the chameleon's equation of motion that ensures that energy is conserved during the rebound.  An analysis of this backreaction term revealed that it effectively introduces a new term to the chameleon's potential that dominates over $V(\phi)$ during the rebound.  This ``dissipative potential" does not significantly change the timescale of the rebound, but it halts the chameleon's climb up its potential while the occupation numbers of the excited modes are still very small.  Nevertheless, numerical calculations confirmed that a large fraction of the chameleon's energy is transferred to fluctuations with $k\simeq (\Delta t)^{-1}$ during the rebound, which significantly alters the trajectory of the spatially averaged field.  

Furthermore, these numerical calculations revealed that a simple derivation of $\Delta t$ using the dissipative potential accurately predicts the fluctuation wave numbers that receive the most energy for a given value of $\dot{\phi}$ after the kicks.  These wave numbers are shown in Figure \ref{Fig:kpeakContour} for an exponential chameleon potential.  Power-law potentials generate fluctuations with even higher wave numbers.  Thus we see that nearly all chameleons generate extremely high-energy perturbations during the rebound, with strongly coupled chameleons generating trans-Planckian fluctuations.  Since we cannot trust chameleon gravity up to such high energies, we cannot predict what impact these fluctuations may have on BBN.  

Our calculation of the excitation of perturbations during the rebound is not a complete treatment.  First, we only included the first-order backreaction of the perturbations on the evolution of the spatially averaged field.  This backreaction term is a one-loop correction to the chameleon's equation of motion, and our omission of higher-order backreaction terms implies that we are neglecting higher-order quantum corrections.  During the rebound, the first-order backreaction term dominates over the bare potential, which indicates that the higher-order backreaction terms are not insignificant.  Since the ``dissipative potential" generated by the first-order backreaction is responsible for the transfer of energy from the perturbations to the spatially averaged field after the rebound, higher-order backreaction terms would probably affect the final distribution of energy.  Nevertheless, the first-order backreaction term alone provides insight into how the transfer of energy to the perturbations affects the rebound.  Since the inclusion of the backreaction term does not change the timescale of the rebound, it does not affect which modes are excited.  The backreaction strongly affects the evolution of the spatially averaged field, however, and it suppresses the occupation numbers of the excited modes. 

Second, we linearized the equation of motion for the perturbation modes, which neglects interactions between modes with different wavelengths.  We explored the importance of these nonlinear interactions by varying the IR cutoff of our perturbations, which shifts the distinction between the perturbations and the spatially averaged field.  Since the equation of motion for the spatially averaged field is not linearized, the IR cutoff determines which nonlinear effects are neglected.  Changing the IR cutoff strongly affects the final state of the chameleon field because it varies how much energy is transferred to the perturbations.  Therefore, a full nonlinear treatment of the perturbation equations and their backreaction is required to determine how the chameleon field evolves after the rebound.  However, changing the IR cutoff does not alter the position of the peak in the perturbation spectrum.  The timescale of the rebound determines which modes are excited, and this timescale is not affected by nonlinear interactions.  Therefore, the utility of a fully nonlinear treatment of the perturbations would be limited by our ignorance of physics at energy scales much greater than 1000 TeV.  

Another source of uncertainty in our calculations is the chameleon potential $V(\phi)$.  The large velocity imparted to the chameleon field by the kicks implies that the chameleon reaches very small values during the rebound; even when its journey is curtailed by the transfer of energy to perturbations, the field reaches values that are smaller than the values obtained in the densest astrophysical objects.  There are reasons to distrust the chameleon potential at these small values of $\phi$: the one-loop Coleman-Weinberg correction to $V(\phi)$ is much larger than $V(\phi)$ itself, and the dimensionless quartic coupling $d^4 V/d\phi^4$ exceeds unity.  These facts alone support our conclusion that chameleon gravity suffers a computational breakdown after the kicks, independently of the severe violations of adiabaticity that trigger particle production.  

These flaws are commonplace in chameleon gravity, however; even for moderate densities ($\rho \simeq 10\,\mbox{g cm}^{-3}$), $d^4 V/d\phi^4>1$ for nearly all values of $n$ and $\beta$, and the one-loop quantum corrections dominate the potential for several chameleon models \cite{Upadhye:2012vh}.  Most analyses of chameleon gravity ignore these difficulties, implicitly assuming that $V(\phi)$ is protected from quantum effects in some way.  We have extended chameleon gravity the same privilege by assuming that $V(\phi)$ is valid for all values of $\phi$ during the rebound, and we have demonstrated that a computational breakdown still occurs prior to BBN.  Yet, it may be possible to avoid this calamity by changing $V(\phi)$ for small $\phi$ values.  It would be interesting to search for well-behaved potentials that enable the chameleon screening mechanism while avoiding particle production after the kicks.  Any such potential would have to be more complicated than the exponential and power-law potentials that we considered, and we leave this investigation for future work.  

Another interesting avenue for further investigation is to determine if any other modified gravity theories suffer from a similar computational breakdown in the early Universe.  Coupled dark energy theories with quintessence fields that couple to dark matter but not baryons \cite{1990PhRvL..64..123D, 2000PhRvD..62d3511A} may also be susceptible; if the dark matter particle is a thermal relic, then the quintessence field will be kicked when the dark matter particle becomes nonrelativistic.  Since these coupled dark energy theories do not need the chameleon mechanism to evade laboratory and Solar System constraints on fifth forces, the scalar's potential function is less constrained than the potential functions we considered.  Nevertheless, the fact that the quintessence field is supposed to drive cosmic acceleration implies that its potential function involves at least one mass scale that is much smaller than the mass of the dark matter particle.  Furthermore, fifth forces exclusive to the dark sector can be constrained by tidal streams \cite{2006PhRvL..97m1303K,2006PhRvD..74h3007K} and anisotropies in the cosmic microwave background \cite{2013PhRvD..88f3519P}, which require $\beta \lsim 0.07$ in the absence of a screening mechanism.  Therefore, these theories can only include gravitational-strength couplings if they employ the chameleon mechanism, which would require a chameleon scalar potential.

In general, chameleon gravity provides a cautionary tale about the dangers of uniting dark energy and high-energy physics in a single theory; the extreme hierarchy of energy scales can produce surprising and uncontrollable effects.  The key ingredients that make chameleon gravity vulnerable are 1) a field that couples to the stress-energy tensor and is initially displaced from the minimum of its potential, and 2) a potential function for which $V''(\phi)$ changes significantly when the scalar field rolls by a small amount ($\Delta \phi \ll$ GeV).  Given these two features, we expect that the field will be accelerated to a GeV-scale velocity in the early Universe, and then that large velocity will induce rapid changes in the field's mass, leading to the excitation of high-energy particles.  An easy way to prevent a violation of adiabaticity may be to impose a shift symmetry; if $V(\phi)$ is insensitive to the value of $\phi$, it does not matter how large $\dot{\phi}$ becomes.  Therefore, the perils faced by chameleon gravity in the early Universe provide additional motivation for including a shift symmetry in any scalar-tensor theory that attempts explain the current epoch of cosmic acceleration.

\acknowledgments
A.E. was supported by the Canadian Institute for Theoretical Astrophysics, the Perimeter Institute for Theoretical Physics, and the Canadian Institute for Advanced Research prior to the final stages of this work.  Research at Perimeter Institute is supported by the Government of Canada through Industry Canada and by the Province of Ontario through the Ministry of Research and Innovation.  N.B. thanks DESY, the University of Geneva, and the Perimeter Institute for their hospitality during the completion of this work.  C.B. is supported by a Royal Society University Research Fellowship.

\appendix
\section{The Kick Function}
\label{App:Sigma}

In this Appendix, we review the thermodynamics of the early Universe and derive the kick function given by Eq.~(\ref{singlekick}) \cite{Damour:1992kf, Damour:1993id, Coc:2006rt}.  We also compute the kick function generated by Standard-Model particles that we use in Section \ref{sec:classicalkicks}.

The energy density $\rho$ and the pressure $p$ of particles in thermal equilibrium are
\begin{align}
\rho &= \frac{g}{2\pi^2} \int_m^\infty \frac{\sqrt{E^2-m^2}}{\exp(E/T)\pm1} E^2 dE; \label{rho}\\
p &= \frac{g}{6\pi^2} \int_m^\infty \frac{(E^2-m^2)^{3/2}}{\exp(E/T)\pm1} dE. \label{press}
\end{align}
where $g$ is the number of degrees of freedom for the particle species; $m$ is the particle's mass; $T$ is the temperature of the radiation bath; and the $+$ sign in the denominator applies to fermions, while the $-$ sign applies to bosons.  We define $\rho_R$ to be the sum of the the energy densities for all particles that are in thermal equilibrium with the radiation bath in the early Universe (including neutrinos), and we define $g_* \equiv \rho_R[(\pi^2/30)T^4]^{-1}$.  To compute the kick function $\Sigma \equiv (\rho_R-3p)/\rho_R$, we first evaluate $\rho-3p$ for each particle species:
\begin{align}
\rho - 3p &= \frac{g m^2}{2\pi^2} \int_m^\infty \frac{\sqrt{E^2-m^2}}{\exp(E/T)\pm1} dE; \\
&= \frac{g}{2\pi^2} T^4 \left(\frac{m}{T}\right)^2  \int_{m/T}^\infty \frac{\sqrt{u^2 -(m/T)^2}}{e^u\pm1}du,
\end{align}
where we have introduced $u = E/T$ as the integration variable in the last line.  Dividing this expression by $\rho_R = g_*(\pi^2/30)T^4$ yields the contribution to $\Sigma$ from a single particle species, as given by Eq.~(\ref{singlekick}):
\beq
\Sigma_i(T) = \frac{15}{\pi^4}\frac{g_i}{g_*(T)} \left(\frac{m_i}{T}\right)^2\int_{m_i/T}^\infty \frac{\sqrt{u^2 - (m_i/T_J)^2}}{e^u \pm 1} du.
\eeq

In Section \ref{sec:classicalkicks}, we evaluate $\Sigma(T)$ for the Standard Model by summing the contributions from the particle species listed in Table \ref{tab:SMparticles}.  Prior calculations of the kick function assumed that $g_*$ was constant during each kick and computed $g_*$ for each particle species by summing the contributions of all particles with $m\leq m_i$:
\beq
g_* = \sum^{\mathrm{bosons}}_i g_i \left(\frac{T_i}{T}\right)^4 + \frac{7}{8}\sum^{\mathrm{fermions}}_i  g_i \left(\frac{T_i}{T}\right)^4 ,
\label{gstar}
\eeq
where $T_i$ is the temperature of the particle species.  At temperatures greater than 1 MeV, $T_i = T$ for all particles included in $\rho_R$, but at lower temperatures, the neutrinos decouple from the photon bath, and $T_\nu \neq T$.  This computation of $g_*$ overestimates $g_*$ during the kick because it treats the particle responsible for the kick as if it were relativistic throughout the kick.  To avoid underestimating $\Sigma$ in this way, we compute $g_*(T)$ by numerically evaluating $\rho(T)$ for all the particles in Table \ref{tab:SMparticles} and adding these energy densities to the energy densities of the relativistic particles.  Prior to the QCD phase transition, the relativistic bosons are gluons ($g=16$) and photons ($g=2$), and the relativistic fermions are light quarks ($g=36$ for u, d, s), muons, electrons, and neutrinos ($g=6$).  Therefore, $g_*(T)$ smoothly decreases from 106.75 to 61.75 prior to the QCD phase transition.   We treat the QCD phase transition as an instantaneous event that occurs at a temperature of 170 MeV.  Below this temperature, the quarks and the gluons are bound into hadrons, and the only particles that contribute to $g_*$ are pions, muons, electrons, neutrinos and photons.  Consequently, $g_*$ discontinuously changes from 61.75 to 17.25 when $T=170$ MeV, which generates the discontinuity in $\Sigma(T)$ seen in Figure \ref{Fig:sigma}.  After the QCD phase transition, $g_*(T)$ decreases smoothly from 17.25 to 10.75 as the temperatures decreases from 170 MeV to 10 MeV.

The calculation of $\Sigma(T)$ for $T\lsim 1$ MeV is complicated by the decoupling of the neutrinos from the radiation bath.  After they decouple, the neutrinos' temperature is proportional to $1/a$, which causes it to differ from the temperature of the radiation bath during the final kick.  To evaluate the neutrino temperature $T_\nu$ at temperatures below 1 MeV, we employ the conservation of entropy to obtain
\beq
T_\nu = \left[\frac{g_{*S}(T)}{10.75}\right]^{1/3}T; \quad T < 1\, \mathrm{MeV},
\eeq
where $g_{*S}$ is the total entropy density of all particles in thermal equilibrium with the radiation bath divided by $(2\pi^2/45)T^3$.  We evaluate $g_{*S}(T)$ by computing the entropy density, $s = (\rho+P)/T$, from Eqs.~(\ref{rho}) and (\ref{press}) in the same way that we used Eq.~(\ref{rho}) to evaluate $g_*(T)$.  In addition to providing the neutrino temperature, this computation of $g_{*S}(T)$ is used to numerically solve Eq.~(\ref{T_J}) for $T_J(a_*)$.  Once we know how $T_\nu/T$ changes as the temperature cools below 1 MeV, we can use Eq.~(\ref{gstar}) to evaluate the neutrinos' contribution to $g_*(T)$ during the final kick, during which $g_*(T)$ decreases from 10.75 to 3.36.  This evolution of $g_*$ significantly enhances the amplitude of the final kick compared to earlier calculations that assumed a fixed value of $g_*=10.75$.

\begin{table}
\begin{tabular} {| c | c | c || c | c | c |}
\multicolumn{6}{c}{Contributions to $\Sigma$}\\
\hline
\multicolumn{3}{| c ||}{fermions}&\multicolumn{3}{ c |}{bosons}\\
\hline
particle & $g$ & m (GeV) & particle & $g$ & m (GeV) \\
\hline
\multicolumn{6}{|c|}{{\it before QCD phase transition}}\\
\hline
top 		& 12		& 172 	& Higgs 	& 1	& 125 \\
bottom	&12		& 4.2		& Z		& 3	& 91   \\
charm 	& 12		& 1.3 	& W$^\pm$ & 6 & 80	  \\
tau		& 4		& 1.8		& 		&	& \\
\hline
\multicolumn{6}{|c|}{{\it after QCD phase transition}}\\
\hline
muon	& 4		& 0.106  	& $\pi^0$ & 1 & 0.140 \\
electron	& 4		& $5.11 \times 10^{-4}$ & $\pi^{\pm}$ & 2 & 0.135 \\
\hline 
\end{tabular}
\caption{The numbers of degrees of freedom ($g$) and the masses ($m$) of the particles that we include in the kick function $\Sigma(T)$.  For the fermions, the contributions from antiparticles are included in the number of degrees of freedom for each species.}
\label{tab:SMparticles}
\end{table}

\section{Effects of the QCD trace anomaly}
\label{App:TA}

The QCD trace anomaly makes a nearly constant contribution to $\Sigma$ at temperatures greater than 100 GeV: $\Sigma_\mathrm{ta} \simeq 10^{-3}$.  At temperatures greater than $\sim500$ GeV, the trace anomaly dominates over the other known contributions to $\Sigma$.  In this appendix, we explore the consequences of adding a nearly constant $\Sigma_\mathrm{ta}$ to $\Sigma$.  This additional contribution will only affect strongly coupled chameleons whose surfing temperatures exceed 200 GeV (see Fig.~\ref{Fig:surfphidot}).   As long as $\beta <  \sqrt{1/(3\Sigma_\mathrm{ta}})$, the surfing solution still exists, but the additional contribution from the trace anomaly increases the surfing temperature.  If $\beta \gsim 5$, including the trace anomaly increases the surfing temperature by more than 5\%.  Since increasing the surfing temperature increases the chameleon's velocity when it reaches the minimum of its effective potential [see Eq.~(\ref{surfingvelocity})], including the QCD trace anomaly increases the impact velocity of strongly coupled chameleons.

The QCD trace anomaly has a more profound effect if $\beta>  \sqrt{1/(3\Sigma_\mathrm{ta}})$ because the surfing solution no longer exists for these chameleons.  However, the nearly constant value of $\Sigma_\mathrm{ta}$ introduces a new solution to the chameleon's equation of motion; integrating Eq.~(\ref{eom}) for a chameleon that is initially at rest implies
\begin{align}
\varphi^\prime(p) & = -3 \beta e^{-p} \int_1^{e^p} \Sigma \,da; \nonumber \\
&\simeq - 3\beta\Sigma_\mathrm{ta} \left( 1- e^{-p} \right).
\end{align}
Thus we see that $|\varphi^\prime(p)|$ approaches a constant value that is greater than the surfing value ($|\varphi^\prime(p)| = 1/\beta$).  An additional integration yields
\beq
\varphi(p) = \varphi_i - 3\beta\Sigma_\mathrm{ta} \left(p + e^{-p} -1\right),
\eeq
which we can insert into Eq.~(\ref{T_J}) to obtain the Jordan-frame temperature.  Assuming that $g_{*S}$ is constant at these high temperatures, 
\beq
T_J = \initT \exp\left[p(3\beta^2\Sigma_\mathrm{ta} -1)+\left(e^{-p}-1\right)3\beta^2\Sigma_\mathrm{ta}\right] .
\eeq
Since $3\beta^2\Sigma_\mathrm{ta} > 1$ for these non-surfing chameleons, $T_J$ increases exponentially with $p$; whereas the surfing solution halts the cooling of the Universe, this solution heats the Universe.  Furthermore, the fact that $\dot\phi \propto T_J^2 \varphi^\prime (p)$ for $\varphi \ll 1$ implies that the chameleon's velocity when it reaches the minimum of its effective potential depends on its initial position, with larger values of $\varphi_i$ giving larger impact velocities.   This solution to the chameleon's equation of motion implies that a nearly constant contribution to $\Sigma$ from the QCD trace anomaly dramatically enhances the impact velocity of chameleons with  $\beta>  \sqrt{1/(3\Sigma_\mathrm{ta}})$.

\section{Particle Production in an Expanding Universe}
\label{App:pp}

In this Appendix, we review how non-adiabatic changes in a field's effective mass result in particle production \cite{BirrellDavies}.  
We begin with a scalar field in a homogeneous expanding universe described by an FRW metric:
\begin{equation}
  ds^2 = -dt^2 + a^2(t) d{\bf x}^2 = a^2(\tau) \left[ -d\tau^2 + d{\bf x}^2 \right].
\end{equation}
The scalar field's equation of motion is
\begin{equation}
\label{KG}
  \ddot{\phi} + 3 H \dot{\phi} - \frac{\nabla^2}{a^2}\phi + V'(\phi) = 0,
\end{equation}
where a dot denotes differentiation with respect to cosmic time $t$, $H = \dot{a}/{a}$, and $\nabla^2$ is the Laplacian with respect to the comoving coordinates ${\bf x}$.

We define $\phibar(t)$ to be the spatial average of $\phi(t,{\bf x})$ over a large volume, and then we define the perturbation $\delta \phi$ through
\begin{equation}
\label{split}
  \phi(t,{\bf x}) = \phibar(t) + \delta\phi(t,{\bf x}).
\end{equation}
In Section \ref{sec:analytics}, we expanded $V'(\phi)$ in Eq.~(\ref{KG}) around $\phibar$ to obtain Eq.~(\ref{fullexpand}).  We then took the spatial average of this equation, which yielded Eq.~(\ref{spatialavg}).  Subtracting Eq.~(\ref{spatialavg}) from Eq.~(\ref{fullexpand}) and keeping only terms linear in $\delta\phi$ provides the linearized perturbation equation:
\begin{equation}
\label{linear_no_br}
   \left[ \partial_t^2 + 3 H \partial_t - \frac{\nabla^2}{a^2} + V^{\prime\prime}(\phibar) \right] \delta \phi \simeq 0.
\end{equation}

We quantize $\delta\phi$ by expressing it in terms of annihilation operators $ \hat{a}_{\bf k} $ and creation operators $\hat{a}^\dagger_{\bf k}$:
\begin{equation}
  \hat{\delta\phi}(\tau,{\bf x}) = \int \frac{d^3k}{(2\pi)^{3}} \left[ \hat{a}_{\bf k} \frac{\phi_k(\tau)}{a(\tau)} e^{i{\bf k}\cdot {\bf x}} +\hat{a}^\dagger_{\bf k} \frac{\phi^*_k(\tau)}{a(\tau)} e^{-i{\bf k}\cdot {\bf x}}  \right].
  \label{operatorquant}
\end{equation}
The annihilation and creation operators obey the standard commutation relations:
\begin{equation}
  \left[\hat{a}_{\bf k} , \hat{a}_{\bf k'}^\dagger \right] = (2\pi)^3\delta^{(3)}\left({\bf k}-{\bf k'}\right),
\end{equation}
and all other combinations commute.  The annihilation operator annihilates the vacuum state: $\hat{a}_{\bf k} |0\rangle = 0$.  

With this decomposition of $\delta\phi$, Eq.~(\ref{linear_no_br}) implies that
\begin{align}
  & \phi_k''(\tau) + \omega_k^2(\tau) \phi_k = 0 \label{mode_eq}, \\
  & \omega_k^2(\tau) = k^2 + a^2 V''(\phibar) - \frac{a''(\tau)}{a} \label{omega}.
\end{align}
We solve this equation by expressing the mode functions $\phi_k(\tau)$ in terms of Bogoliubov coefficients $\alpha_k(\tau)$ and $\beta_k(\tau)$:
\beq
\label{WKB1} 
  \phi_k(\tau) =  \frac{\alpha_k(\tau)}{\sqrt{2\omega_k(\tau)}} e^{-i\int^\tau \omega_k(\tau')d\tau'} +
                            \frac{\beta_k(\tau)}{\sqrt{2\omega_k(\tau)}} e^{+i\int^\tau \omega_k(\tau')d\tau'}.
\eeq
This expression solves Eq.~(\ref{mode_eq}) provided that \cite{Kofman:1997yn}
\begin{subequations}\label{alphbeta}
\begin{align}
  \alpha_k'(\tau) &= \frac{\omega_k'(\tau)}{2\omega_k(\tau)} \beta_k(\tau) e^{+2i\int^\tau \omega_k(\tau')d\tau'} \label{alpha}, \\
  \beta_k'(\tau) &= \frac{\omega_k'(\tau)}{2\omega_k(\tau)} \alpha_k(\tau) e^{-2i\int^\tau \omega_k(\tau')d\tau'} \label{beta}.
\end{align}
\end{subequations}
To give $\hat{\delta\phi}$ the proper commutation relation with its conjugate momentum, the Bogoliubov coefficients must satisfy
\begin{equation}
  |\alpha_k(\tau)|^2 - |\beta_k(\tau)|^2 = 1.
  \label{Bogonorm}
\end{equation}

The energy density of the fluctuations is defined as
\begin{equation}
  \langle \rho_{\mathrm{fluct}} \rangle = \frac{1}{2 a^4} \int \frac{d^3k}{(2\pi)^3} \left[  |\phi_k'|^2 + \omega_k^2 |\phi_k|^2  \right].
  \label{rhofluc}
\end{equation}
Like the zero-point energy of an infinite tower of harmonic oscillators, this integral is divergent.   We can expose this divergence by substituting Eq.~(\ref{WKB1}) for $\phi_k$ and using Eqs.~(\ref{alphbeta}) and (\ref{Bogonorm}), which gives
\begin{equation}
\langle \rho_{\mathrm{fluct}} \rangle = \frac{1}{a^4} \int \frac{d^3k}{(2\pi)^3}\omega_k \left[|\beta_k|^2+\frac12 \right].
\label{rhofluc2}
\end{equation}
Thus we see that $|\beta_k|^2=0$ corresponds to the lowest possible energy density, and the divergent portion of the integral corresponds to the infinite energy of this ground state.
Therefore, the $|\beta_k|^2=0$ state is interpreted as the vacuum state, and we regularize $\langle\rho_{\mathrm{fluct}}\rangle$ by subtracting the energy density of the vacuum state \cite{BirrellDavies}, which gives
\begin{equation}
  {\langle\rho_{\mathrm{fluct}}\rangle}_{\mathrm{reg}} = \frac{1}{a^4} \int {\frac{d^3k}{(2\pi)^3} n_k(\tau) \omega_k(\tau)},
  \label{rho_fluct}
\end{equation}
where we have introduced the occupation number $n_k$:
\begin{equation}
  n_k(\tau) \equiv \frac{1}{2\omega_k(\tau)} \left[  |\phi_k'|^2 + \omega_k^2 |\phi_k|^2  \right] - \frac{1}{2}.
  \label{nkdef}
\end{equation}
We see from Eq.~(\ref{rhofluc2}) that
\begin{equation}
  n_k(\tau) =|\beta_k(\tau)|^2.
\end{equation}

If we start from a vacuum state, then Eq.~(\ref{beta}) implies that we will remain in a vacuum state while the adiabaticity condition,
\begin{equation}
  \frac{\omega_k'(\tau)}{\omega^2_k} \ll 1,
  \label{adiabatcond}
\end{equation}
is satisfied.   Conversely, particle production occurs when the effective mass of the perturbations varies sufficiently rapidly that ${\omega_k'(\tau)}/{\omega^2_k} \simeq 1$.  In that case, Eq.~(\ref{beta}) states that the vacuum state will evolve to a state with $n_k = |\beta_k|^2 \neq 0$.


\begin{thebibliography}{85}
\expandafter\ifx\csname natexlab\endcsname\relax\def\natexlab#1{#1}\fi
\expandafter\ifx\csname bibnamefont\endcsname\relax
  \def\bibnamefont#1{#1}\fi
\expandafter\ifx\csname bibfnamefont\endcsname\relax
  \def\bibfnamefont#1{#1}\fi
\expandafter\ifx\csname citenamefont\endcsname\relax
  \def\citenamefont#1{#1}\fi
\expandafter\ifx\csname url\endcsname\relax
  \def\url#1{\texttt{#1}}\fi
\expandafter\ifx\csname urlprefix\endcsname\relax\def\urlprefix{URL }\fi
\providecommand{\bibinfo}[2]{#2}
\providecommand{\eprint}[2][]{\url{#2}}

\bibitem[{\citenamefont{Wetterich}(1995)}]{Wetterich:1994bg}
\bibinfo{author}{\bibfnamefont{C.}~\bibnamefont{Wetterich}},
  \bibinfo{journal}{Astron.Astrophys.} \textbf{\bibinfo{volume}{301}},
  \bibinfo{pages}{321} (\bibinfo{year}{1995}), \eprint{hep-th/9408025}.

\bibitem[{\citenamefont{Zlatev et~al.}(1999)\citenamefont{Zlatev, Wang, and
  Steinhardt}}]{Zlatev:1998tr}
\bibinfo{author}{\bibfnamefont{I.}~\bibnamefont{Zlatev}},
  \bibinfo{author}{\bibfnamefont{L.-M.} \bibnamefont{Wang}}, \bibnamefont{and}
  \bibinfo{author}{\bibfnamefont{P.~J.} \bibnamefont{Steinhardt}},
  \bibinfo{journal}{Phys.Rev.Lett.} \textbf{\bibinfo{volume}{82}},
  \bibinfo{pages}{896} (\bibinfo{year}{1999}), \eprint{astro-ph/9807002}.

\bibitem[{\citenamefont{Amendola}(2000)}]{Amendola:1999er}
\bibinfo{author}{\bibfnamefont{L.}~\bibnamefont{Amendola}},
  \bibinfo{journal}{Phys.Rev.} \textbf{\bibinfo{volume}{D62}},
  \bibinfo{pages}{043511} (\bibinfo{year}{2000}), \eprint{astro-ph/9908023}.

\bibitem[{\citenamefont{Caldwell and Kamionkowski}(2009)}]{Caldwell:2009ix}
\bibinfo{author}{\bibfnamefont{R.~R.} \bibnamefont{Caldwell}} \bibnamefont{and}
  \bibinfo{author}{\bibfnamefont{M.}~\bibnamefont{Kamionkowski}},
  \bibinfo{journal}{Ann.Rev.Nucl.Part.Sci.} \textbf{\bibinfo{volume}{59}},
  \bibinfo{pages}{397} (\bibinfo{year}{2009}), \eprint{0903.0866}.

\bibitem[{\citenamefont{Copeland et~al.}(2006)\citenamefont{Copeland, Sami, and
  Tsujikawa}}]{Copeland:2006wr}
\bibinfo{author}{\bibfnamefont{E.~J.} \bibnamefont{Copeland}},
  \bibinfo{author}{\bibfnamefont{M.}~\bibnamefont{Sami}}, \bibnamefont{and}
  \bibinfo{author}{\bibfnamefont{S.}~\bibnamefont{Tsujikawa}},
  \bibinfo{journal}{Int.J.Mod.Phys.} \textbf{\bibinfo{volume}{D15}},
  \bibinfo{pages}{1753} (\bibinfo{year}{2006}), \eprint{hep-th/0603057}.

\bibitem[{\citenamefont{Carroll}(1998)}]{Carroll:1998zi}
\bibinfo{author}{\bibfnamefont{S.~M.} \bibnamefont{Carroll}},
  \bibinfo{journal}{Phys.Rev.Lett.} \textbf{\bibinfo{volume}{81}},
  \bibinfo{pages}{3067} (\bibinfo{year}{1998}), \eprint{astro-ph/9806099}.

\bibitem[{\citenamefont{Adelberger et~al.}(2009)\citenamefont{Adelberger,
  Gundlach, Heckel, Hoedl, and Schlamminger}}]{Adelberger:2009zz}
\bibinfo{author}{\bibfnamefont{E.}~\bibnamefont{Adelberger}},
  \bibinfo{author}{\bibfnamefont{J.}~\bibnamefont{Gundlach}},
  \bibinfo{author}{\bibfnamefont{B.}~\bibnamefont{Heckel}},
  \bibinfo{author}{\bibfnamefont{S.}~\bibnamefont{Hoedl}}, \bibnamefont{and}
  \bibinfo{author}{\bibfnamefont{S.}~\bibnamefont{Schlamminger}},
  \bibinfo{journal}{Prog.Part.Nucl.Phys.} \textbf{\bibinfo{volume}{62}},
  \bibinfo{pages}{102} (\bibinfo{year}{2009}).

\bibitem[{\citenamefont{Khoury and
  Weltman}(2004{\natexlab{a}})}]{Khoury:2003aq}
\bibinfo{author}{\bibfnamefont{J.}~\bibnamefont{Khoury}} \bibnamefont{and}
  \bibinfo{author}{\bibfnamefont{A.}~\bibnamefont{Weltman}},
  \bibinfo{journal}{Phys.Rev.Lett.} \textbf{\bibinfo{volume}{93}},
  \bibinfo{pages}{171104} (\bibinfo{year}{2004}{\natexlab{a}}),
  \eprint{astro-ph/0309300}.

\bibitem[{\citenamefont{Khoury and
  Weltman}(2004{\natexlab{b}})}]{Khoury:2003rn}
\bibinfo{author}{\bibfnamefont{J.}~\bibnamefont{Khoury}} \bibnamefont{and}
  \bibinfo{author}{\bibfnamefont{A.}~\bibnamefont{Weltman}},
  \bibinfo{journal}{Phys.Rev.} \textbf{\bibinfo{volume}{D69}},
  \bibinfo{pages}{044026} (\bibinfo{year}{2004}{\natexlab{b}}),
  \eprint{astro-ph/0309411}.

\bibitem[{\citenamefont{Jain and Khoury}(2010)}]{Jain:2010ka}
\bibinfo{author}{\bibfnamefont{B.}~\bibnamefont{Jain}} \bibnamefont{and}
  \bibinfo{author}{\bibfnamefont{J.}~\bibnamefont{Khoury}},
  \bibinfo{journal}{Annals Phys.} \textbf{\bibinfo{volume}{325}},
  \bibinfo{pages}{1479} (\bibinfo{year}{2010}), \eprint{1004.3294}.

\bibitem[{\citenamefont{Carroll et~al.}(2004)\citenamefont{Carroll, Duvvuri,
  Trodden, and Turner}}]{Carroll:2003wy}
\bibinfo{author}{\bibfnamefont{S.~M.} \bibnamefont{Carroll}},
  \bibinfo{author}{\bibfnamefont{V.}~\bibnamefont{Duvvuri}},
  \bibinfo{author}{\bibfnamefont{M.}~\bibnamefont{Trodden}}, \bibnamefont{and}
  \bibinfo{author}{\bibfnamefont{M.~S.} \bibnamefont{Turner}},
  \bibinfo{journal}{Phys.Rev.} \textbf{\bibinfo{volume}{D70}},
  \bibinfo{pages}{043528} (\bibinfo{year}{2004}), \eprint{astro-ph/0306438}.

\bibitem[{\citenamefont{Chiba}(2003)}]{Chiba:2003ir}
\bibinfo{author}{\bibfnamefont{T.}~\bibnamefont{Chiba}},
  \bibinfo{journal}{Phys.Lett.} \textbf{\bibinfo{volume}{B575}},
  \bibinfo{pages}{1} (\bibinfo{year}{2003}), \eprint{astro-ph/0307338}.

\bibitem[{\citenamefont{Chiba et~al.}(2007)\citenamefont{Chiba, Smith, and
  Erickcek}}]{Chiba:2006jp}
\bibinfo{author}{\bibfnamefont{T.}~\bibnamefont{Chiba}},
  \bibinfo{author}{\bibfnamefont{T.~L.} \bibnamefont{Smith}}, \bibnamefont{and}
  \bibinfo{author}{\bibfnamefont{A.~L.} \bibnamefont{Erickcek}},
  \bibinfo{journal}{Phys.Rev.} \textbf{\bibinfo{volume}{D75}},
  \bibinfo{pages}{124014} (\bibinfo{year}{2007}), \eprint{astro-ph/0611867}.

\bibitem[{\citenamefont{Faulkner et~al.}(2007)\citenamefont{Faulkner, Tegmark,
  Bunn, and Mao}}]{Faulkner:2006ub}
\bibinfo{author}{\bibfnamefont{T.}~\bibnamefont{Faulkner}},
  \bibinfo{author}{\bibfnamefont{M.}~\bibnamefont{Tegmark}},
  \bibinfo{author}{\bibfnamefont{E.~F.} \bibnamefont{Bunn}}, \bibnamefont{and}
  \bibinfo{author}{\bibfnamefont{Y.}~\bibnamefont{Mao}},
  \bibinfo{journal}{Phys.Rev.} \textbf{\bibinfo{volume}{D76}},
  \bibinfo{pages}{063505} (\bibinfo{year}{2007}), \eprint{astro-ph/0612569}.

\bibitem[{\citenamefont{Hu and Sawicki}(2007)}]{Hu:2007nk}
\bibinfo{author}{\bibfnamefont{W.}~\bibnamefont{Hu}} \bibnamefont{and}
  \bibinfo{author}{\bibfnamefont{I.}~\bibnamefont{Sawicki}},
  \bibinfo{journal}{Phys.Rev.} \textbf{\bibinfo{volume}{D76}},
  \bibinfo{pages}{064004} (\bibinfo{year}{2007}), \eprint{0705.1158}.

\bibitem[{\citenamefont{Brax et~al.}(2008)\citenamefont{Brax, van~de Bruck,
  Davis, and Shaw}}]{Brax:2008hh}
\bibinfo{author}{\bibfnamefont{P.}~\bibnamefont{Brax}},
  \bibinfo{author}{\bibfnamefont{C.}~\bibnamefont{van~de Bruck}},
  \bibinfo{author}{\bibfnamefont{A.-C.} \bibnamefont{Davis}}, \bibnamefont{and}
  \bibinfo{author}{\bibfnamefont{D.~J.} \bibnamefont{Shaw}},
  \bibinfo{journal}{Phys.Rev.} \textbf{\bibinfo{volume}{D78}},
  \bibinfo{pages}{104021} (\bibinfo{year}{2008}), \eprint{0806.3415}.

\bibitem[{\citenamefont{Mota and Shaw}(2006)}]{Mota:2006ed}
\bibinfo{author}{\bibfnamefont{D.~F.} \bibnamefont{Mota}} \bibnamefont{and}
  \bibinfo{author}{\bibfnamefont{D.~J.} \bibnamefont{Shaw}},
  \bibinfo{journal}{Phys.Rev.Lett.} \textbf{\bibinfo{volume}{97}},
  \bibinfo{pages}{151102} (\bibinfo{year}{2006}), \eprint{hep-ph/0606204}.

\bibitem[{\citenamefont{Mota and Shaw}(2007)}]{Mota:2006fz}
\bibinfo{author}{\bibfnamefont{D.~F.} \bibnamefont{Mota}} \bibnamefont{and}
  \bibinfo{author}{\bibfnamefont{D.~J.} \bibnamefont{Shaw}},
  \bibinfo{journal}{Phys.Rev.} \textbf{\bibinfo{volume}{D75}},
  \bibinfo{pages}{063501} (\bibinfo{year}{2007}), \eprint{hep-ph/0608078}.

\bibitem[{\citenamefont{Brax et~al.}(2007{\natexlab{a}})\citenamefont{Brax,
  van~de Bruck, and Davis}}]{Brax:2007ak}
\bibinfo{author}{\bibfnamefont{P.}~\bibnamefont{Brax}},
  \bibinfo{author}{\bibfnamefont{C.}~\bibnamefont{van~de Bruck}},
  \bibnamefont{and} \bibinfo{author}{\bibfnamefont{A.-C.} \bibnamefont{Davis}},
  \bibinfo{journal}{Phys.Rev.Lett.} \textbf{\bibinfo{volume}{99}},
  \bibinfo{pages}{121103} (\bibinfo{year}{2007}{\natexlab{a}}),
  \eprint{hep-ph/0703243}.

\bibitem[{\citenamefont{Brax et~al.}(2007{\natexlab{b}})\citenamefont{Brax,
  van~de Bruck, Davis, Mota, and Shaw}}]{Brax:2007hi}
\bibinfo{author}{\bibfnamefont{P.}~\bibnamefont{Brax}},
  \bibinfo{author}{\bibfnamefont{C.}~\bibnamefont{van~de Bruck}},
  \bibinfo{author}{\bibfnamefont{A.-C.} \bibnamefont{Davis}},
  \bibinfo{author}{\bibfnamefont{D.~F.} \bibnamefont{Mota}}, \bibnamefont{and}
  \bibinfo{author}{\bibfnamefont{D.~J.} \bibnamefont{Shaw}},
  \bibinfo{journal}{Phys.Rev.} \textbf{\bibinfo{volume}{D76}},
  \bibinfo{pages}{085010} (\bibinfo{year}{2007}{\natexlab{b}}),
  \eprint{0707.2801}.

\bibitem[{\citenamefont{Ahlers et~al.}(2008)\citenamefont{Ahlers, Lindner,
  Ringwald, Schrempp, and Weniger}}]{Ahlers:2007st}
\bibinfo{author}{\bibfnamefont{M.}~\bibnamefont{Ahlers}},
  \bibinfo{author}{\bibfnamefont{A.}~\bibnamefont{Lindner}},
  \bibinfo{author}{\bibfnamefont{A.}~\bibnamefont{Ringwald}},
  \bibinfo{author}{\bibfnamefont{L.}~\bibnamefont{Schrempp}}, \bibnamefont{and}
  \bibinfo{author}{\bibfnamefont{C.}~\bibnamefont{Weniger}},
  \bibinfo{journal}{Phys.Rev.} \textbf{\bibinfo{volume}{D77}},
  \bibinfo{pages}{015018} (\bibinfo{year}{2008}), \eprint{0710.1555}.

\bibitem[{\citenamefont{Gies et~al.}(2008)\citenamefont{Gies, Mota, and
  Shaw}}]{Gies:2007su}
\bibinfo{author}{\bibfnamefont{H.}~\bibnamefont{Gies}},
  \bibinfo{author}{\bibfnamefont{D.~F.} \bibnamefont{Mota}}, \bibnamefont{and}
  \bibinfo{author}{\bibfnamefont{D.~J.} \bibnamefont{Shaw}},
  \bibinfo{journal}{Phys.Rev.} \textbf{\bibinfo{volume}{D77}},
  \bibinfo{pages}{025016} (\bibinfo{year}{2008}), \eprint{0710.1556}.

\bibitem[{\citenamefont{Steffen et~al.}(2010)}]{Steffen:2010ze}
\bibinfo{author}{\bibfnamefont{J.~H.} \bibnamefont{Steffen}}
  \bibnamefont{et~al.} (\bibinfo{collaboration}{GammeV Collaboration}),
  \bibinfo{journal}{Phys.Rev.Lett.} \textbf{\bibinfo{volume}{105}},
  \bibinfo{pages}{261803} (\bibinfo{year}{2010}), \eprint{1010.0988}.

\bibitem[{\citenamefont{Brax and Burrage}(2010)}]{Brax:2010jk}
\bibinfo{author}{\bibfnamefont{P.}~\bibnamefont{Brax}} \bibnamefont{and}
  \bibinfo{author}{\bibfnamefont{C.}~\bibnamefont{Burrage}},
  \bibinfo{journal}{Phys.Rev.} \textbf{\bibinfo{volume}{D82}},
  \bibinfo{pages}{095014} (\bibinfo{year}{2010}), \eprint{1009.1065}.

\bibitem[{\citenamefont{Brax and Pignol}(2011)}]{Brax:2011hb}
\bibinfo{author}{\bibfnamefont{P.}~\bibnamefont{Brax}} \bibnamefont{and}
  \bibinfo{author}{\bibfnamefont{G.}~\bibnamefont{Pignol}},
  \bibinfo{journal}{Phys.Rev.Lett.} \textbf{\bibinfo{volume}{107}},
  \bibinfo{pages}{111301} (\bibinfo{year}{2011}), \eprint{1105.3420}.

\bibitem[{\citenamefont{Pokotilovski}(2013)}]{Pokotilovski:2012js}
\bibinfo{author}{\bibfnamefont{Y.}~\bibnamefont{Pokotilovski}},
  \bibinfo{journal}{Phys.Lett.} \textbf{\bibinfo{volume}{B719}},
  \bibinfo{pages}{341} (\bibinfo{year}{2013}), \eprint{1203.5017}.

\bibitem[{\citenamefont{Brax et~al.}(2013{\natexlab{a}})\citenamefont{Brax,
  Pignol, and Roulier}}]{Brax:2013cfa}
\bibinfo{author}{\bibfnamefont{P.}~\bibnamefont{Brax}},
  \bibinfo{author}{\bibfnamefont{G.}~\bibnamefont{Pignol}}, \bibnamefont{and}
  \bibinfo{author}{\bibfnamefont{D.}~\bibnamefont{Roulier}},
  \bibinfo{journal}{Phys.Rev.} \textbf{\bibinfo{volume}{D88}},
  \bibinfo{pages}{083004} (\bibinfo{year}{2013}{\natexlab{a}}),
  \eprint{1306.6536}.

\bibitem[{\citenamefont{Brax and Burrage}(2011)}]{Brax:2010gp}
\bibinfo{author}{\bibfnamefont{P.}~\bibnamefont{Brax}} \bibnamefont{and}
  \bibinfo{author}{\bibfnamefont{C.}~\bibnamefont{Burrage}},
  \bibinfo{journal}{Phys.Rev.} \textbf{\bibinfo{volume}{D83}},
  \bibinfo{pages}{035020} (\bibinfo{year}{2011}), \eprint{1010.5108}.

\bibitem[{\citenamefont{Brax et~al.}(2007{\natexlab{c}})\citenamefont{Brax,
  van~de Bruck, Davis, Mota, and Shaw}}]{Brax:2007vm}
\bibinfo{author}{\bibfnamefont{P.}~\bibnamefont{Brax}},
  \bibinfo{author}{\bibfnamefont{C.}~\bibnamefont{van~de Bruck}},
  \bibinfo{author}{\bibfnamefont{A.-C.} \bibnamefont{Davis}},
  \bibinfo{author}{\bibfnamefont{D.~F.} \bibnamefont{Mota}}, \bibnamefont{and}
  \bibinfo{author}{\bibfnamefont{D.~J.} \bibnamefont{Shaw}},
  \bibinfo{journal}{Phys.Rev.} \textbf{\bibinfo{volume}{D76}},
  \bibinfo{pages}{124034} (\bibinfo{year}{2007}{\natexlab{c}}),
  \eprint{0709.2075}.

\bibitem[{\citenamefont{Brax et~al.}(2010{\natexlab{a}})\citenamefont{Brax,
  van~de Bruck, Davis, Shaw, and Iannuzzi}}]{Brax:2010xx}
\bibinfo{author}{\bibfnamefont{P.}~\bibnamefont{Brax}},
  \bibinfo{author}{\bibfnamefont{C.}~\bibnamefont{van~de Bruck}},
  \bibinfo{author}{\bibfnamefont{A.}~\bibnamefont{Davis}},
  \bibinfo{author}{\bibfnamefont{D.}~\bibnamefont{Shaw}}, \bibnamefont{and}
  \bibinfo{author}{\bibfnamefont{D.}~\bibnamefont{Iannuzzi}},
  \bibinfo{journal}{Phys.Rev.Lett.} \textbf{\bibinfo{volume}{104}},
  \bibinfo{pages}{241101} (\bibinfo{year}{2010}{\natexlab{a}}),
  \eprint{1003.1605}.

\bibitem[{\citenamefont{Rybka et~al.}(2010)\citenamefont{Rybka, Hotz,
  Rosenberg, Asztalos, Carosi et~al.}}]{Rybka:2010ah}
\bibinfo{author}{\bibfnamefont{G.}~\bibnamefont{Rybka}},
  \bibinfo{author}{\bibfnamefont{M.}~\bibnamefont{Hotz}},
  \bibinfo{author}{\bibfnamefont{L.}~\bibnamefont{Rosenberg}},
  \bibinfo{author}{\bibfnamefont{S.}~\bibnamefont{Asztalos}},
  \bibinfo{author}{\bibfnamefont{G.}~\bibnamefont{Carosi}},
  \bibnamefont{et~al.}, \bibinfo{journal}{Phys.Rev.Lett.}
  \textbf{\bibinfo{volume}{105}}, \bibinfo{pages}{051801}
  (\bibinfo{year}{2010}), \eprint{1004.5160}.

\bibitem[{\citenamefont{Brax et~al.}(2009)\citenamefont{Brax, Burrage, Davis,
  Seery, and Weltman}}]{Brax:2009aw}
\bibinfo{author}{\bibfnamefont{P.}~\bibnamefont{Brax}},
  \bibinfo{author}{\bibfnamefont{C.}~\bibnamefont{Burrage}},
  \bibinfo{author}{\bibfnamefont{A.-C.} \bibnamefont{Davis}},
  \bibinfo{author}{\bibfnamefont{D.}~\bibnamefont{Seery}}, \bibnamefont{and}
  \bibinfo{author}{\bibfnamefont{A.}~\bibnamefont{Weltman}},
  \bibinfo{journal}{JHEP} \textbf{\bibinfo{volume}{0909}}, \bibinfo{pages}{128}
  (\bibinfo{year}{2009}), \eprint{0904.3002}.

\bibitem[{\citenamefont{Brax et~al.}(2010{\natexlab{b}})\citenamefont{Brax,
  Burrage, Davis, Seery, and Weltman}}]{Brax:2009ey}
\bibinfo{author}{\bibfnamefont{P.}~\bibnamefont{Brax}},
  \bibinfo{author}{\bibfnamefont{C.}~\bibnamefont{Burrage}},
  \bibinfo{author}{\bibfnamefont{A.-C.} \bibnamefont{Davis}},
  \bibinfo{author}{\bibfnamefont{D.}~\bibnamefont{Seery}}, \bibnamefont{and}
  \bibinfo{author}{\bibfnamefont{A.}~\bibnamefont{Weltman}},
  \bibinfo{journal}{Phys.Rev.} \textbf{\bibinfo{volume}{D81}},
  \bibinfo{pages}{103524} (\bibinfo{year}{2010}{\natexlab{b}}),
  \eprint{0911.1267}.

\bibitem[{\citenamefont{Baker et~al.}(2012{\natexlab{a}})\citenamefont{Baker,
  Lindner, Semertzidis, Upadhye, and Zioutas}}]{Baker:2012ah}
\bibinfo{author}{\bibfnamefont{O.}~\bibnamefont{Baker}},
  \bibinfo{author}{\bibfnamefont{A.}~\bibnamefont{Lindner}},
  \bibinfo{author}{\bibfnamefont{Y.}~\bibnamefont{Semertzidis}},
  \bibinfo{author}{\bibfnamefont{A.}~\bibnamefont{Upadhye}}, \bibnamefont{and}
  \bibinfo{author}{\bibfnamefont{K.}~\bibnamefont{Zioutas}}
  (\bibinfo{year}{2012}{\natexlab{a}}), \eprint{1201.6508}.

\bibitem[{\citenamefont{Baker et~al.}(2012{\natexlab{b}})\citenamefont{Baker,
  Lindner, Upadhye, and Zioutas}}]{Baker:2012nq}
\bibinfo{author}{\bibfnamefont{K.}~\bibnamefont{Baker}},
  \bibinfo{author}{\bibfnamefont{A.}~\bibnamefont{Lindner}},
  \bibinfo{author}{\bibfnamefont{A.}~\bibnamefont{Upadhye}}, \bibnamefont{and}
  \bibinfo{author}{\bibfnamefont{K.}~\bibnamefont{Zioutas}}
  (\bibinfo{year}{2012}{\natexlab{b}}), \eprint{1201.0079}.

\bibitem[{\citenamefont{Burrage
  et~al.}(2009{\natexlab{a}})\citenamefont{Burrage, Davis, and
  Shaw}}]{Burrage:2009mj}
\bibinfo{author}{\bibfnamefont{C.}~\bibnamefont{Burrage}},
  \bibinfo{author}{\bibfnamefont{A.-C.} \bibnamefont{Davis}}, \bibnamefont{and}
  \bibinfo{author}{\bibfnamefont{D.~J.} \bibnamefont{Shaw}},
  \bibinfo{journal}{Phys.Rev.Lett.} \textbf{\bibinfo{volume}{102}},
  \bibinfo{pages}{201101} (\bibinfo{year}{2009}{\natexlab{a}}),
  \eprint{0902.2320}.

\bibitem[{\citenamefont{Burrage
  et~al.}(2009{\natexlab{b}})\citenamefont{Burrage, Davis, and
  Shaw}}]{Burrage:2008ii}
\bibinfo{author}{\bibfnamefont{C.}~\bibnamefont{Burrage}},
  \bibinfo{author}{\bibfnamefont{A.-C.} \bibnamefont{Davis}}, \bibnamefont{and}
  \bibinfo{author}{\bibfnamefont{D.~J.} \bibnamefont{Shaw}},
  \bibinfo{journal}{Phys.Rev.} \textbf{\bibinfo{volume}{D79}},
  \bibinfo{pages}{044028} (\bibinfo{year}{2009}{\natexlab{b}}),
  \eprint{0809.1763}.

\bibitem[{\citenamefont{{Levshakov} et~al.}(2010)\citenamefont{{Levshakov},
  {Lapinov}, {Henkel}, {Molaro}, {Reimers}, {Kozlov}, and
  {Agafonova}}}]{Levshakov:2010tj}
\bibinfo{author}{\bibfnamefont{S.~A.} \bibnamefont{{Levshakov}}},
  \bibinfo{author}{\bibfnamefont{A.~V.} \bibnamefont{{Lapinov}}},
  \bibinfo{author}{\bibfnamefont{C.}~\bibnamefont{{Henkel}}},
  \bibinfo{author}{\bibfnamefont{P.}~\bibnamefont{{Molaro}}},
  \bibinfo{author}{\bibfnamefont{D.}~\bibnamefont{{Reimers}}},
  \bibinfo{author}{\bibfnamefont{M.~G.} \bibnamefont{{Kozlov}}},
  \bibnamefont{and} \bibinfo{author}{\bibfnamefont{I.~I.}
  \bibnamefont{{Agafonova}}}, \bibinfo{journal}{Astron. \& Astrophys.}
  \textbf{\bibinfo{volume}{524}}, \bibinfo{eid}{A32} (\bibinfo{year}{2010}),
  \eprint{1008.1160}.

\bibitem[{\citenamefont{Brax and Davis}(2013)}]{Brax:2013uh}
\bibinfo{author}{\bibfnamefont{P.}~\bibnamefont{Brax}} \bibnamefont{and}
  \bibinfo{author}{\bibfnamefont{A.-C.} \bibnamefont{Davis}}
  (\bibinfo{year}{2013}), \eprint{1301.5587}.

\bibitem[{\citenamefont{Schelpe}(2010)}]{Schelpe:2010he}
\bibinfo{author}{\bibfnamefont{C.~A.} \bibnamefont{Schelpe}},
  \bibinfo{journal}{Phys.Rev.} \textbf{\bibinfo{volume}{D82}},
  \bibinfo{pages}{044033} (\bibinfo{year}{2010}), \eprint{1003.0232}.

\bibitem[{\citenamefont{Davis et~al.}(2009)\citenamefont{Davis, Schelpe, and
  Shaw}}]{Davis:2009vk}
\bibinfo{author}{\bibfnamefont{A.-C.} \bibnamefont{Davis}},
  \bibinfo{author}{\bibfnamefont{C.~A.} \bibnamefont{Schelpe}},
  \bibnamefont{and} \bibinfo{author}{\bibfnamefont{D.~J.} \bibnamefont{Shaw}},
  \bibinfo{journal}{Phys.Rev.} \textbf{\bibinfo{volume}{D80}},
  \bibinfo{pages}{064016} (\bibinfo{year}{2009}), \eprint{0907.2672}.

\bibitem[{\citenamefont{Hu et~al.}(2013)\citenamefont{Hu, Liguori, Bartolo, and
  Matarrese}}]{Hu:2013aqa}
\bibinfo{author}{\bibfnamefont{B.}~\bibnamefont{Hu}},
  \bibinfo{author}{\bibfnamefont{M.}~\bibnamefont{Liguori}},
  \bibinfo{author}{\bibfnamefont{N.}~\bibnamefont{Bartolo}}, \bibnamefont{and}
  \bibinfo{author}{\bibfnamefont{S.}~\bibnamefont{Matarrese}},
  \bibinfo{journal}{Phys.Rev.} \textbf{\bibinfo{volume}{D88}},
  \bibinfo{pages}{123514} (\bibinfo{year}{2013}), \eprint{1307.5276}.

\bibitem[{\citenamefont{Brax et~al.}(2013{\natexlab{b}})\citenamefont{Brax,
  Clesse, and Davis}}]{Brax:2012cr}
\bibinfo{author}{\bibfnamefont{P.}~\bibnamefont{Brax}},
  \bibinfo{author}{\bibfnamefont{S.}~\bibnamefont{Clesse}}, \bibnamefont{and}
  \bibinfo{author}{\bibfnamefont{A.-C.} \bibnamefont{Davis}},
  \bibinfo{journal}{JCAP} \textbf{\bibinfo{volume}{1301}}, \bibinfo{pages}{003}
  (\bibinfo{year}{2013}{\natexlab{b}}), \eprint{1207.1273}.

\bibitem[{\citenamefont{{Gannouji} et~al.}(2010)\citenamefont{{Gannouji},
  {Moraes}, {Mota}, {Polarski}, {Tsujikawa}, and {Winther}}}]{Gannouji:2010fc}
\bibinfo{author}{\bibfnamefont{R.}~\bibnamefont{{Gannouji}}},
  \bibinfo{author}{\bibfnamefont{B.}~\bibnamefont{{Moraes}}},
  \bibinfo{author}{\bibfnamefont{D.~F.} \bibnamefont{{Mota}}},
  \bibinfo{author}{\bibfnamefont{D.}~\bibnamefont{{Polarski}}},
  \bibinfo{author}{\bibfnamefont{S.}~\bibnamefont{{Tsujikawa}}},
  \bibnamefont{and} \bibinfo{author}{\bibfnamefont{H.~A.}
  \bibnamefont{{Winther}}}, \bibinfo{journal}{\prd}
  \textbf{\bibinfo{volume}{82}}, \bibinfo{eid}{124006} (\bibinfo{year}{2010}),
  \eprint{1010.3769}.

\bibitem[{\citenamefont{Upadhye}(2012)}]{Upadhye:2012qu}
\bibinfo{author}{\bibfnamefont{A.}~\bibnamefont{Upadhye}},
  \bibinfo{journal}{Phys.Rev.} \textbf{\bibinfo{volume}{D86}},
  \bibinfo{pages}{102003} (\bibinfo{year}{2012}), \eprint{1209.0211}.

\bibitem[{\citenamefont{Brax and Valageas}(2012)}]{Brax:2012sy}
\bibinfo{author}{\bibfnamefont{P.}~\bibnamefont{Brax}} \bibnamefont{and}
  \bibinfo{author}{\bibfnamefont{P.}~\bibnamefont{Valageas}},
  \bibinfo{journal}{Phys.Rev.} \textbf{\bibinfo{volume}{D86}},
  \bibinfo{pages}{063512} (\bibinfo{year}{2012}), \eprint{1205.6583}.

\bibitem[{\citenamefont{{Li} et~al.}(2013)\citenamefont{{Li}, {Hellwing},
  {Koyama}, {Zhao}, {Jennings}, and {Baugh}}}]{Li:2012by}
\bibinfo{author}{\bibfnamefont{B.}~\bibnamefont{{Li}}},
  \bibinfo{author}{\bibfnamefont{W.~A.} \bibnamefont{{Hellwing}}},
  \bibinfo{author}{\bibfnamefont{K.}~\bibnamefont{{Koyama}}},
  \bibinfo{author}{\bibfnamefont{G.-B.} \bibnamefont{{Zhao}}},
  \bibinfo{author}{\bibfnamefont{E.}~\bibnamefont{{Jennings}}},
  \bibnamefont{and} \bibinfo{author}{\bibfnamefont{C.~M.}
  \bibnamefont{{Baugh}}}, \bibinfo{journal}{Mon.Not.Roy.Astron.Soc.}
  \textbf{\bibinfo{volume}{428}}, \bibinfo{pages}{743} (\bibinfo{year}{2013}),
  \eprint{1206.4317}.

\bibitem[{\citenamefont{{Hellwing} et~al.}(2013)\citenamefont{{Hellwing}, {Li},
  {Frenk}, and {Cole}}}]{Hellwing:2013rxa}
\bibinfo{author}{\bibfnamefont{W.~A.} \bibnamefont{{Hellwing}}},
  \bibinfo{author}{\bibfnamefont{B.}~\bibnamefont{{Li}}},
  \bibinfo{author}{\bibfnamefont{C.~S.} \bibnamefont{{Frenk}}},
  \bibnamefont{and} \bibinfo{author}{\bibfnamefont{S.}~\bibnamefont{{Cole}}},
  \bibinfo{journal}{Mon.Not.Roy.Astron.Soc.} \textbf{\bibinfo{volume}{435}},
  \bibinfo{pages}{2806} (\bibinfo{year}{2013}), \eprint{1305.7486}.

\bibitem[{\citenamefont{Brax et~al.}(2013{\natexlab{c}})\citenamefont{Brax,
  Davis, Li, Winther, and Zhao}}]{Brax:2013mua}
\bibinfo{author}{\bibfnamefont{P.}~\bibnamefont{Brax}},
  \bibinfo{author}{\bibfnamefont{A.-C.} \bibnamefont{Davis}},
  \bibinfo{author}{\bibfnamefont{B.}~\bibnamefont{Li}},
  \bibinfo{author}{\bibfnamefont{H.~A.} \bibnamefont{Winther}},
  \bibnamefont{and} \bibinfo{author}{\bibfnamefont{G.-B.} \bibnamefont{Zhao}},
  \bibinfo{journal}{JCAP} \textbf{\bibinfo{volume}{1304}}, \bibinfo{pages}{029}
  (\bibinfo{year}{2013}{\natexlab{c}}), \eprint{1303.0007}.

\bibitem[{\citenamefont{{Cardone} et~al.}(2013)\citenamefont{{Cardone},
  {Camera}, {Mainini}, {Romano}, {Diaferio}, {Maoli}, and
  {Scaramella}}}]{Cardone:2012zn}
\bibinfo{author}{\bibfnamefont{V.~F.} \bibnamefont{{Cardone}}},
  \bibinfo{author}{\bibfnamefont{S.}~\bibnamefont{{Camera}}},
  \bibinfo{author}{\bibfnamefont{R.}~\bibnamefont{{Mainini}}},
  \bibinfo{author}{\bibfnamefont{A.}~\bibnamefont{{Romano}}},
  \bibinfo{author}{\bibfnamefont{A.}~\bibnamefont{{Diaferio}}},
  \bibinfo{author}{\bibfnamefont{R.}~\bibnamefont{{Maoli}}}, \bibnamefont{and}
  \bibinfo{author}{\bibfnamefont{R.}~\bibnamefont{{Scaramella}}},
  \bibinfo{journal}{Mon.Not.Roy.Astron.Soc.} \textbf{\bibinfo{volume}{430}},
  \bibinfo{pages}{2896} (\bibinfo{year}{2013}), \eprint{1204.3148}.

\bibitem[{\citenamefont{Jain et~al.}(2013)\citenamefont{Jain, Vikram, and
  Sakstein}}]{Jain:2012tn}
\bibinfo{author}{\bibfnamefont{B.}~\bibnamefont{Jain}},
  \bibinfo{author}{\bibfnamefont{V.}~\bibnamefont{Vikram}}, \bibnamefont{and}
  \bibinfo{author}{\bibfnamefont{J.}~\bibnamefont{Sakstein}},
  \bibinfo{journal}{Astrophys.J.} \textbf{\bibinfo{volume}{779}},
  \bibinfo{pages}{39} (\bibinfo{year}{2013}), \eprint{1204.6044}.

\bibitem[{\citenamefont{Hinterbichler et~al.}(2011)\citenamefont{Hinterbichler,
  Khoury, and Nastase}}]{Hinterbichler:2010wu}
\bibinfo{author}{\bibfnamefont{K.}~\bibnamefont{Hinterbichler}},
  \bibinfo{author}{\bibfnamefont{J.}~\bibnamefont{Khoury}}, \bibnamefont{and}
  \bibinfo{author}{\bibfnamefont{H.}~\bibnamefont{Nastase}},
  \bibinfo{journal}{JHEP} \textbf{\bibinfo{volume}{1103}}, \bibinfo{pages}{061}
  (\bibinfo{year}{2011}), \eprint{1012.4462}.

\bibitem[{\citenamefont{Nastase and Weltman}(2013)}]{Nastase:2013ik}
\bibinfo{author}{\bibfnamefont{H.}~\bibnamefont{Nastase}} \bibnamefont{and}
  \bibinfo{author}{\bibfnamefont{A.}~\bibnamefont{Weltman}},
  \bibinfo{journal}{JHEP} \textbf{\bibinfo{volume}{1308}}, \bibinfo{pages}{059}
  (\bibinfo{year}{2013}), \eprint{1301.7120}.

\bibitem[{\citenamefont{{Hinterbichler}
  et~al.}(2013)\citenamefont{{Hinterbichler}, {Khoury}, {Nastase}, and
  {Rosenfeld}}}]{2013JHEP...08..053H}
\bibinfo{author}{\bibfnamefont{K.}~\bibnamefont{{Hinterbichler}}},
  \bibinfo{author}{\bibfnamefont{J.}~\bibnamefont{{Khoury}}},
  \bibinfo{author}{\bibfnamefont{H.}~\bibnamefont{{Nastase}}},
  \bibnamefont{and}
  \bibinfo{author}{\bibfnamefont{R.}~\bibnamefont{{Rosenfeld}}},
  \bibinfo{journal}{JHEP} \textbf{\bibinfo{volume}{8}}, \bibinfo{pages}{53}
  (\bibinfo{year}{2013}), \eprint{1301.6756}.

\bibitem[{\citenamefont{Brax et~al.}(2004{\natexlab{a}})\citenamefont{Brax,
  van~de Bruck, and Davis}}]{Brax:2004ym}
\bibinfo{author}{\bibfnamefont{P.}~\bibnamefont{Brax}},
  \bibinfo{author}{\bibfnamefont{C.}~\bibnamefont{van~de Bruck}},
  \bibnamefont{and} \bibinfo{author}{\bibfnamefont{A.}~\bibnamefont{Davis}},
  \bibinfo{journal}{JCAP} \textbf{\bibinfo{volume}{0411}}, \bibinfo{pages}{004}
  (\bibinfo{year}{2004}{\natexlab{a}}), \eprint{astro-ph/0408464}.

\bibitem[{\citenamefont{Brax and Martin}(2007)}]{Brax:2006np}
\bibinfo{author}{\bibfnamefont{P.}~\bibnamefont{Brax}} \bibnamefont{and}
  \bibinfo{author}{\bibfnamefont{J.}~\bibnamefont{Martin}},
  \bibinfo{journal}{Phys.Lett.} \textbf{\bibinfo{volume}{B647}},
  \bibinfo{pages}{320} (\bibinfo{year}{2007}), \eprint{hep-th/0612208}.

\bibitem[{\citenamefont{Upadhye et~al.}(2012)\citenamefont{Upadhye, Hu, and
  Khoury}}]{Upadhye:2012vh}
\bibinfo{author}{\bibfnamefont{A.}~\bibnamefont{Upadhye}},
  \bibinfo{author}{\bibfnamefont{W.}~\bibnamefont{Hu}}, \bibnamefont{and}
  \bibinfo{author}{\bibfnamefont{J.}~\bibnamefont{Khoury}},
  \bibinfo{journal}{Phys.Rev.Lett.} \textbf{\bibinfo{volume}{109}},
  \bibinfo{pages}{041301} (\bibinfo{year}{2012}), \eprint{1204.3906}.

\bibitem[{\citenamefont{Arbuzova et~al.}(2012)\citenamefont{Arbuzova, Dolgov,
  and Reverberi}}]{Arbuzova:2011fu}
\bibinfo{author}{\bibfnamefont{E.}~\bibnamefont{Arbuzova}},
  \bibinfo{author}{\bibfnamefont{A.}~\bibnamefont{Dolgov}}, \bibnamefont{and}
  \bibinfo{author}{\bibfnamefont{L.}~\bibnamefont{Reverberi}},
  \bibinfo{journal}{JCAP} \textbf{\bibinfo{volume}{1202}}, \bibinfo{pages}{049}
  (\bibinfo{year}{2012}), \eprint{1112.4995}.

\bibitem[{\citenamefont{Arbuzova et~al.}(2013)\citenamefont{Arbuzova, Dolgov,
  and Reverberi}}]{Arbuzova:2013ina}
\bibinfo{author}{\bibfnamefont{E.}~\bibnamefont{Arbuzova}},
  \bibinfo{author}{\bibfnamefont{A.}~\bibnamefont{Dolgov}}, \bibnamefont{and}
  \bibinfo{author}{\bibfnamefont{L.}~\bibnamefont{Reverberi}},
  \bibinfo{journal}{Phys. Rev.} \textbf{\bibinfo{volume}{D88}}, \bibinfo{pages}{024035}
  (\bibinfo{year}{2013}), \eprint{1305.5668}.

\bibitem[{\citenamefont{{Erickcek} et~al.}(2013)\citenamefont{{Erickcek},
  {Barnaby}, {Burrage}, and {Huang}}}]{ourprl}
\bibinfo{author}{\bibfnamefont{A.~L.} \bibnamefont{{Erickcek}}},
  \bibinfo{author}{\bibfnamefont{N.}~\bibnamefont{{Barnaby}}},
  \bibinfo{author}{\bibfnamefont{C.}~\bibnamefont{{Burrage}}},
  \bibnamefont{and} \bibinfo{author}{\bibfnamefont{Z.}~\bibnamefont{{Huang}}},
  \bibinfo{journal}{Physical Review Letters} \textbf{\bibinfo{volume}{110}},
  \bibinfo{eid}{171101} (\bibinfo{year}{2013}), \eprint{1304.0009}.

\bibitem[{\citenamefont{Brax et~al.}(2004{\natexlab{b}})\citenamefont{Brax,
  van~de Bruck, Davis, Khoury, and Weltman}}]{Brax:2004qh}
\bibinfo{author}{\bibfnamefont{P.}~\bibnamefont{Brax}},
  \bibinfo{author}{\bibfnamefont{C.}~\bibnamefont{van~de Bruck}},
  \bibinfo{author}{\bibfnamefont{A.-C.} \bibnamefont{Davis}},
  \bibinfo{author}{\bibfnamefont{J.}~\bibnamefont{Khoury}}, \bibnamefont{and}
  \bibinfo{author}{\bibfnamefont{A.}~\bibnamefont{Weltman}},
  \bibinfo{journal}{Phys.Rev.} \textbf{\bibinfo{volume}{D70}},
  \bibinfo{pages}{123518} (\bibinfo{year}{2004}{\natexlab{b}}),
  \eprint{astro-ph/0408415}.

\bibitem[{\citenamefont{{Coc} et~al.}(2012)\citenamefont{{Coc}, {Descouvemont},
  {Olive}, {Uzan}, and {Vangioni}}}]{2012PhRvD..86d3529C}
\bibinfo{author}{\bibfnamefont{A.}~\bibnamefont{{Coc}}},
  \bibinfo{author}{\bibfnamefont{P.}~\bibnamefont{{Descouvemont}}},
  \bibinfo{author}{\bibfnamefont{K.~A.} \bibnamefont{{Olive}}},
  \bibinfo{author}{\bibfnamefont{J.-P.} \bibnamefont{{Uzan}}},
  \bibnamefont{and}
  \bibinfo{author}{\bibfnamefont{E.}~\bibnamefont{{Vangioni}}},
  \bibinfo{journal}{\prd} \textbf{\bibinfo{volume}{86}}, \bibinfo{eid}{043529}
  (\bibinfo{year}{2012}), \eprint{1206.1139}.

\bibitem[{\citenamefont{{Berengut} et~al.}(2013)\citenamefont{{Berengut},
  {Epelbaum}, {Flambaum}, {Hanhart}, {Mei{\ss}ner}, {Nebreda}, and
  {Pel{\'a}ez}}}]{2013PhRvD..87h5018B}
\bibinfo{author}{\bibfnamefont{J.~C.} \bibnamefont{{Berengut}}},
  \bibinfo{author}{\bibfnamefont{E.}~\bibnamefont{{Epelbaum}}},
  \bibinfo{author}{\bibfnamefont{V.~V.} \bibnamefont{{Flambaum}}},
  \bibinfo{author}{\bibfnamefont{C.}~\bibnamefont{{Hanhart}}},
  \bibinfo{author}{\bibfnamefont{U.-G.} \bibnamefont{{Mei{\ss}ner}}},
  \bibinfo{author}{\bibfnamefont{J.}~\bibnamefont{{Nebreda}}},
  \bibnamefont{and} \bibinfo{author}{\bibfnamefont{J.~R.}
  \bibnamefont{{Pel{\'a}ez}}}, \bibinfo{journal}{\prd}
  \textbf{\bibinfo{volume}{87}}, \bibinfo{eid}{085018} (\bibinfo{year}{2013}),
  \eprint{1301.1738}.

\bibitem[{\citenamefont{Mota and Schelpe}(2012)}]{Mota:2011nh}
\bibinfo{author}{\bibfnamefont{D.~F.} \bibnamefont{Mota}} \bibnamefont{and}
  \bibinfo{author}{\bibfnamefont{C.~A.} \bibnamefont{Schelpe}},
  \bibinfo{journal}{Phys.Rev.} \textbf{\bibinfo{volume}{D86}},
  \bibinfo{pages}{123002} (\bibinfo{year}{2012}), \eprint{1108.0892}.

\bibitem[{\citenamefont{Brax et~al.}(2012)\citenamefont{Brax, Davis, and
  Li}}]{Brax:2011aw}
\bibinfo{author}{\bibfnamefont{P.}~\bibnamefont{Brax}},
  \bibinfo{author}{\bibfnamefont{A.-C.} \bibnamefont{Davis}}, \bibnamefont{and}
  \bibinfo{author}{\bibfnamefont{B.}~\bibnamefont{Li}},
  \bibinfo{journal}{Phys.Lett.} \textbf{\bibinfo{volume}{B715}},
  \bibinfo{pages}{38} (\bibinfo{year}{2012}), \eprint{1111.6613}.

\bibitem[{\citenamefont{{Wang} et~al.}(2012)\citenamefont{{Wang}, {Hui}, and
  {Khoury}}}]{Wang:2012kj}
\bibinfo{author}{\bibfnamefont{J.}~\bibnamefont{{Wang}}},
  \bibinfo{author}{\bibfnamefont{L.}~\bibnamefont{{Hui}}}, \bibnamefont{and}
  \bibinfo{author}{\bibfnamefont{J.}~\bibnamefont{{Khoury}}},
  \bibinfo{journal}{Physical Review Letters} \textbf{\bibinfo{volume}{109}},
  \bibinfo{eid}{241301} (\bibinfo{year}{2012}), \eprint{1208.4612}.

\bibitem[{\citenamefont{Damour and
  Nordtvedt}(1993{\natexlab{a}})}]{Damour:1992kf}
\bibinfo{author}{\bibfnamefont{T.}~\bibnamefont{Damour}} \bibnamefont{and}
  \bibinfo{author}{\bibfnamefont{K.}~\bibnamefont{Nordtvedt}},
  \bibinfo{journal}{Phys.Rev.Lett.} \textbf{\bibinfo{volume}{70}},
  \bibinfo{pages}{2217} (\bibinfo{year}{1993}{\natexlab{a}}).

\bibitem[{\citenamefont{Damour and
  Nordtvedt}(1993{\natexlab{b}})}]{Damour:1993id}
\bibinfo{author}{\bibfnamefont{T.}~\bibnamefont{Damour}} \bibnamefont{and}
  \bibinfo{author}{\bibfnamefont{K.}~\bibnamefont{Nordtvedt}},
  \bibinfo{journal}{Phys.Rev.} \textbf{\bibinfo{volume}{D48}},
  \bibinfo{pages}{3436} (\bibinfo{year}{1993}{\natexlab{b}}).

\bibitem[{\citenamefont{Coc et~al.}(2006)\citenamefont{Coc, Olive, Uzan, and
  Vangioni}}]{Coc:2006rt}
\bibinfo{author}{\bibfnamefont{A.}~\bibnamefont{Coc}},
  \bibinfo{author}{\bibfnamefont{K.~A.} \bibnamefont{Olive}},
  \bibinfo{author}{\bibfnamefont{J.-P.} \bibnamefont{Uzan}}, \bibnamefont{and}
  \bibinfo{author}{\bibfnamefont{E.}~\bibnamefont{Vangioni}},
  \bibinfo{journal}{Phys.Rev.} \textbf{\bibinfo{volume}{D73}},
  \bibinfo{pages}{083525} (\bibinfo{year}{2006}), \eprint{astro-ph/0601299}.

\bibitem[{\citenamefont{Bazavov et~al.}(2009)\citenamefont{Bazavov,
  Bhattacharya, Cheng, Christ, DeTar et~al.}}]{Bazavov:2009zn}
\bibinfo{author}{\bibfnamefont{A.}~\bibnamefont{Bazavov}},
  \bibinfo{author}{\bibfnamefont{T.}~\bibnamefont{Bhattacharya}},
  \bibinfo{author}{\bibfnamefont{M.}~\bibnamefont{Cheng}},
  \bibinfo{author}{\bibfnamefont{N.}~\bibnamefont{Christ}},
  \bibinfo{author}{\bibfnamefont{C.}~\bibnamefont{DeTar}},
  \bibnamefont{et~al.}, \bibinfo{journal}{Phys.Rev.}
  \textbf{\bibinfo{volume}{D80}}, \bibinfo{pages}{014504}
  (\bibinfo{year}{2009}), \eprint{0903.4379}.

\bibitem[{\citenamefont{{Bors{\'a}nyi}
  et~al.}(2010)\citenamefont{{Bors{\'a}nyi}, {Endr{\H o}di}, {Fodor},
  {Jakov{\'a}c}, {Katz}, {Krieg}, {Ratti}, and {Szab{\'o}}}}]{Borsanyi:2010cj}
\bibinfo{author}{\bibfnamefont{S.}~\bibnamefont{{Bors{\'a}nyi}}},
  \bibinfo{author}{\bibfnamefont{G.}~\bibnamefont{{Endr{\H o}di}}},
  \bibinfo{author}{\bibfnamefont{Z.}~\bibnamefont{{Fodor}}},
  \bibinfo{author}{\bibfnamefont{A.}~\bibnamefont{{Jakov{\'a}c}}},
  \bibinfo{author}{\bibfnamefont{S.~D.} \bibnamefont{{Katz}}},
  \bibinfo{author}{\bibfnamefont{S.}~\bibnamefont{{Krieg}}},
  \bibinfo{author}{\bibfnamefont{C.}~\bibnamefont{{Ratti}}}, \bibnamefont{and}
  \bibinfo{author}{\bibfnamefont{K.~K.} \bibnamefont{{Szab{\'o}}}},
  \bibinfo{journal}{JHEP} \textbf{\bibinfo{volume}{1011}}, \bibinfo{pages}{077}
  (\bibinfo{year}{2010}), \eprint{1007.2580}.

\bibitem[{\citenamefont{Caldwell and Gubser}(2013)}]{Caldwell:2013mox}
\bibinfo{author}{\bibfnamefont{R.~R.} \bibnamefont{Caldwell}} \bibnamefont{and}
  \bibinfo{author}{\bibfnamefont{S.~S.} \bibnamefont{Gubser}},
  \bibinfo{journal}{Phys.Rev.} \textbf{\bibinfo{volume}{D87}},
  \bibinfo{pages}{063523} (\bibinfo{year}{2013}), \eprint{1302.1201}.

\bibitem[{\citenamefont{Arnold and Espinosa}(1993)}]{Arnold:1992rz}
\bibinfo{author}{\bibfnamefont{P.~B.} \bibnamefont{Arnold}} \bibnamefont{and}
  \bibinfo{author}{\bibfnamefont{O.}~\bibnamefont{Espinosa}},
  \bibinfo{journal}{Phys.Rev.} \textbf{\bibinfo{volume}{D47}},
  \bibinfo{pages}{3546} (\bibinfo{year}{1993}), \eprint{hep-ph/9212235}.

\bibitem[{\citenamefont{Kajantie et~al.}(2003)\citenamefont{Kajantie, Laine,
  Rummukainen, and Schroder}}]{Kajantie:2002wa}
\bibinfo{author}{\bibfnamefont{K.}~\bibnamefont{Kajantie}},
  \bibinfo{author}{\bibfnamefont{M.}~\bibnamefont{Laine}},
  \bibinfo{author}{\bibfnamefont{K.}~\bibnamefont{Rummukainen}},
  \bibnamefont{and} \bibinfo{author}{\bibfnamefont{Y.}~\bibnamefont{Schroder}},
  \bibinfo{journal}{Phys.Rev.} \textbf{\bibinfo{volume}{D67}},
  \bibinfo{pages}{105008} (\bibinfo{year}{2003}), \eprint{hep-ph/0211321}.

\bibitem[{\citenamefont{Davoudiasl et~al.}(2004)\citenamefont{Davoudiasl,
  Kitano, Kribs, Murayama, and Steinhardt}}]{Davoudiasl:2004gf}
\bibinfo{author}{\bibfnamefont{H.}~\bibnamefont{Davoudiasl}},
  \bibinfo{author}{\bibfnamefont{R.}~\bibnamefont{Kitano}},
  \bibinfo{author}{\bibfnamefont{G.~D.} \bibnamefont{Kribs}},
  \bibinfo{author}{\bibfnamefont{H.}~\bibnamefont{Murayama}}, \bibnamefont{and}
  \bibinfo{author}{\bibfnamefont{P.~J.} \bibnamefont{Steinhardt}},
  \bibinfo{journal}{Phys.Rev.Lett.} \textbf{\bibinfo{volume}{93}},
  \bibinfo{pages}{201301} (\bibinfo{year}{2004}), \eprint{hep-ph/0403019}.

\bibitem[{\citenamefont{{Kofman} et~al.}(2004)\citenamefont{{Kofman}, {Linde},
  {Liu}, {Maloney}, {McAllister}, and {Silverstein}}}]{Kofman:2004yc}
\bibinfo{author}{\bibfnamefont{L.}~\bibnamefont{{Kofman}}},
  \bibinfo{author}{\bibfnamefont{A.}~\bibnamefont{{Linde}}},
  \bibinfo{author}{\bibfnamefont{X.}~\bibnamefont{{Liu}}},
  \bibinfo{author}{\bibfnamefont{A.}~\bibnamefont{{Maloney}}},
  \bibinfo{author}{\bibfnamefont{L.}~\bibnamefont{{McAllister}}},
  \bibnamefont{and}
  \bibinfo{author}{\bibfnamefont{E.}~\bibnamefont{{Silverstein}}},
  \bibinfo{journal}{JHEP} \textbf{\bibinfo{volume}{0405}}, \bibinfo{pages}{030}
  (\bibinfo{year}{2004}), \eprint{hep-th/0403001}.

\bibitem[{\citenamefont{Braden et~al.}(2010)\citenamefont{Braden, Kofman, and
  Barnaby}}]{Braden:2010wd}
\bibinfo{author}{\bibfnamefont{J.}~\bibnamefont{Braden}},
  \bibinfo{author}{\bibfnamefont{L.}~\bibnamefont{Kofman}}, \bibnamefont{and}
  \bibinfo{author}{\bibfnamefont{N.}~\bibnamefont{Barnaby}},
  \bibinfo{journal}{JCAP} \textbf{\bibinfo{volume}{1007}}, \bibinfo{pages}{016}
  (\bibinfo{year}{2010}), \eprint{1005.2196}.

\bibitem[{\citenamefont{Boyanovsky et~al.}(1995)\citenamefont{Boyanovsky,
  de~Vega, Holman, Lee, and Singh}}]{Boyanovsky:1994me}
\bibinfo{author}{\bibfnamefont{D.}~\bibnamefont{Boyanovsky}},
  \bibinfo{author}{\bibfnamefont{H.}~\bibnamefont{de~Vega}},
  \bibinfo{author}{\bibfnamefont{R.}~\bibnamefont{Holman}},
  \bibinfo{author}{\bibfnamefont{D.}~\bibnamefont{Lee}}, \bibnamefont{and}
  \bibinfo{author}{\bibfnamefont{A.}~\bibnamefont{Singh}},
  \bibinfo{journal}{Phys.Rev.} \textbf{\bibinfo{volume}{D51}},
  \bibinfo{pages}{4419} (\bibinfo{year}{1995}), \eprint{hep-ph/9408214}.

\bibitem[{\citenamefont{{Damour} et~al.}(1990)\citenamefont{{Damour},
  {Gibbons}, and {Gundlach}}}]{1990PhRvL..64..123D}
\bibinfo{author}{\bibfnamefont{T.}~\bibnamefont{{Damour}}},
  \bibinfo{author}{\bibfnamefont{G.~W.} \bibnamefont{{Gibbons}}},
  \bibnamefont{and}
  \bibinfo{author}{\bibfnamefont{C.}~\bibnamefont{{Gundlach}}},
  \bibinfo{journal}{Physical Review Letters} \textbf{\bibinfo{volume}{64}},
  \bibinfo{pages}{123} (\bibinfo{year}{1990}).

\bibitem[{\citenamefont{{Amendola}}(2000)}]{2000PhRvD..62d3511A}
\bibinfo{author}{\bibfnamefont{L.}~\bibnamefont{{Amendola}}},
  \bibinfo{journal}{\prd} \textbf{\bibinfo{volume}{62}}, \bibinfo{eid}{043511}
  (\bibinfo{year}{2000}), \eprint{astro-ph/9908023}.

\bibitem[{\citenamefont{{Kesden} and
  {Kamionkowski}}(2006{\natexlab{a}})}]{2006PhRvL..97m1303K}
\bibinfo{author}{\bibfnamefont{M.}~\bibnamefont{{Kesden}}} \bibnamefont{and}
  \bibinfo{author}{\bibfnamefont{M.}~\bibnamefont{{Kamionkowski}}},
  \bibinfo{journal}{Physical Review Letters} \textbf{\bibinfo{volume}{97}},
  \bibinfo{eid}{131303} (\bibinfo{year}{2006}{\natexlab{a}}),
  \eprint{astro-ph/0606566}.

\bibitem[{\citenamefont{{Kesden} and
  {Kamionkowski}}(2006{\natexlab{b}})}]{2006PhRvD..74h3007K}
\bibinfo{author}{\bibfnamefont{M.}~\bibnamefont{{Kesden}}} \bibnamefont{and}
  \bibinfo{author}{\bibfnamefont{M.}~\bibnamefont{{Kamionkowski}}},
  \bibinfo{journal}{\prd} \textbf{\bibinfo{volume}{74}}, \bibinfo{eid}{083007}
  (\bibinfo{year}{2006}{\natexlab{b}}), \eprint{astro-ph/0608095}.

\bibitem[{\citenamefont{{Pettorino}}(2013)}]{2013PhRvD..88f3519P}
\bibinfo{author}{\bibfnamefont{V.}~\bibnamefont{{Pettorino}}},
  \bibinfo{journal}{\prd} \textbf{\bibinfo{volume}{88}}, \bibinfo{eid}{063519}
  (\bibinfo{year}{2013}), \eprint{1305.7457}.

\bibitem[{\citenamefont{Birrell and Davies}(1984)}]{BirrellDavies}
\bibinfo{author}{\bibfnamefont{N.~D.} \bibnamefont{Birrell}} \bibnamefont{and}
  \bibinfo{author}{\bibfnamefont{P.~C.~W.} \bibnamefont{Davies}},
  \emph{\bibinfo{title}{Quantum Fields in Curved Space}}, Cambridge Monographs
  on Mathematical Physics (\bibinfo{publisher}{Cambridge University Press},
  \bibinfo{year}{1984}).

\bibitem[{\citenamefont{Kofman et~al.}(1997)\citenamefont{Kofman, Linde, and
  Starobinsky}}]{Kofman:1997yn}
\bibinfo{author}{\bibfnamefont{L.}~\bibnamefont{Kofman}},
  \bibinfo{author}{\bibfnamefont{A.~D.} \bibnamefont{Linde}}, \bibnamefont{and}
  \bibinfo{author}{\bibfnamefont{A.~A.} \bibnamefont{Starobinsky}},
  \bibinfo{journal}{Phys.Rev.} \textbf{\bibinfo{volume}{D56}},
  \bibinfo{pages}{3258} (\bibinfo{year}{1997}), \eprint{hep-ph/9704452}.

\end{thebibliography}
\end{document}